\documentclass[mathleft
]{an}
\usepackage{graphicx}
\usepackage{times}
\def\msol{M$_{\odot}$ }
\def\kms{$\rm km\, s^{-1}$}
\def\cm3{$\rm cm^{-3}$}
\def\Ts{$\rm T_{*}$}
\def\Vs{$\rm V_{s}$}
\def\n0{$\rm n_{0}$}
\def\B0{$\rm B_{0}$}

\def\Fh{$\rm F_{h}$~}
\def\erg{$\rm erg\, cm^{-2}\, s^{-1}$}
\def\mum{$\mu$m}

\def\agr{a$_{gr}$}

\def\ff{$\it ff$}
\def\sf{\sqrt{\it ff$}}

\begin{document}

\Pagespan{1}{19}
\Yearpublication{}%
\Yearsubmission{}%
\Month{}%
\Volume{}%
\Issue{}%

\title{Comparison of dust-to-gas ratios in luminous,\\
 ultraluminous, and hyperluminous infrared galaxies}

\author{M. Contini\inst{1}\fnmsep\thanks{Corresponding author:
  \email{contini@post.tau.ac.il}\newline}
\and  T. Contini\inst{2}
}
\titlerunning{Dust-to-gas ratios in LIRGs,ULIRGs, and HLIRGs}
\authorrunning{M. Contini \& T. Contini}
\institute{
School of Physics and  Astronomy, Tel-Aviv University, Tel-Aviv, 69978 Israel
\and
Laboratoire d'Astrophysique de Toulouse et Tarbe (LA2T - UMR 5572), 
Observatoire Midi-Pyr\'en\'ees, 14 avenue E.Belin, F-31400 Toulouse, France}

\received{}
\accepted{}
\publonline{later}

\keywords{galaxies: active -  galaxies: starburst - infrared: galaxies - shock waves}

\abstract{%
The dust-to-gas ratios in three different samples of  luminous, ultraluminous, and
hyperluminous  infrared galaxies are calculated by modelling their
 radio to soft X-ray spectral energy distributions (SED)
 using composite models which account for the photoionizing radiation
from HII regions, starbursts, or AGNs, and for shocks.
The models are  limited to a set which broadly reproduces the mid-IR fine structure
line ratios of local, IR bright, starburst galaxies.
The results show that two types of clouds contribute to
the IR emission. Those characterized by low shock velocities and low preshock de
nsities
explain the far-IR dust emission, while those with higher  velocities and densities
contribute to the mid-IR dust  emission.
 Clouds with shock velocities of 500 \kms prevail in hyperluminous infrared galaxies.
 An AGN  is  found in nearly all of the ultraluminous  infrared galaxies  and in half of the luminous
infrared galaxies of the sample.
High IR luminosities depend on dust-to-gas ratios as high as   $\sim$  0.1  by mass, however
most hyperluminous IR galaxies show dust-to-gas ratios much  lower than those
calculated for the   luminous  and ultraluminous IR galaxies.
}

\maketitle

\section{Introduction}

Observations  in the infrared (IR) have revealed  
 galaxies with very high IR luminosities, up to L$_{IR}$ $\sim 10^{13}$ L$_{\odot}$ 
(Lutz et al. 1998; Laurent et al. 2000; Tran et al. 2001;  Verma et al. 2002).
 The origin of this high IR luminosity  
is under investigation (Genzel et al. 1998; Rowan-Robinson 1992, 2000; 
Rigopoulou et al. 1999; Farrah et al. 2002; Verma et al. 2003; etc.). 
The main questions  concern the type (such as HII galaxy, 
starburst nucleus galaxy, or active galactic nucleus (AGN)) of the  IR bright and luminous galaxies, 
the source of dust illumination and heating by a photoionizing flux and/or collisional 
processes, the composition of dust ranging from relatively large silicate grains 
to small graphite, PAH, ices, etc. (Spoon et al. 2002). 

Dust plays   an important role in the understanding of  the near and
distant Universe (Schneider, Ferrara \& Salvaterra 2004). 
Formation of grains in the ISM
is extremely inefficient, the preferred sites  being the atmospheres
of evolved low mass (M $<$ 8 M$_\odot$) stars from where it is transported into the ISM
through stellar winds (Whittet 1992). However, at high redshifts, e.g. $z >$ 5,
 the evolutionary time-scales
of low-mass stars ($0.1-1$ Gyr) start to be comparable with the age of the Universe.
Type II supernovae are the only known dust sources with evolutionary time-scales
shorter than the Hubble time. Dust enrichment of the ISM must have occurred primarily
on considerably shorter time-scales in the ejecta of supernova explosions.

In  the starburst environments  and in the  outflow
regions  of AGN,
dust  dominates the IR emission  by reprocessing radiation  from the primary source
(stars and/or an active nucleus).
Collisional processes which  were found to be so important
to explain the emission-line and continuum spectra 
(e.g. the strong high-ionization level line fluxes, the soft
X-ray emission, strong heating of dust, radio synchrotron emission, etc.)
from the nuclear and
circumnuclear regions of AGNs, starbursts and HII regions
(e.g. Contini \& Viegas 2000)   dominate the emitted radiation flux in mergers.
Particularly, it was demonstrated that collisional heating of dust grains by gas
in supersonic velocity regimes, leads to relatively high temperatures 
which could explain the IR  flux between 3 and 1000 \mum\
(the "IR bump") in the continuum spectral energy distribution (SED) of starburst galaxies
(Contini \& Contini 2003) and AGNs (Contini, Viegas \& Prieto 2004).

Our aim is to test  whether the IR luminosity  is
directly related to the  amount of dust relative to  gas, namely the dust-to-gas
ratio ($d/g$) which
can  be  considered as a key parameter in modelling galaxies 
 revealing  high dust formation  regions and  tracing  stellar evolution.

We investigate three
luminosity classes commonly used for IR luminous galaxies:  hyper-, ultra-, and luminous
infrared galaxies (HLIRGs, ULIRGs, and LIRGs, respectively) 
which correspond to
$>$ 10$^{13}$, $>$ 10$^{12}$ and $>$ 10$^{11}$ h$_{65}^{-2}$ L$_{\odot}$, 
according to the definition of Sanders \& Mirabel (1996). 
Three samples are chosen as  representative of the different
luminosity classes :  Farrah et al. (2002),
Rigopoulou et al (1999), and Verma  et al. (2003).
Only three galaxies from the Verma et al. (2003) sample
correspond to the definition of Sanders \& Mirabel, the other  showing lower IR
luminosities belong to the  IR bright galaxies.
However,  in the following we will refer to them together with the luminous IR galaxies
among the LIRGs.
The samples are briefly presented in Sect. 2.

The  galaxies in the different samples are   chosen  only because they are  IR luminous and not
by their  type (e.g. starburst, HII regions, AGN, etc) which, viceversa,
we will try to   find out  by modelling.
Galaxies are generally classified through
 the analysis of the SED of the continuum
and of the absorption and emission bands of  dust main components, e.g. silicates and PAHs
(Spoon et al. 2002), on top of it.
Diagnostic diagrams  are generally used to distinguish between AGN and star forming regions
on the basis of  PAH to continuum intensity ratios (Genzel et al. 1998, Peeters et al 2004),
of the ratio of the continuum flux at different
wavelengths (Genzel \& Cesarsky 2000), PAH to fine structure lines  (FSL)
(Genzel et al 1998) and FSL diagrams (Sturm et al 2002).
However, recent observations of mid-IR  AGN spectra  by 
 the {\it Spitzer} Space Telescope (Weedman et al. 2005)
found no spectral parameters  that unambiguously distinguish
AGNs and starbursts based only on the slopes of the continuum spectra.

The  IR luminosity of galaxies depends on both  dust  reprocessed  radiation
in the IR and  bremsstrahlung from cool gas.
Dust IR emission is  usually analysed independently of
 bremsstrahlung which  covers the whole radio -  X-ray frequency range.
Indeed, in the IR domain dust emission dominates, but  the modelling of the SED
requires the consistent calculation of gas and dust  spectra (Sect. 3).
Previous modelling of AGNs and starburst galaxies (Viegas, Contini \& Contini 1999;
Contini, Viegas \& Prieto 2004) has shown that many different conditions
 coexist in each galaxy. The observed continuum and emission-line
spectra account for all of them.
Modelling the continuum SED by multi-cloud models, we will be able to determine  the $d/g$ ratios
and the relative contribution
of the different mechanisms which are at work in the IR luminous "active" galaxies :  AGN,
starbursts, supernova remnants, etc.
 The results are presented in Sect. 4,  discussed in Sect. 5
and summarized in Sect. 6.

\section{The  samples}

\begin{table}
\caption{Characteristics of the galaxies}
\begin{tabular}{lllllll}\\ \hline \hline
        &type$^a$ &$d^b$&$ log L_{\rm IR}$ \\
        &    &  (Mpc)     &(L$_{\odot}$)  \\
\hline
LIRGs :&&&&\\
NGC 253  &HII& 3.27&10.23\\
IC 342   &HII&0.45&7.57\\
II Zw 40  &HII&10.5&9.3 \\
M82    &HII&2.7&10.18  \\
NGC 3256  &HII&37.95&11.4 \\
NGC 3690   &HII&41.64&11.53 \\
NGC 4038/9 &HII&21.9&10.65\\
NGC 4945   &HII,Sy2&7.47&10.66 \\
NGC 5236  &HII&6.92&10.11 \\
NGC 5253&HII&5.39&9.21 \\
NGC 7552   &HII,Liner&21.15&10.83   \\
ULIRGs : &&&&\\
Mrk 1014  &QSO&652.3&12.50       \\
UGC 5101    &IrS&160.&11.94     \\
Mrk 231   &IrS&168.8&12.48         \\
Mrk 273   &IrS&148.&12.07        \\
Arp 220    &IrS&72.&12.11        \\
NGC 6240  &IrS&96. &11.78        \\
HLIRGs :&&&&&&\\
F00235+1024 &nl&2320.& 13.04$^c$ \\
07380-2342 &nl & 1160.&13.34$^c$ \\
F10026+4949&Sy1& 4480.&13.81$^c$ \\
F12509+3122  &QSO& 3120.&13.26$^c$\\
13279+3401 &QSO& 1714. &12.88$^c$ \\
14026+4341&Sy1& 1280. &12.90$^c$\\
F14218+3845&QSO& 4840.&13.06$^c$ \\
F16124+3241 &nl&3381.  &13.02$^c$\\
EJ1640+41&QSO&4400.   &12.90$^c$\\
18216+6418 &Sy1&1200.  &13.14$^c$\\
\hline\\
\end{tabular}

\flushleft
$^a$  from Verma  et al. (2003) for LIRGs, Rigopoulou  et al. (1999) for ULIRGs, and from
Farrah et al. (2002) for HLIRGs.

$^b$ adopting H$_0$=75 \kms Mpc$^{-1}$

$^c$ 1-1000 \mum ~ luminosities obtained from the best-fitting combined starburst-AGN 
models (Farrah et al. 2002, table 3)
   
\end{table}

The sample of Verma et al. (2003)  contains starburst regions that are optically
obscured, therefore similar to dense star forming regions in ULIRGs and HLIRGs.
The IR luminosities are  between  10$^9$ and  10$^{12}$ L$_{\odot}$,   except for IC 342
corresponding to 3.74 10$^7$ L$_{\odot}$ which will be used for comparison.
 We will  check whether  an
active nucleus  cannot be  excluded in some of these starburst galaxies,
 leading to  hybrid types.

ULIRGs 
dominate the local luminosity function of galaxies with L$>$ 10$^{12} L_{\odot}$
(Sanders \& Mirabel 1996 and references therein).
 Most of their
bolometric luminosity is emitted in the far-IR (8-1000 \mum) (Rigopoulou et al. 1999)
and they show more prominent FIR emission than either starburst galaxies or AGNs
(Peeters et al. 2004). Most ULIRGs are very rich in dust and gas, and they are
interacting systems with distorted morphologies. Their optical spectra mimic
those of Seyfert galaxies (Rigopoulou et al. 1999, Lutz et al. 1999). 
The MIR spectra of ULIRGs
are more complicated than the spectra of other galaxies due to the presence of
copious amount of dust in their nuclear regions (Genzel et al. 1998), leading to
strong absorption features which may distort any emission component (Sturm et al. 2002).

Infrared, millimeter and radio characteristics of ULIRGs are  similar to those
of starburst galaxies, while nuclear optical emission-line spectra are  similar
to those of Seyfert galaxies. A central AGN and circumnuclear star formation may
both play a role in the FIR emission (Rigopoulou et al 1999). 
At intermediate luminosities, some ULIRG spectra
are PAH dominated, while others show signs of a 8 \mum\  broad peak, indicating
the presence of deeply dust-enshrouded ionizing sources (Rigopoulou et al. 1999). 
On the other hand, the spectra
at high FIR luminosities bear close resemblance to AGN hot dust continua, showing
little or no sign of PAH features (Clavel et al. 2000). Of the ultraluminous IRAS
 selected in Sect. 3,
galaxies, 70-80\% are predominantly powered by recently formed massive stars,
20-30\% by AGN (Genzel et al. 1998; Laurent et al. 2000).
From multi-band HST imaging data Borne et al. (2001)  find that  ULIRGs
show a significant population of very bright knots, which are most likely
produced during the starburst/merger event.

At least some HST observations (Farrah et al. 2002) show that  a small fraction 
of HLIRGs are merging
galaxies. Mergers between evolved galaxies trigger dust enshrouded starburst and
AGN activity (Sanders \& Mirabel 1996). Alternatively, HLIRGs may be very young
or even ``primeval'' galaxies (Rowan-Robinson 2000, Farrah  et al. 2002). Rowan-Robinson  argues
that the bulk of emission in HLIRGs is due to starburst activity, implying star
formation rates $>$ 1000 \msol yr$^{-1}$. Verma et al. (2002) and Farrah et al (2002) claim that both
starburst and AGN activities are required to explain the total IR emission,
the average starburst fraction being 35\%, within a range spanning from 80\%
starburst-dominated to 80\% AGN-dominated.

The types, the distances from Earth
calculated  adopting H$_0$ = 75 \kms Mpc$^{-1}$, and the
luminosities of the galaxies in the different samples are given
in Table 1. 

\begin{figure*}
\centering
\includegraphics[width=75mm]{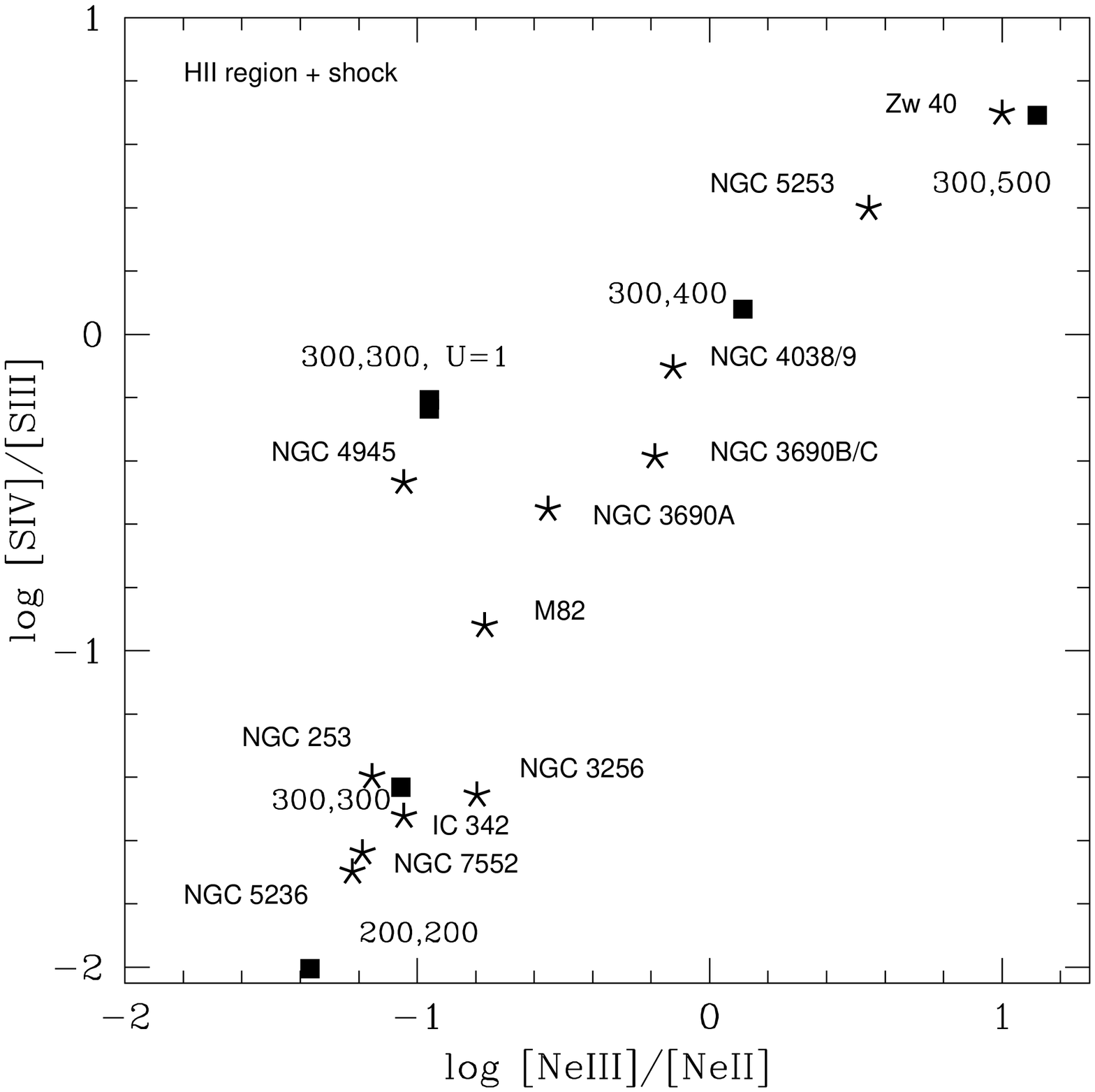}
\includegraphics[width=75mm]{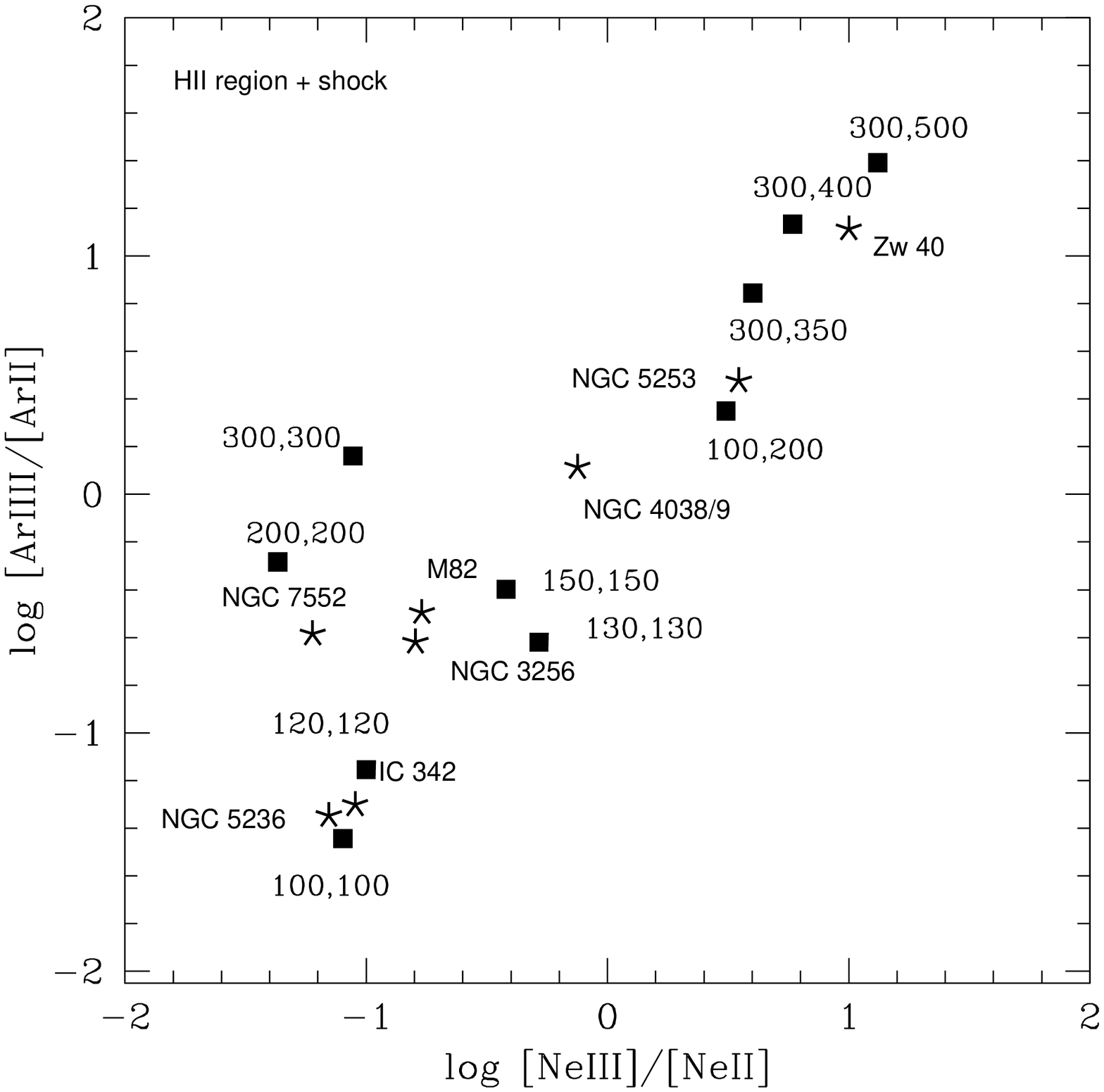}   
\includegraphics[width=75mm]{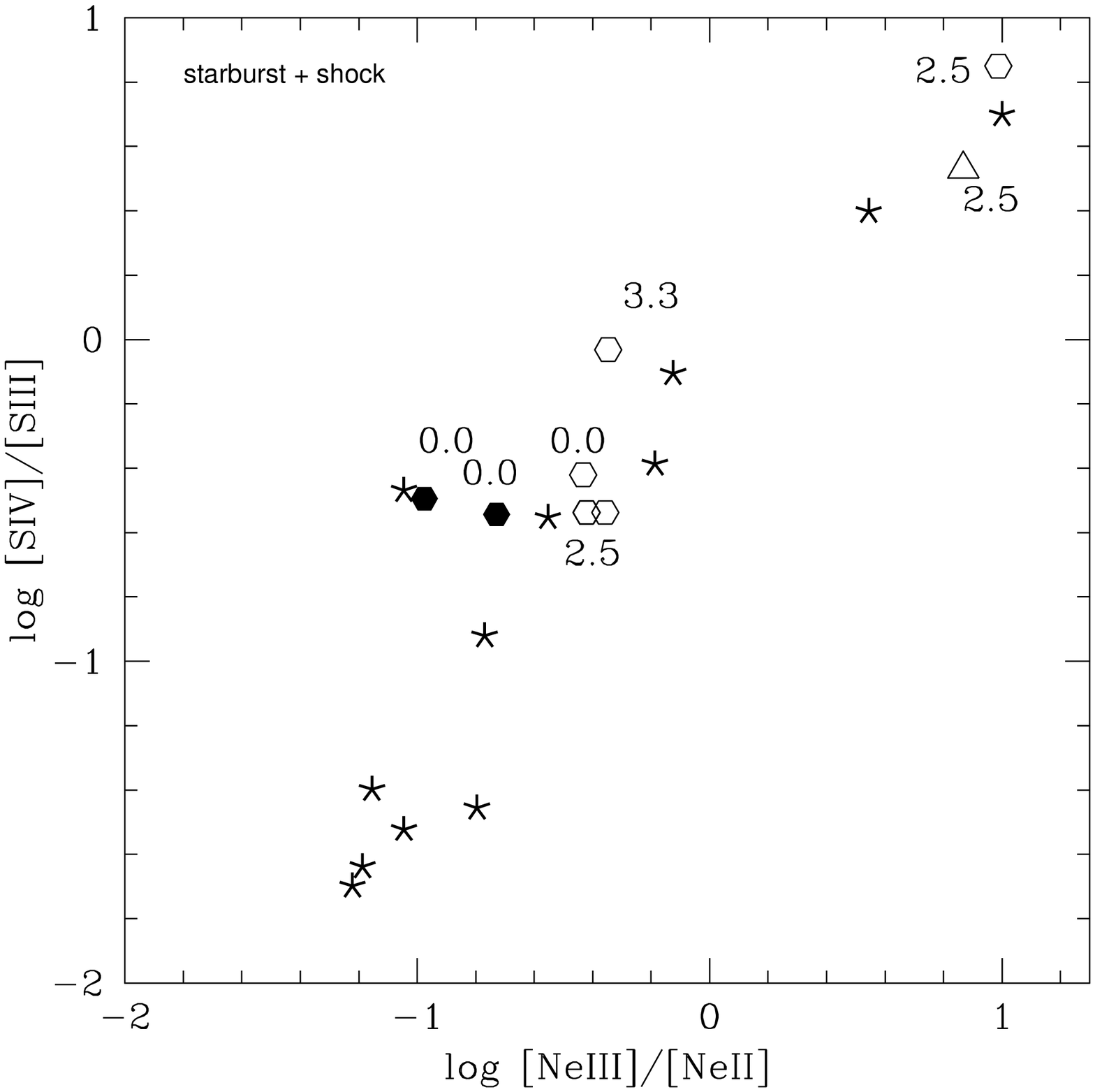} 
\includegraphics[width=75mm]{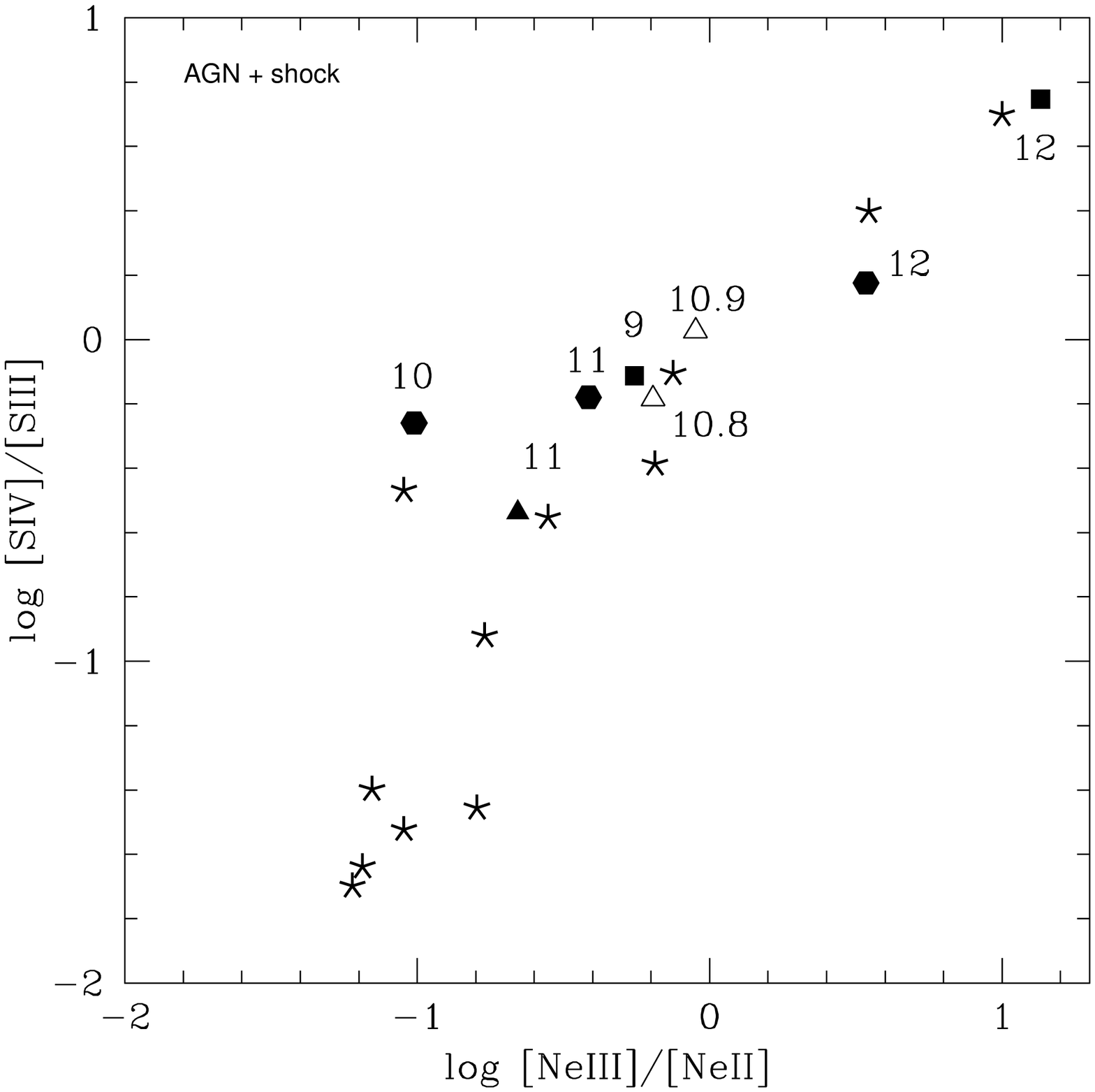}    
\caption
{{\it Top left:} comparison of observed line ratios [SIV]/[SIII] vs. [NeIII]/[NeII]
(asterisks) with models for HII regions + shocks (filled squares). Values of \Vs\ and
\n0\ are indicated near each model.
{\it Top right:} comparison of observed line ratios [ArIII]/[ArII] vs. [NeIII]/NeII]
with models for HII regions + shocks. Same symbols as for top left diagram.
{\it Bottom left:} comparison of observed line ratios [SIV]/[SIII] vs. [NeIII]/[NeII]
(asterisks) with models for starbursts + shocks.
Numbers refer to the starburst age in Myr.
Open symbols represent geometrically
thin clouds with $D=0.03$ pc, filled  symbols  $D=3$ pc; hexagons correspond to
\Vs\ = 300 \kms, \n0\ = 300 \cm3, triangles to \Vs\ = 200 \kms, \n0\ = 200 \cm3.
{\it Bottom right:} comparison of observed line ratios [SIV]/[SIII] vs. [NeIII]/[NeII]
(asterisks) with models for AGN + shocks.
 Numbers represent log \Fh.
filled triangles: \Vs\ =200 \kms, \n0\ = 200, $D=3$ pc;
empty triangles: \Vs\ =200 \kms, \n0\ = 200, $D=0.3$ pc;
black hexagons: \Vs\ =300 \kms, \n0\ = 300, $D=3$ pc;
black squares: \Vs\ =500 \kms, \n0\ = 300, $D=3$ pc.
All models except where indicated correspond to U=10.}
\end{figure*}

\section{Modelling the spectra of starburst galaxies}

We start by modelling the IR emission-line  spectra observed by
Verma et al. (2003) for a sample of starburst galaxies with relatively high IR
fluxes and select the models (Sect. 3.1)
which best explain the trend of the line ratios (Sect. 3.2), because
line spectra  are far more constraining than the SEDs.
So we  derive the grid of models  suitable to model the SEDs (Sect. 3.3)
by the method of
Contini \& Contini (2003) and Contini, Viegas \& Prieto (2004) for different samples
of luminous IR galaxies, and AGNs, respectively.
Namely, gas and dust are  heated and ionized consistently by  photoionization
and  shocks (Sect. 3.4), so
the ratio of  dust reprocessed radiation  to  bremsstrahlung in the IR depends
on $d/g$ (Sect. 3.5).

\subsection{The models}

\begin{table*}
\centering
\caption{Single-cloud  models (top), the fluxes  (in \erg) calculated in the different ranges
(middle) and the maximum temperature of the grains (bottom); 
\Vs : the shock velocity, \n0 : the preshock density,
$D$ : the geometrical thickness of the cloud, $U$ : the ionization parameter,
$d/g$ : the dust-to-gas ratio, \Fh : the  ionization  power-law ($\alpha$=1.5)
flux (in number of photons cm$^{-2}$ s$^{-1}$ eV$^{-1}$ at 1 Ryd),
 $F_{radio_{br}}$  : the  radio flux calculated  by bremsstrahlung,
 $F_{IR_{br}}$  : the IR flux calculated  by bremsstrahlung,
 $F_{opt_{br}}$  : the  optical flux calculated  by bremsstrahlung,
 $F_{UV_{br}}$  : the UV flux calculated  by bremsstrahlung,
 $F_{IR_d}$ : the IR flux calculated  by dust,
 $T_{d}$(max) (K) : the maximum temperature of grains.
 \Ts=10$^4$ K is used for models m1-m5.
}
\begin{tabular}{llllllllll}\\ \hline\hline
 &\multicolumn{2}{|c|}{\em under- gal}&\multicolumn{4}{c|}{\em starbursts}&\multicolumn{1}{c|}{\em winds}&
\multicolumn{1}{c|}{\em AGN}\\ \hline
        &m0&  m1  & m2& m3&m4     &m5 & m6 &m7$^a$&   \\
\hline
\ \Vs (\kms)    &100& 150 & 200& 250 & 300 & 500 & 1000&1000    \\
\ \n0 (\cm3)   &1000& 100 & 200& 300 & 300 & 300& 300 &1000\\
\ $D$ (10$^{19}$ cm)&1.e-4&1.e-4& 1. & 1.  &0.5& 1.& 0.01&0.1 \\
\ $U$             & - &  0.1  & 10 & 10 & 10 & 10& -&log\Fh =11\\
\ $d/g$ (0.04)$^b$&0.01& 0.003  & 1.& 1.& 1.& 1.&1.&0.1\\
\hline
\ $F_{radio_{br}} $ & 7.6e-6 & 3.2e-9 & 0.023 & 0.14 & 0.037 & 0.22 & 6.3e-6 & 6.6\\
\ $F_{IR_{dust}}$ & 0.01& 3.2e-4 & 1.5e3 & 5.8e3 & 694. & 1.23e4 & 819.& 3.6e4\\
\ $F_{IR_{br}}$ & 5.4e-3 & 5.8e-6 & 13.2 & 71. & 32.4 & 100.& 0.01 & 6.0e3\\
\ $F_{opt_{br}}$ &  2.3e-4 & 5.5e-6 & 1.7 & 2.9 & 14.2 & 0.2 & 0.015 & 2.7e3\\
\ $F_{UV_{br}}$ & 6.e-4 & 2.e-5 & 0.195 & 0.14 & 11.6 & 31. & 0.076 & 4.9e3\\
\ $F_{X_{br}}$&-     & 4.3e-7 & 2.2e-3 & 2.6e-3 & 0.018 & 1.& 8.27 & 1.36\\
\hline
\ $T_{d}$(max) (K)  & 63.2   & 51.5&  50.   & 82.  & 90. & 119. & 174. & 155.\\
\hline

\end{tabular}

\flushleft
$^a$ calculated with $a_{gr}$= 1 \mum ;
$^b$ by mass 

\end{table*}

\begin{figure*}
\centering
\includegraphics[width=56mm]{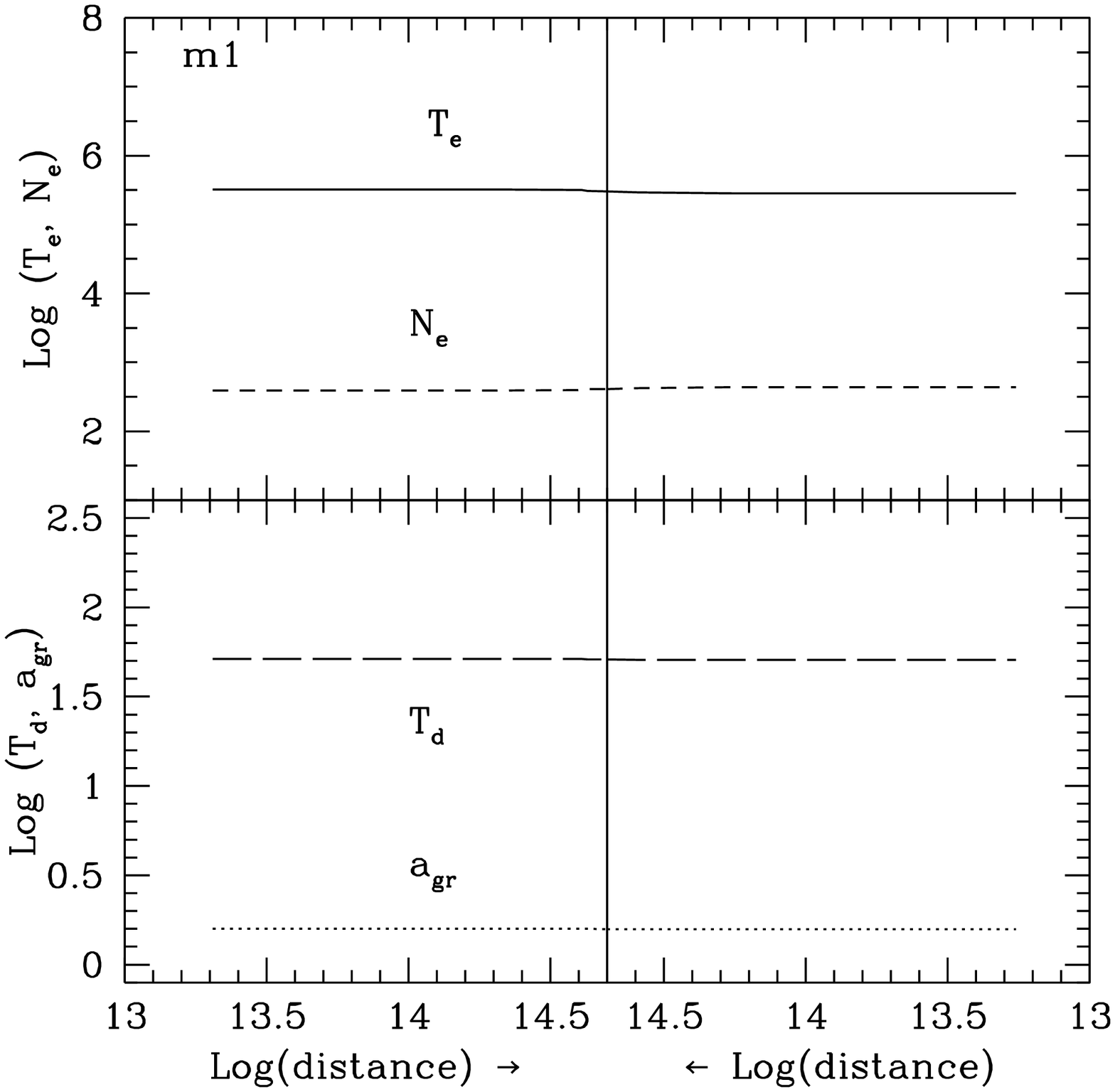} 
\includegraphics[width=56mm]{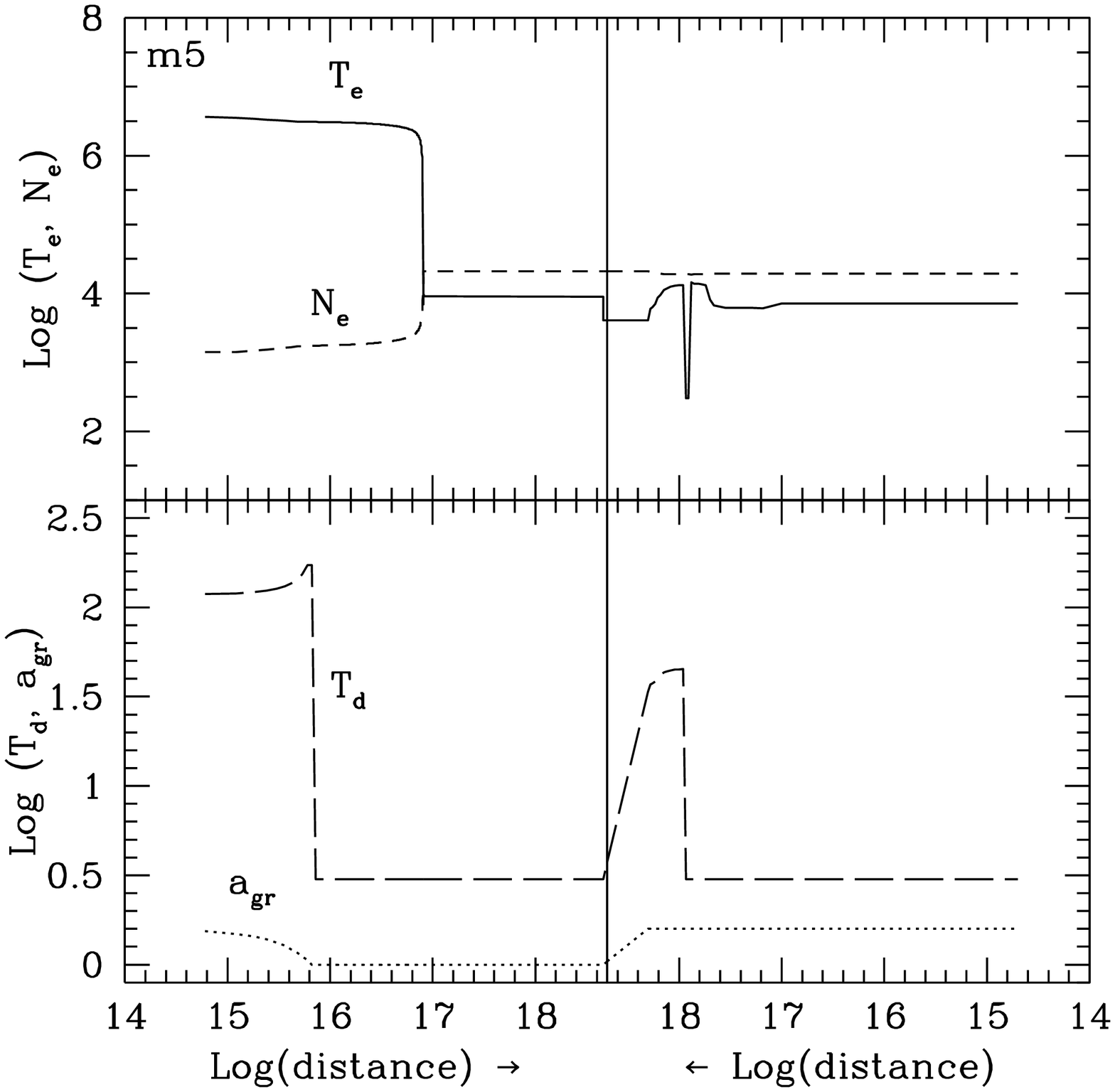}
\includegraphics[width=56mm]{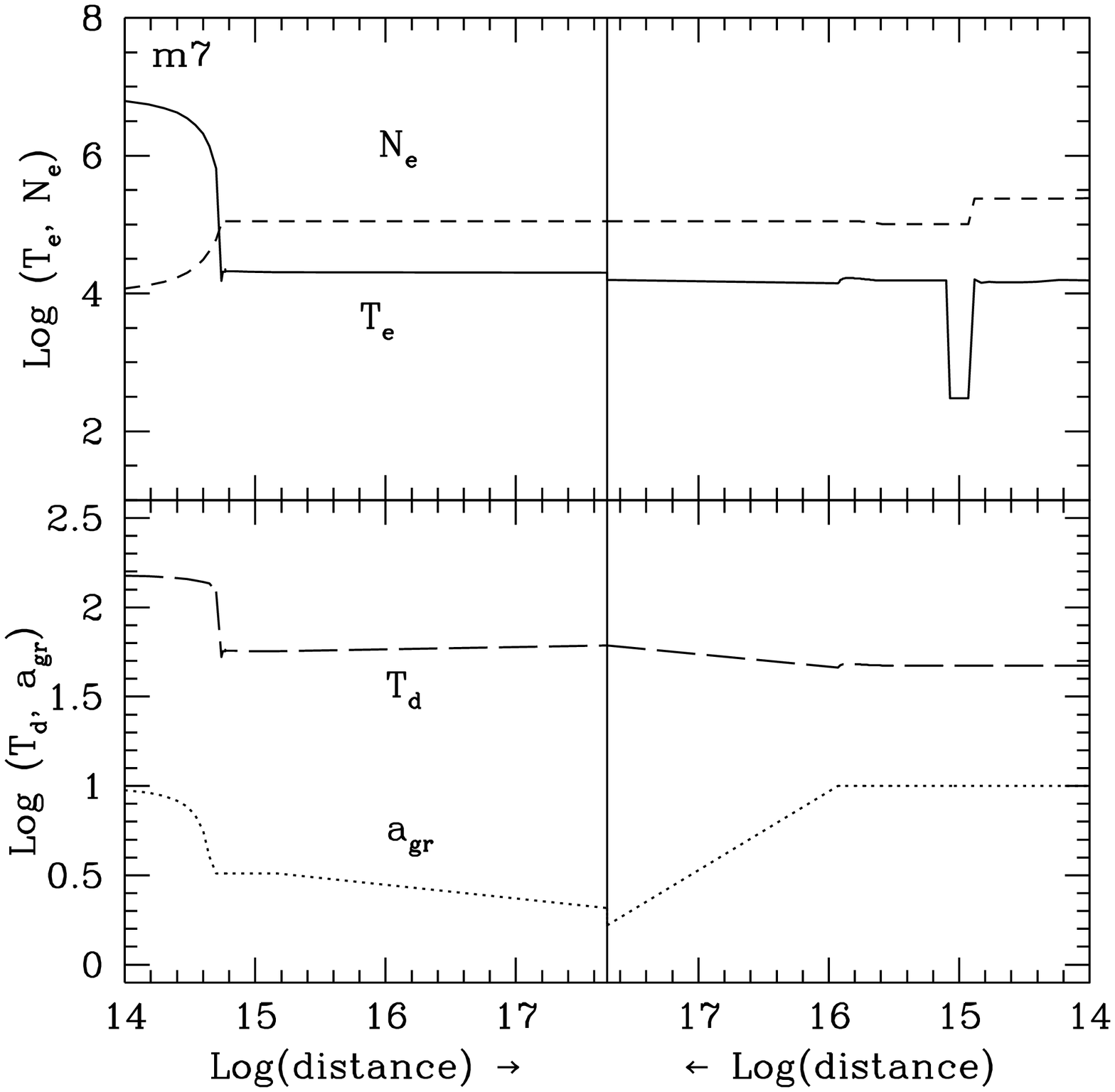}
\caption
{The profile of the electron temperature, $T_{\rm e}$,
of the electron density, $N_{\rm e}$, of the grain temperature, $T_{\rm d}$,
and of the grain radius, $a_{\rm gr}$, throughout a cloud for
models m1 ({\it left}), m5 ({\it middle}), and m7 ({\it right}).
}
\end{figure*}

We use composite models which
account consistently for photoionization from an external 
source and for shocks, in a plane-parallel geometry 
(see Contini \& Contini 2003, Contini et.al. 2004).  
 Richtmyer-Meshkov instability between two fluids
at different densities driven by shock waves
and  Kelvin-Helmholtz instability  lead to  perturbed circular material which  is
broken  up into   relatively small filaments.
(Graham \& Zhang 2000, Contini \& Formiggini 2001). 
The code SUMA (Viegas \& Contini 1994,
Contini \& Contini 2003,  Contini et. al. 2004, and references therein).
is adopted for the calculations of the spectra. 
 SUMA includes  shock hydrodynamics
 which implies  a leading role  of collisional
processes, as well as  excitation and destruction of  dust grains.

Other codes
deal with dust emission in the IR,  in particular  GRASIL
(Silva et al. 1998) and MAPPINGS (Groves et al. 2006 and references therein).
GRASIL predicts time dependent SEDs of galaxies
from X-ray to radio, including the effects of age selective extinction
of stellar populations.
 The $d/g$ ratio, assumed to scale linearly
with the metallicity of the residual gas, 
 determines the optical depth of the clouds.
MAPPINGS adopts stochastic quantum heating of dust
and a power-law size distribution of grains  with index arising from
shattering. 

In our models the photoionizing flux from an AGN, \Fh
(in photons cm$^{-2}$ s$^{-1}$ eV$^{-1}$ at 1 Ryd), corresponds to a power-law.
For starburst galaxies, the radiation flux from a star cluster of  age ($t$)
is characterized by the ionization parameter $U$. For HII regions, the radiation
flux is a black body corresponding to the stellar colour temperature \Ts.
The shocks, characterized by the shock velocity, \Vs, the preshock density, \n0, and the
preshock magnetic field, \B0\ (= 10$^{-4}$ gauss for all models), are created by collision
of the ejected matter with the interstellar clouds.
The  abundance of He, C, N, O, Ne, Mg, Si, S, A, Cl, and Fe relative to H and
the dust-to-gas ratio, $d/g$ are also input parameters.

The clouds move outwards from the galaxy centre in the NLR of AGN
(see e.g. Heckman et al. 1981),
and in the neighbourhood of  starburts, therefore,
the photoionizing flux and the shock act on the opposite edges of the clouds
whose  geometrical thickness, $D$,  plays an important role in  model results.
The structure of the gas downstream is determined by the shock.
 Grains are sputtered 
depending on  \Vs and  \n0 (Contini et al. 2004). 
The sputtering rate  determines the distribution of the grain radius throughout a cloud.
 An initial grain radius $a_{\rm gr}$ = 0.2 \mum\ is generally adopted.
Small grains ($\leq$ 0.2 \mum)  deriving from  sputtering of
larger grains contribute in the integrated continuum SED.

It was found (Schneider et al. 2004) that for all but the smallest progenitor stars,
silicate grains are the dominant dust compound. The characteristic grain sizes range
between 10$^{-3}$ \mum\ and a few $\times$ 0.1 \mum, depending on the grain species
and on the mass of the stellar progenitor. 
Very small grains are destroyed in regions of high UV radiation intensity (Rowan-Robinson 1992),
so the  maximum grain temperature which would account  for the near-IR emission
is limited by sublimation. 
Most of the grain survive in the narrow line region (NLR), where the
 the cloud velocities  are $\sim$  300-1000 \kms, while
small grains ($<$ 0.1 \mum) are easily
sputtered for shock velocities $\geq$ 200 \kms. 

PAHs are responsible for no
more than 22\% of the total 3-1100 \mum ~emission  in normal galaxies (Dale et al. 2001).
In the  shock dominated  starburst and AGN regions,  small grains such as graphite and 
PAH  are  rapidly destroyed by sputtering,
leading to an even lower percentage.
Moreover, a strong  radiation flux  dilutes the PAHs which are hardly seen in AGN
(Sturm et al. 2000, 2002 and Laurent et al. 2000).
 PAH,  with radius $<$0.01
\mum\ survive only in the extended  "cirrus" regions of   starburst galaxies
(see Contini \& Contini 2003), and very seldom in the NLR of AGNs.
Stochastic heating leads to temperature fluctuations around 20 K  of  \agr =200 \AA
~grains, but small grains (\agr= 25 \AA) can reach 50 K  and even
1000 K for \agr=5 \AA (Draine 2003).
However, their  contribution to the IR is negligible because
the cirrus region  is located
in the edges of the extended underlying galaxy, where  $d/g$ is even lower than
in the  ISM.
So only silicate  are accounted for in our models.

\subsection{The line ratios}

\begin{figure}
\includegraphics[width=78mm]{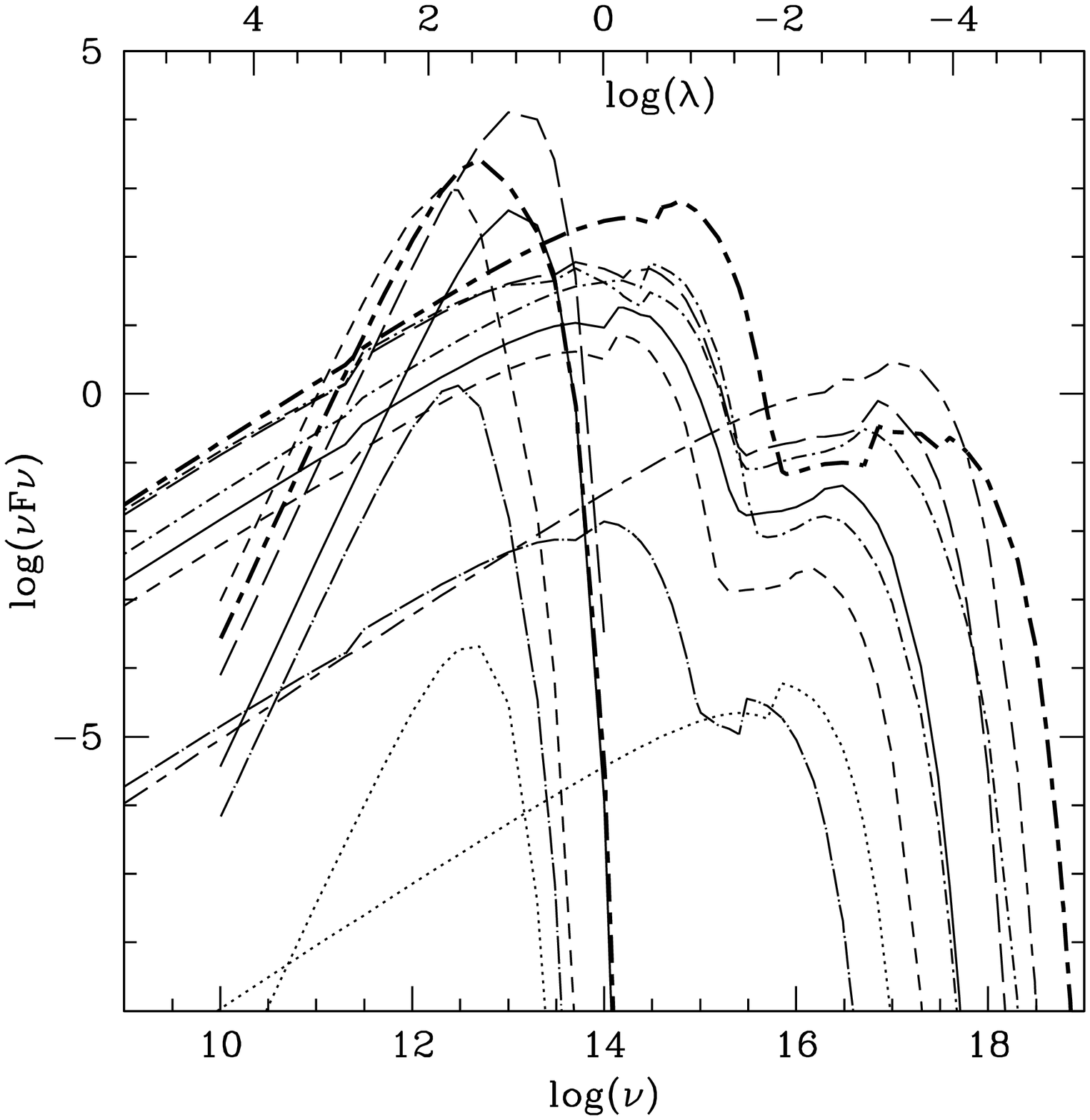} 

\caption{Comparison of the model SEDs :
 m0 (long dash - dotted),
m1 (dotted lines),
m2 (short dashed), m3 (dot - short dashed), m4 (solid),
m5 (long dashed), m6 (thin long dash - short dashed),
and m7 (thick long dash - short dashed).
}

\end{figure}

\begin{figure}
\includegraphics[width=78mm]{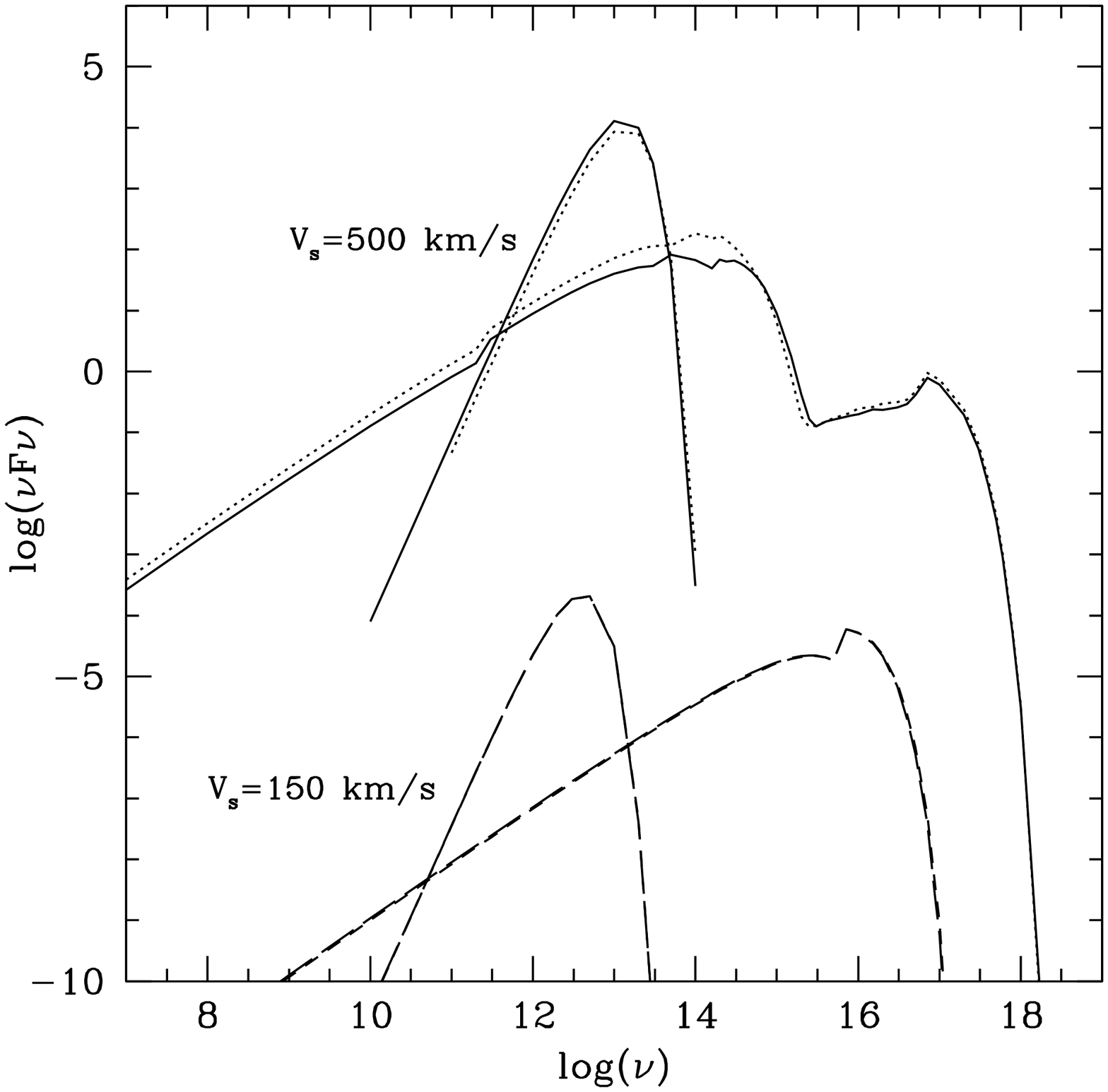}
\caption{Comparison of continua calculated using models which account
i) for a black-body flux with $U=10$ (long-dashed lines for \Vs\ = 150 \kms,
and solid lines for \Vs\ = 500 \kms), and ii) for a power-law flux with log(\Fh) = 10
(short-dashed lines for \Vs\ = 150 \kms and dotted lines for \Vs\ = 500 \kms)
}
\end{figure}

Verma et al. (2003) present ISO Short Wavelength Spectrometer (SWS) spectra 
between 2.38 and 45 \mum\ for 12 starburst regions located in eleven galaxies
with 10$^9$ L$_{\odot}$   $<$ L$_{\rm IR}$ 10$^{12}$ L$_{\odot}$.  
 Radio and infrared measurements demonstrate 
that some of the most active star-forming regions are optically obscured, 
with most of the bolometric luminosity emerging in the IR.

We analyse the spectra observed by Verma et al. (2003, table 2) using
 the grids of models presented by Contini \& Viegas (2001a,b).
We  focus on  the   ratios of MIR fine structure line transitions
 relative to  Ne, S, and Ar.
These elements, in fact,   appear in at least two ionization levels.
 C, Si, and Fe etc., which are important in tracing dusty regions,
as  they  can be trapped into dust grains and depleted from the gaseous phase,
correspond to  only  one  level.

We have chosen  line ratios belonging to the same element
and   corresponding 
to different ionization levels,  which do not depend on the relative abundances,
but only on the physical parameters.
Indeed the relative abundances of the most coolant elements can influence
the line ratios calculated with a given model, changing the distribution of the
ion fractional abundance downstream. However, we are dealing
here with neon, sulfur, and argon which are not important coolants.
Moreover, Verma et al. (2003) conclude that Ne and Ar are overabundant
by no more than a factor of $\sim$3 relative to the solar ones, while S 
is underabundant by a similar factor (see also Contini \& Contini 2003).
We adopt  cosmic abundances (Allen 1973).

The [SIV]10.5/[SIII]18.7 versus 
[NeIII]15.55/[NeII]12.8
emission-line ratios observed in starburst galaxies are compared  in Fig. 1 (top left) with models 
calculated using a black body radiation flux which corresponds to HII regions and shocks.
The models which best reproduce the data  were calculated (Contini \& Viegas 2001b) 
with a stellar colour temperatures
\Ts=10,000 K, a ionization parameter $U=10$, and a geometrical thickness of 3 pc.
The relatively high $U$ indicates that the emitting clouds are close to the stars. 
Only NGC 4945 is explained by lower fluxes ($U \sim$ 1).
The observed line ratios (asterisks) follow the trend of models with increasing \n0\ and \Vs\,
which are indicated near each model.
Higher \Vs\ in young starbursts, e.g. for II Zw 40, can be related to strong stellar winds 
produced by young massive stars (e.g. Wolf-Rayet stars).

In Fig. 1  (bottom left), the [SIV] 10.5 /[SIII]18.7 versus 
[NeIII] 15.55 /[NeII] 12.8 observed
line ratios  are  compared  with models (Contini \& Viegas 2001b) 
which  take into  account black-body radiation from  stars 
with a temperature distribution corresponding to that of a cluster  with a certain age.
The emitting gas is ionized and heated by radiation from   the stars and by  shocks.
Only galaxies showing [SIV]/[SIII] $>$ 0.15
are  explained by  starburst models. The  emitting clouds  are 
mosty  geometrically thin ($D \leq 0.03$ pc), and 
the best fitting models were calculated with $U \leq 0.1$, indicating that the  clouds
 are rather far  from the radiation source, and/or that the
sources are relatively weak. 
The starburst ages in Myr are indicated near each model. 
 Galaxies with line ratios $>$ 0.1 correspond generally to young ages ($t=0.0-2.5$ Myr),
with a maximum of $t=3.3$ Myr for NGC 4038/4039.
However, there is no clear trend between the observed line ratios and starburst ages or 
shock velocities. The shock velocities are generally of the order of  200-300 \kms.

In Fig. 1 (bottom  right) the galaxy sample is compared with models which account for
radiation from an AGN and shocks. Also in these case the galaxies with
low [SIV]/[SIII]  cannot be explained by the models,  while
there is a rough trend of higher line ratios  corresponding to higher flux intensity, \Fh\ 
from the active centre.

Fig. 1 (top  left) shows that models calculated for HII regions explain the line ratios of
all the sample galaxies. For sake of consistency, we have applied the same models
to [ArIII]8.99 / [ArII]6.99 versus [NeIII] 15.5 /[NeII]12.8 diagrams (Fig. 1 top right).
Indeed, the models  fit the observational data well.
The distribution of the galaxies in the diagram is not very different from
that corresponding to [SIV]/[SIII] versus [NeIII]/[NeII], and can
be explained by the  same  shock velocity trend.

 The  starburst galaxies observed by Verma et al. (2003) 
also appear in the sample of  Lutz at al (1998).
In a previous analysis by Viegas, Contini, \& Contini  (1999),
the modelling of II Zw 40  and NGC 5253 led to gas with low  \Vs\ and \n0, in agreement with their young
age,  on the basis of
high [OIV]/([NeII]+0.44[NeIII]) versus [NeIII]/[NeII] line ratios.
However, in the analysis of   
[SIV]/[SIII], we find that the same two galaxies
are explained by HII region  clouds with relatively high \Vs\ (500 \kms),
 by  clouds in starbursts with an age of $\sim$ 2.5 Myr and  velocities of 200-300 \kms,
and by  clouds in  AGNs  with velocities of $\sim$ 300 \kms. 

Recall that ionization potentials of sulphur are  lower than  those of the
corresponding levels of oxygen, therefore the stratification downstream
of  S$^{+3}$ and  S$^{+2}$  ions  is different than  that of  O$^{+3}$, Ne$^{+2}$, and  Ne$^{+1}$.
The  ionization potentials of sulphur are also  lower than  those of the
corresponding levels of argon, therefore  the stratification downstream
of the   S$^{+3}$ and  S$^{+2}$  ions corresponds roughly to  that of  Ar$^{+2}$ and
 Ar$^{+1}$.  Therefore, the two top diagrams of Fig. 1 show similar results,
which are reasonably different from those obtained from the [OIV]/([NeII]+0.44[NeIII]) analysis.
 This indicates that 
 clouds in different physical conditions contribute to the different lines, e.g. [SIV] and [OIV],
with various importance.
Multi-cloud models  were indeed  suggested by Viegas et al. (1999, table 2) by modelling
 other galaxies of the sample, M 82, NGC 3256, and NGC 253, on the basis 
of infrared line ratios.
Particularly,  for M 82 the contribution of the
models  corresponding to  \Vs=400 \kms and 100 \kms   is  0.82 and 0.13
to   the [OIV]25.9 line, but it is  0.40 and 0.45, respectively, to  [NeIII] 15.5. 

Summarizing, the lines which appear in Fig. 1 diagrams are emitted by clouds
located near the
stars, which are characterized by  temperatures of  \Ts\ $\sim 10\,000$ K. 
Model results   show   that   HII regions explain the
[SIV]/[SIII] line ratios for all the galaxies of the sample,
in agreement with the galaxy spectral types given by Verma et al (2003, Table 1). 
Cluster of stars   located in different regions of the galaxies
 also contribute, as well as an active nucleus  in some galaxies.
More particularly, the galaxies of the Verma et al. (2003) sample show that II Zw 40,
NGC 5253, NGC 4038/4039, NGC 3690A and NGC 3690 B/C present a multiple
nature of HII regions, AGN, and starburst galaxies. 

\begin{figure*}
\centering
\includegraphics[width=56mm]{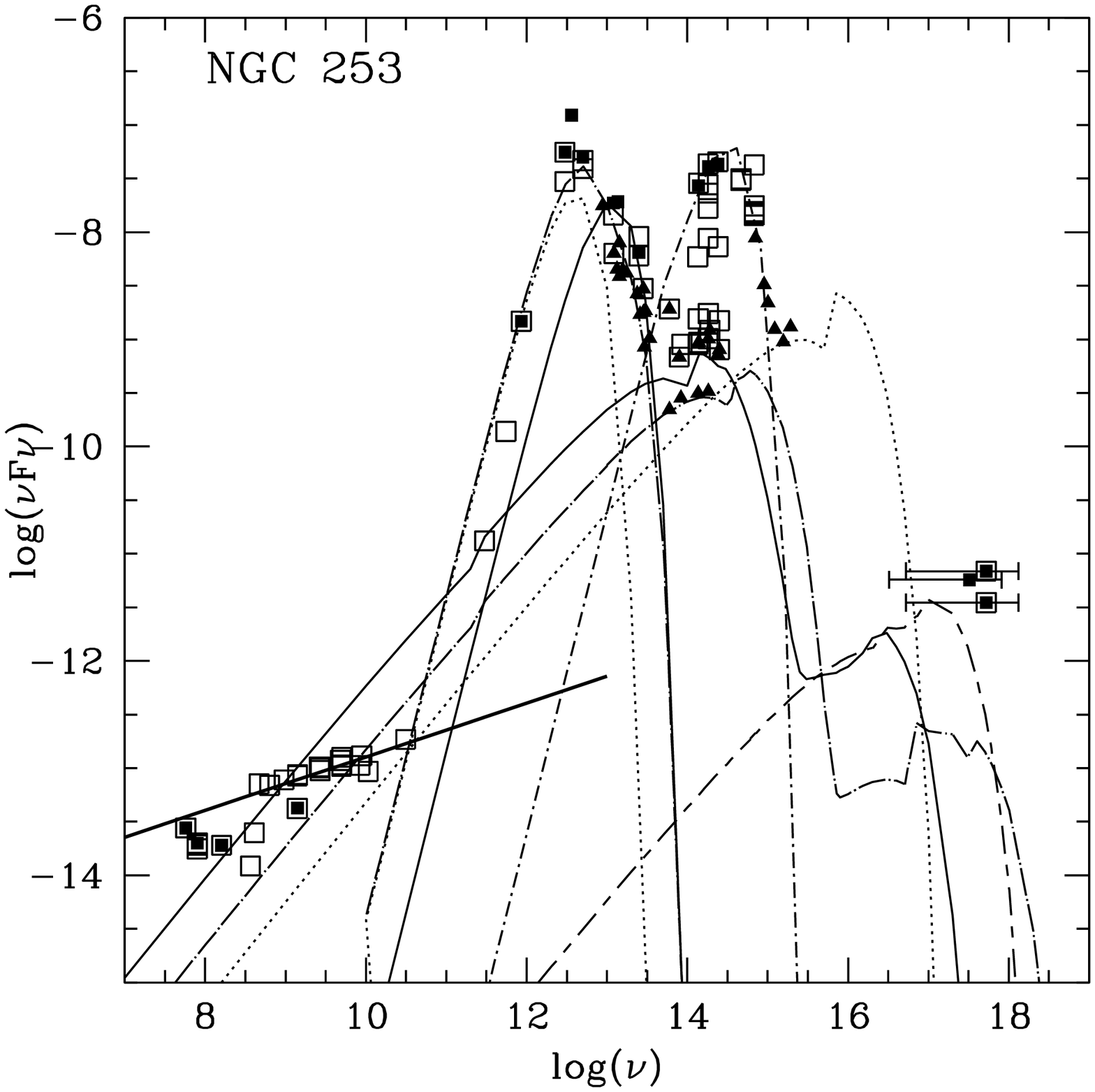}    
\includegraphics[width=56mm]{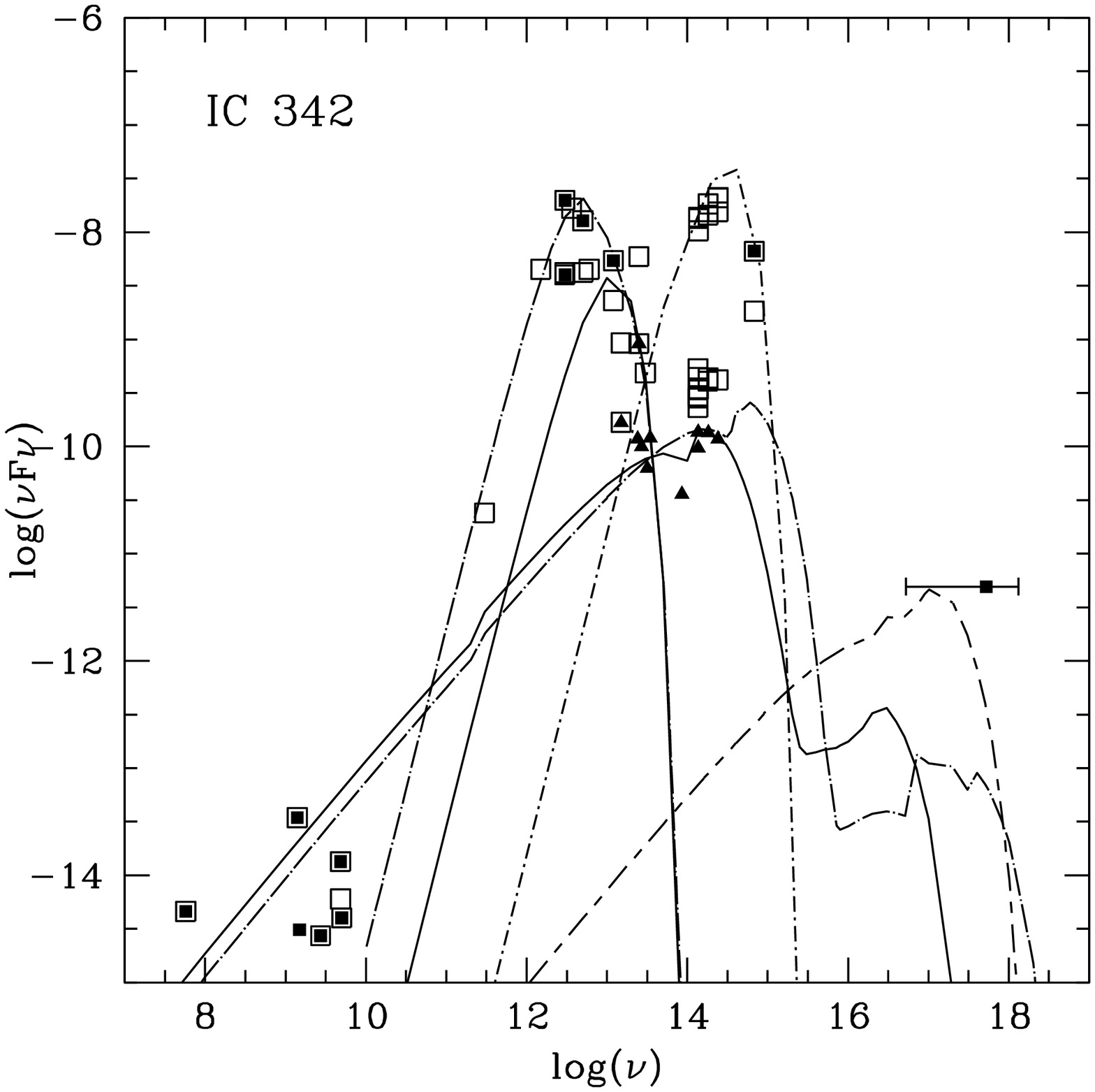}   
\includegraphics[width=56mm]{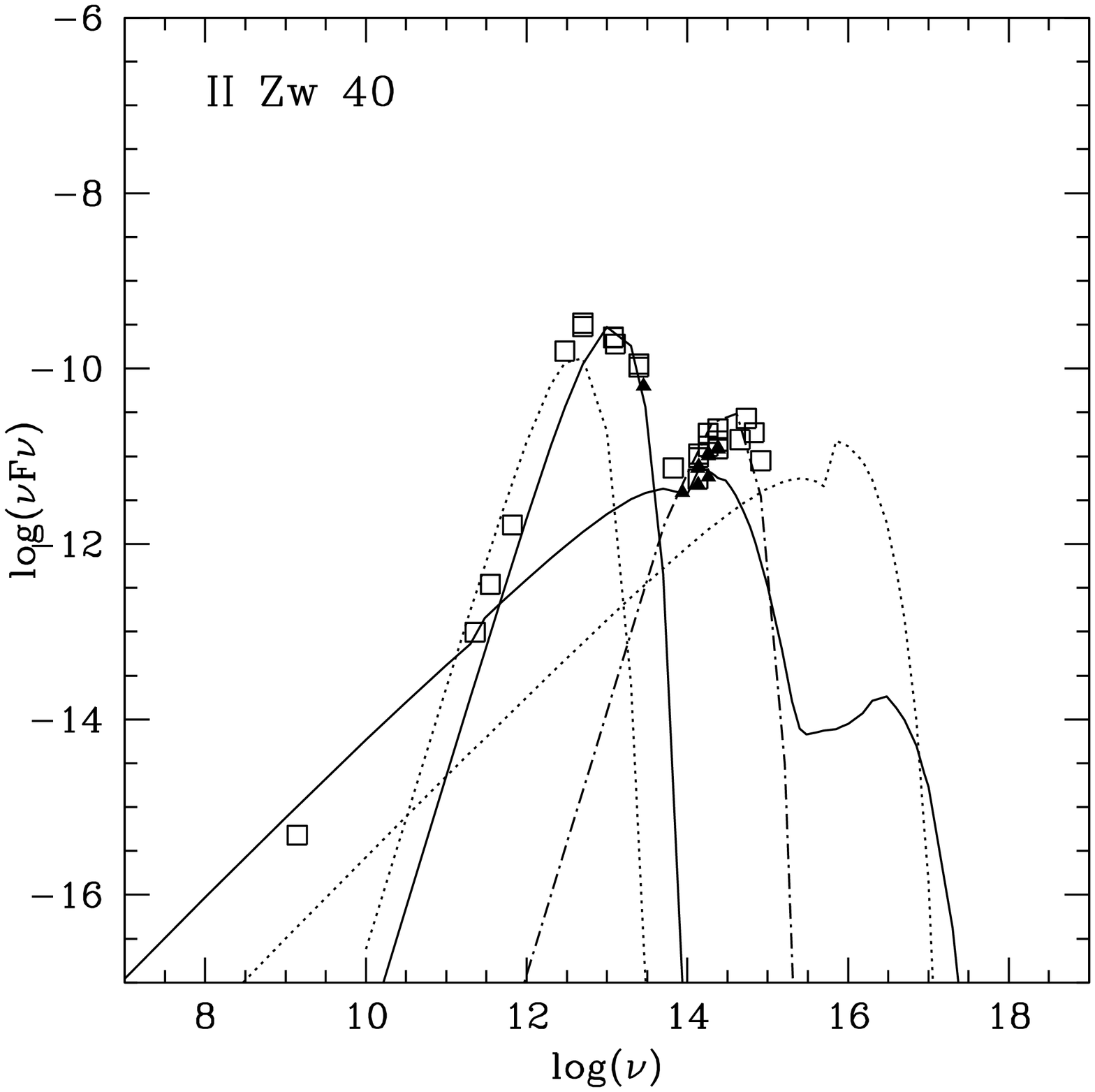}    
\includegraphics[width=56mm]{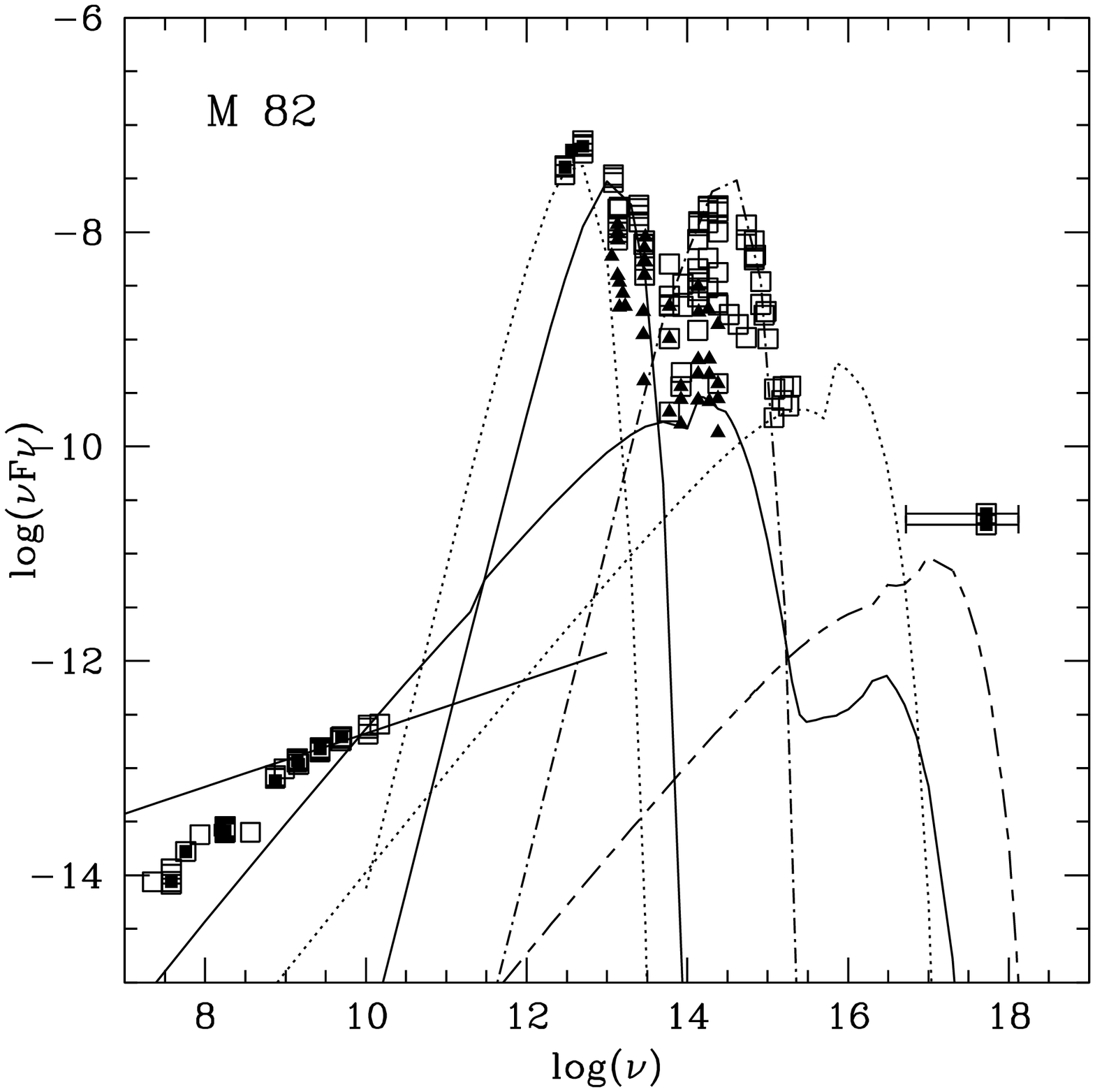} 
\includegraphics[width=56mm]{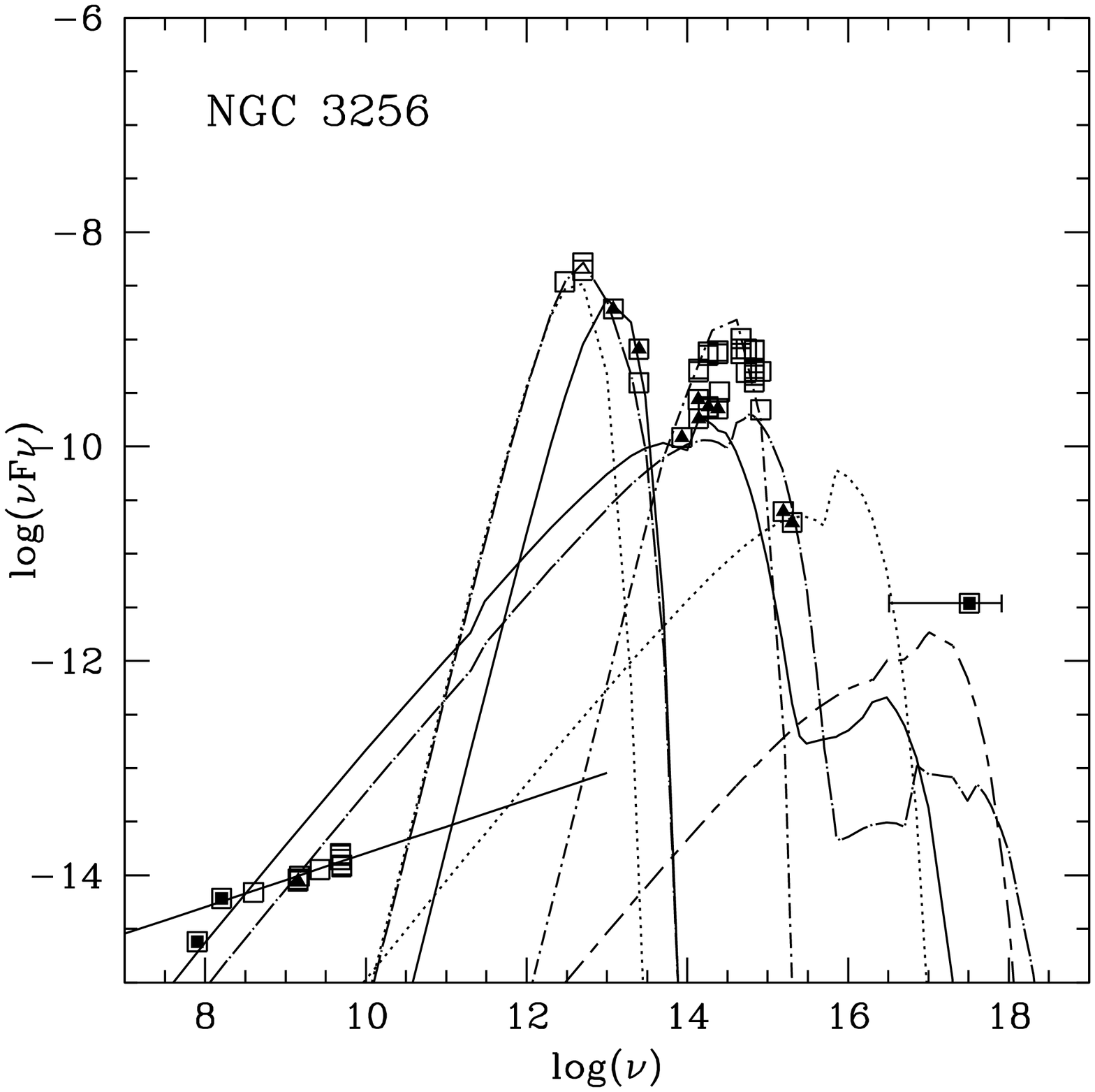}  
\includegraphics[width=56mm]{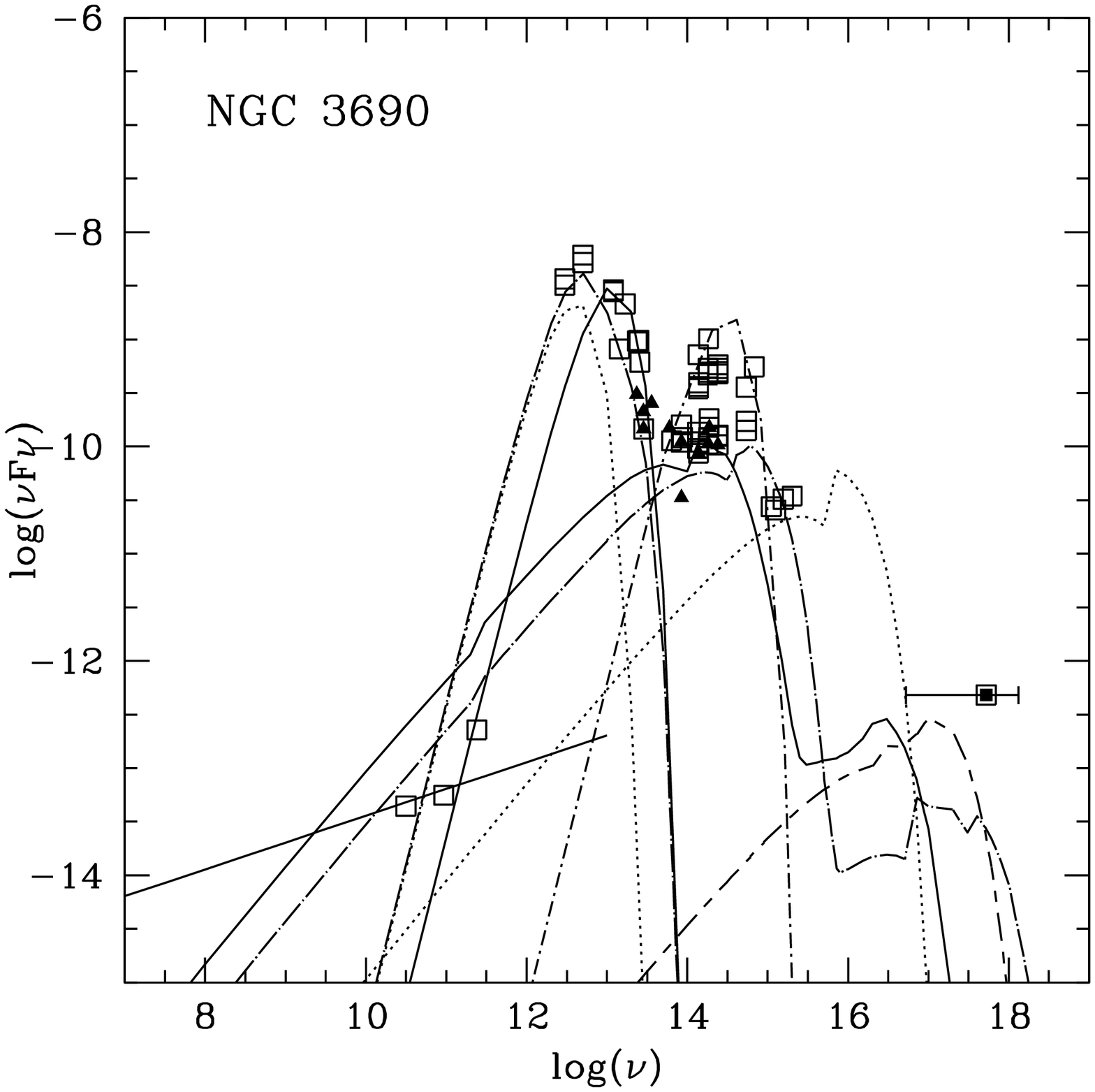}  
\centering
\includegraphics[width=56mm]{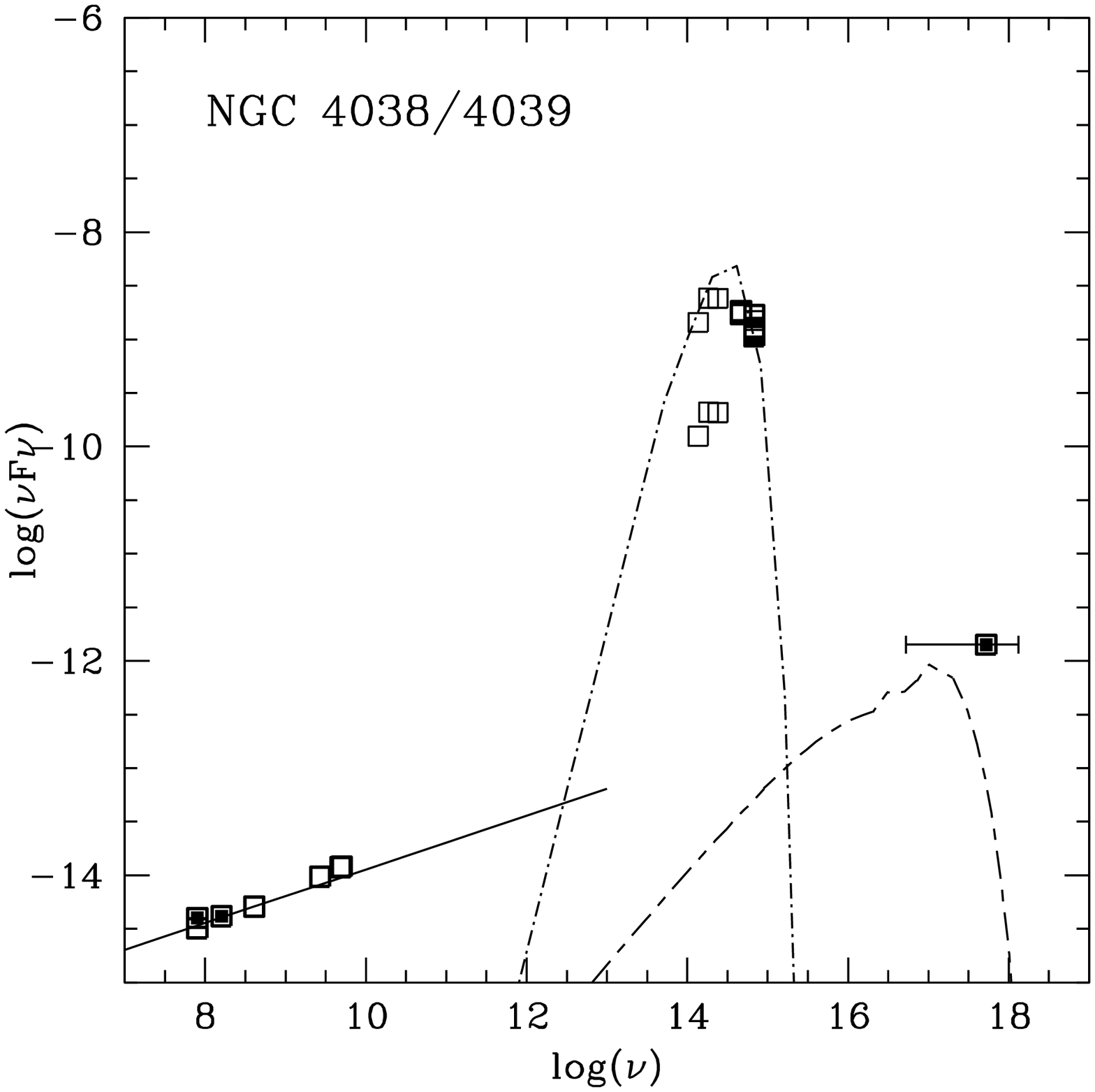}  
\includegraphics[width=56mm]{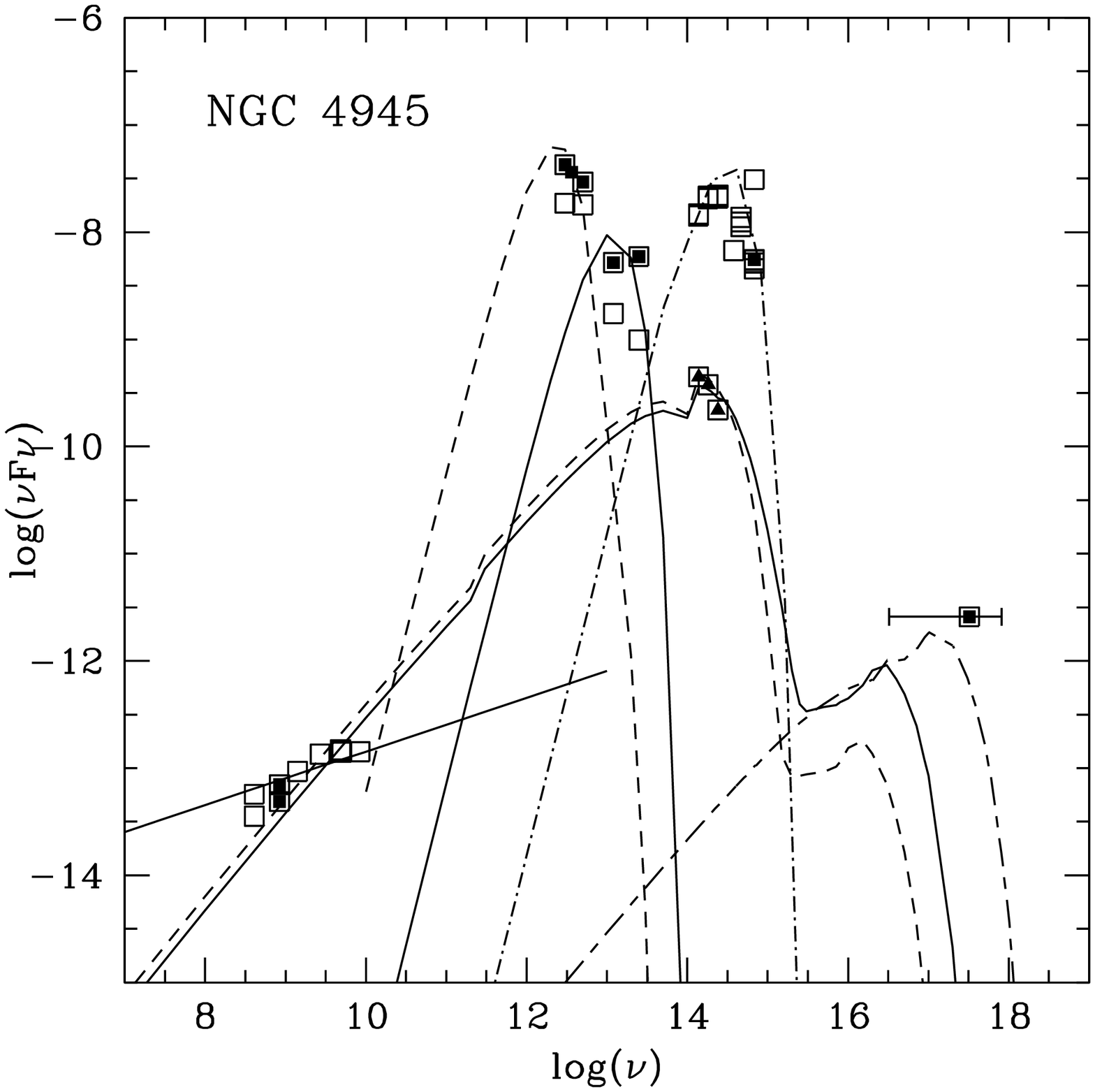}  
\includegraphics[width=56mm]{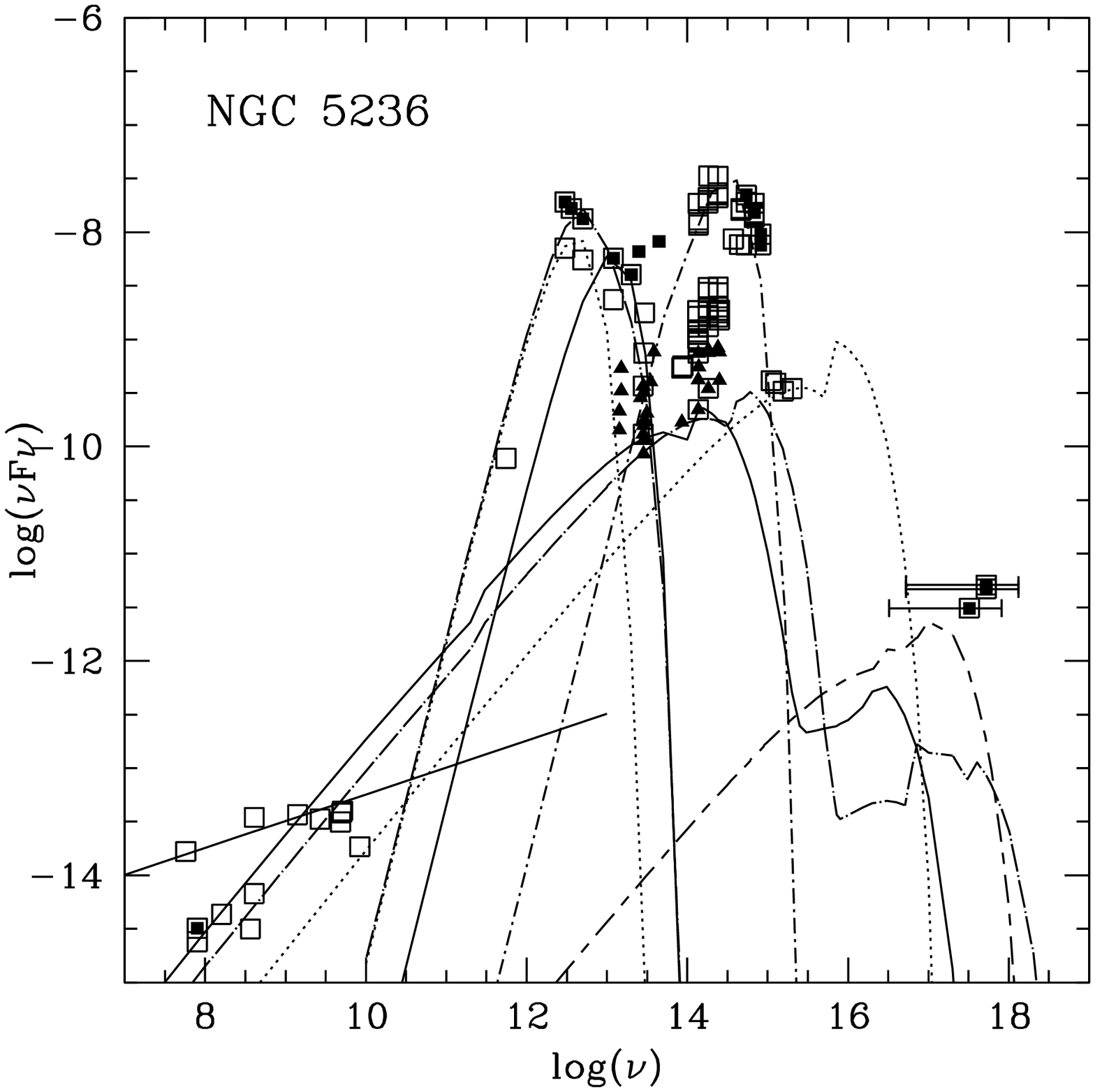}  
\includegraphics[width=56mm]{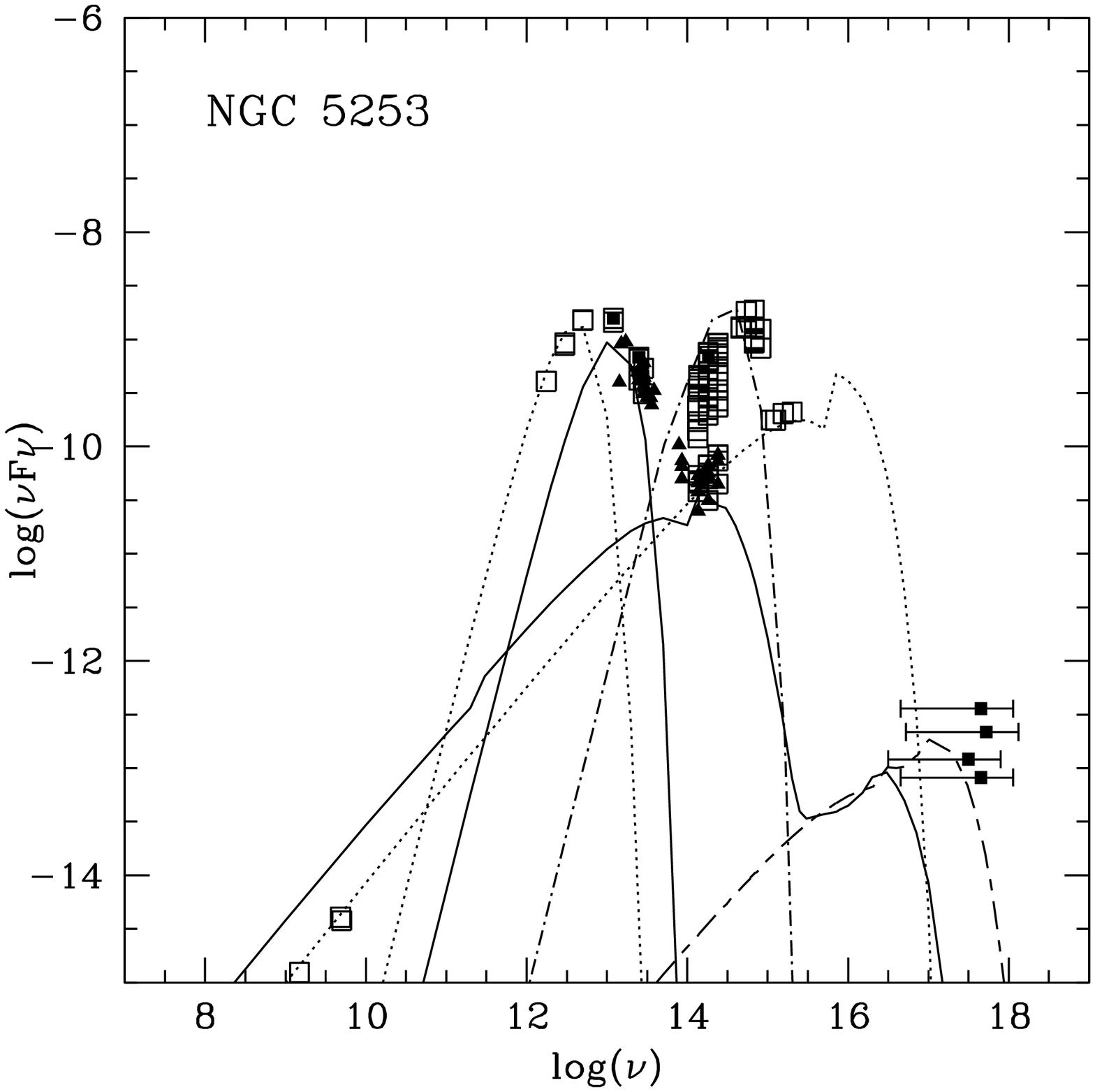}  
\includegraphics[width=56mm]{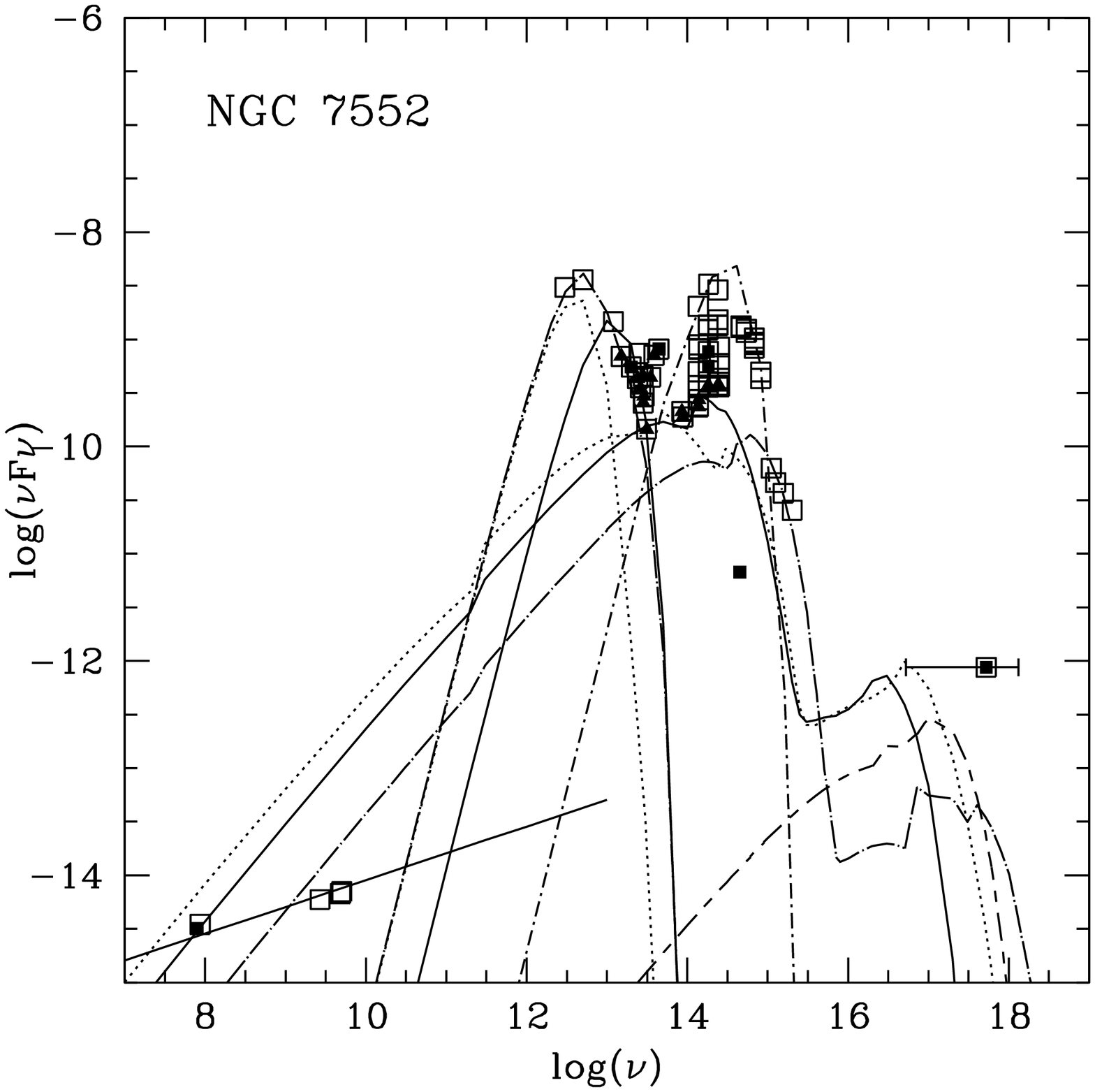}  
\caption
{Modelling the SEDs of the  Verma et al. (2003) luminous IR galaxies sample.
Black triangles: data from the NED. Black squares in the II Zw 40 diagram : data from
Galliano (2004).
Dotted lines: m1; short-dashed lines: m2-m3; solid lines: m4;
long-dashed lines: m5; long dash-dot lines: m7; short-long-dashed lines: m6;
dash-dotted line: bb model for the old star population;
solid line: synchrotron radiation.
}
\end{figure*}

\begin{figure*}
\centering
\includegraphics[width=56mm]{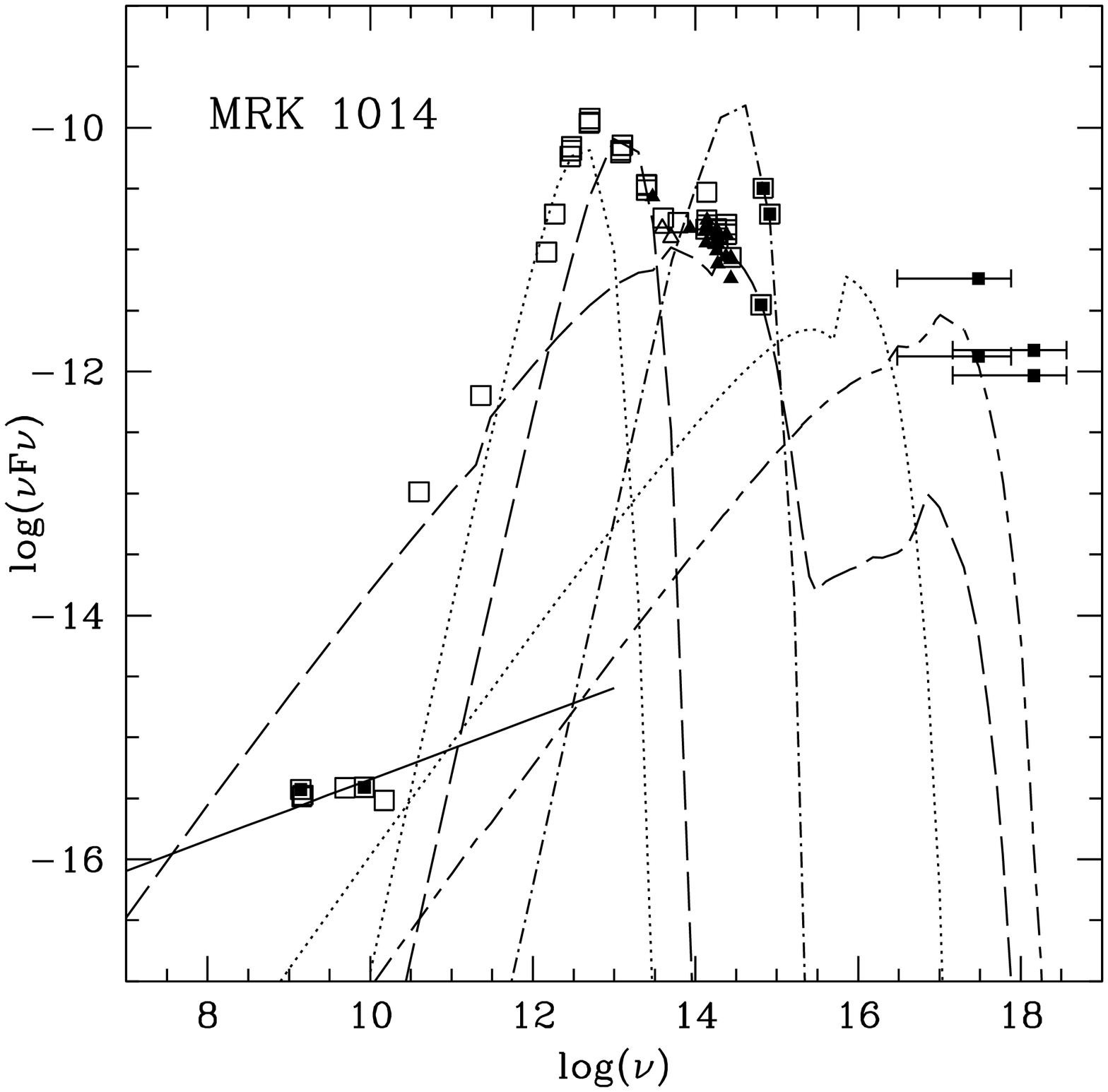}  
\includegraphics[width=56mm]{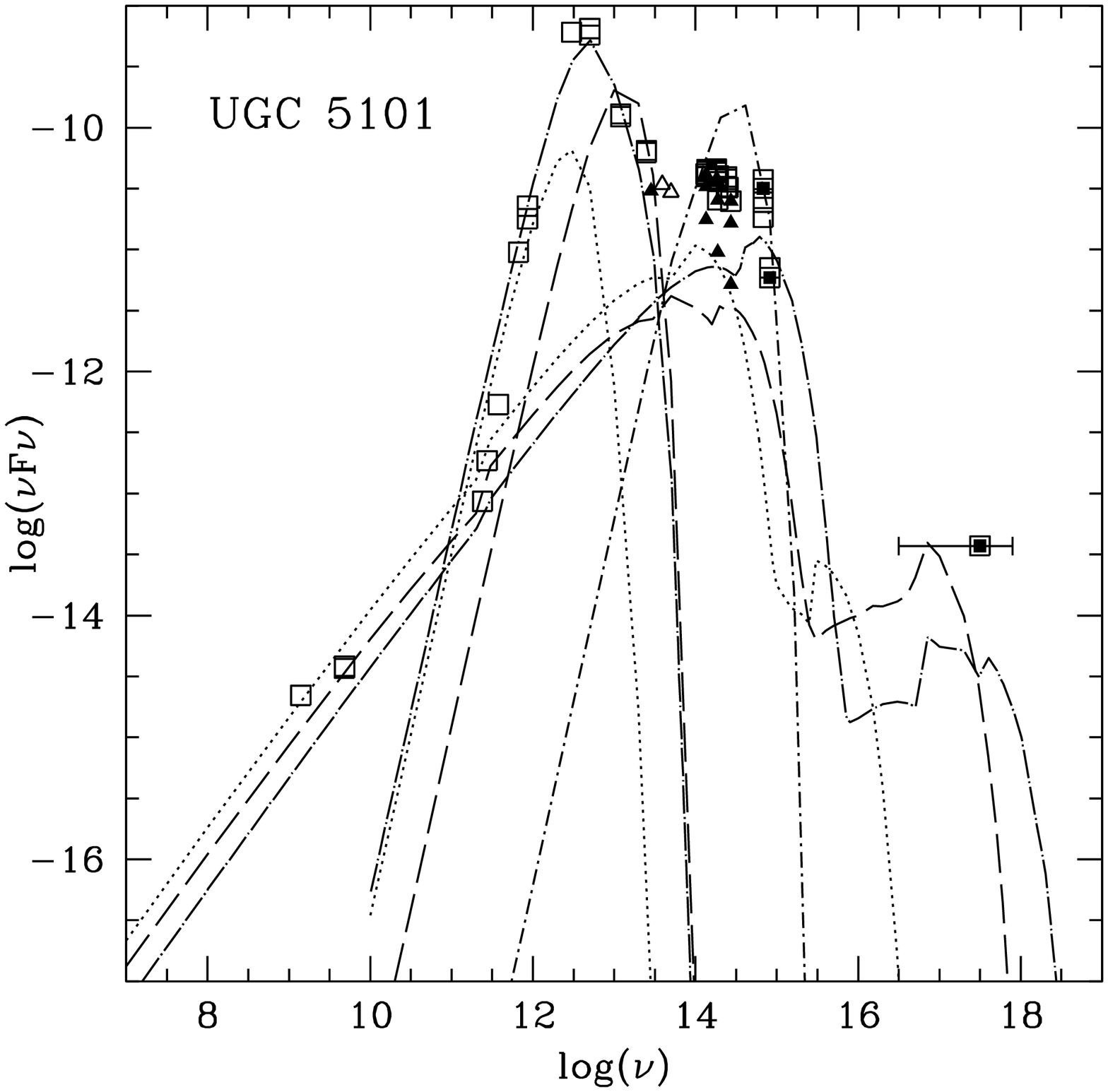}   
\includegraphics[width=56mm]{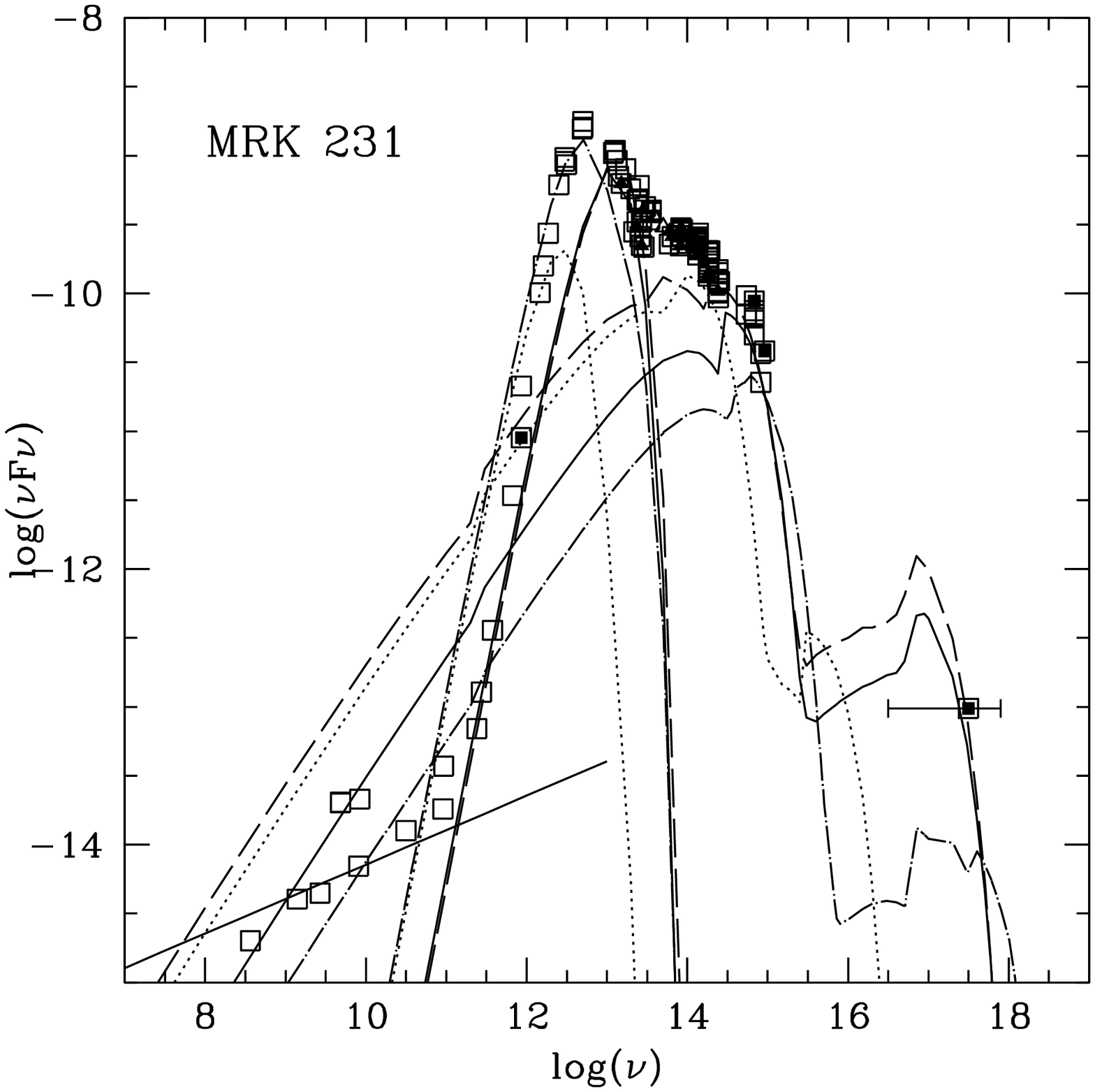} 
\includegraphics[width=56mm]{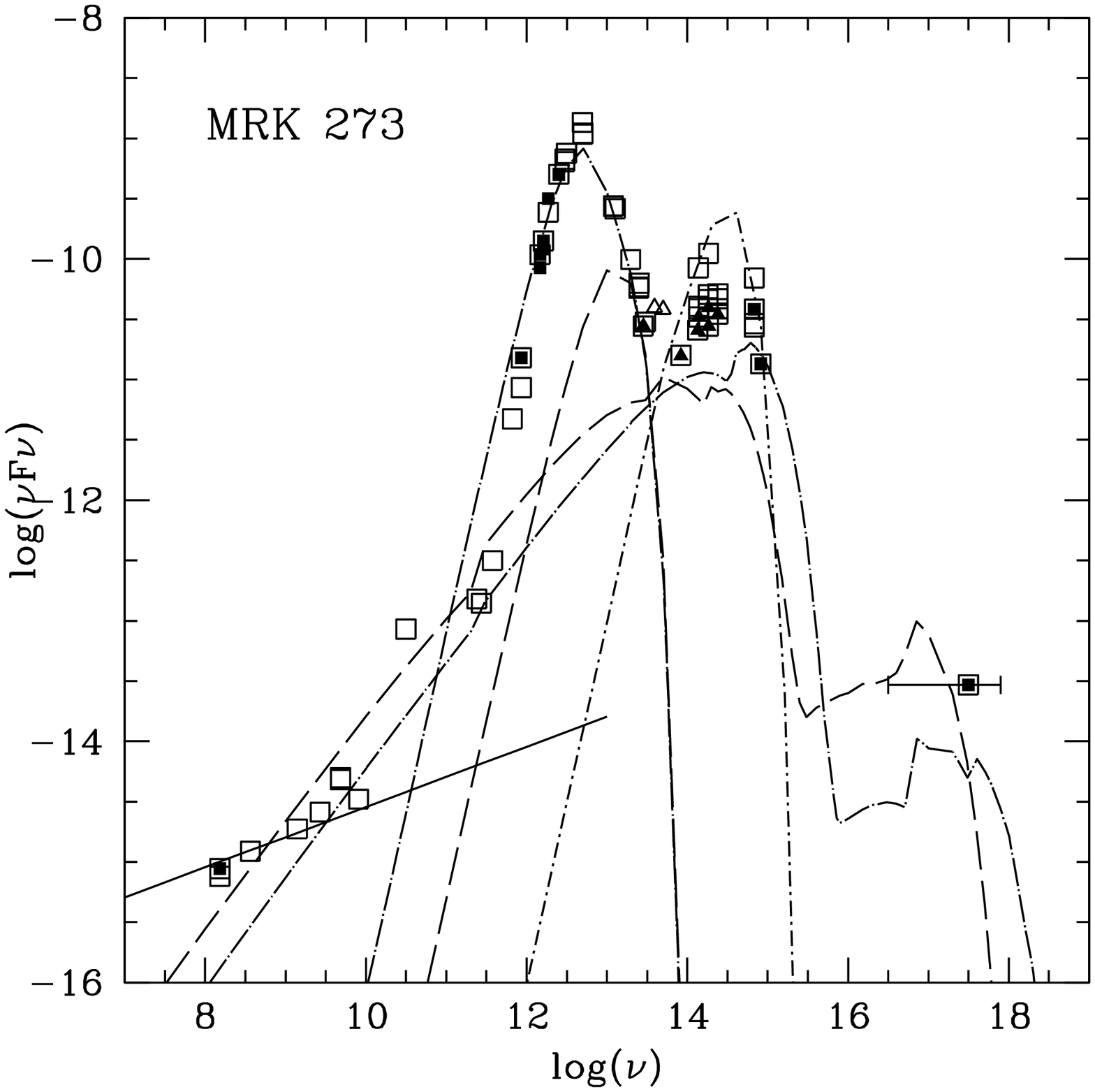} 
\includegraphics[width=56mm]{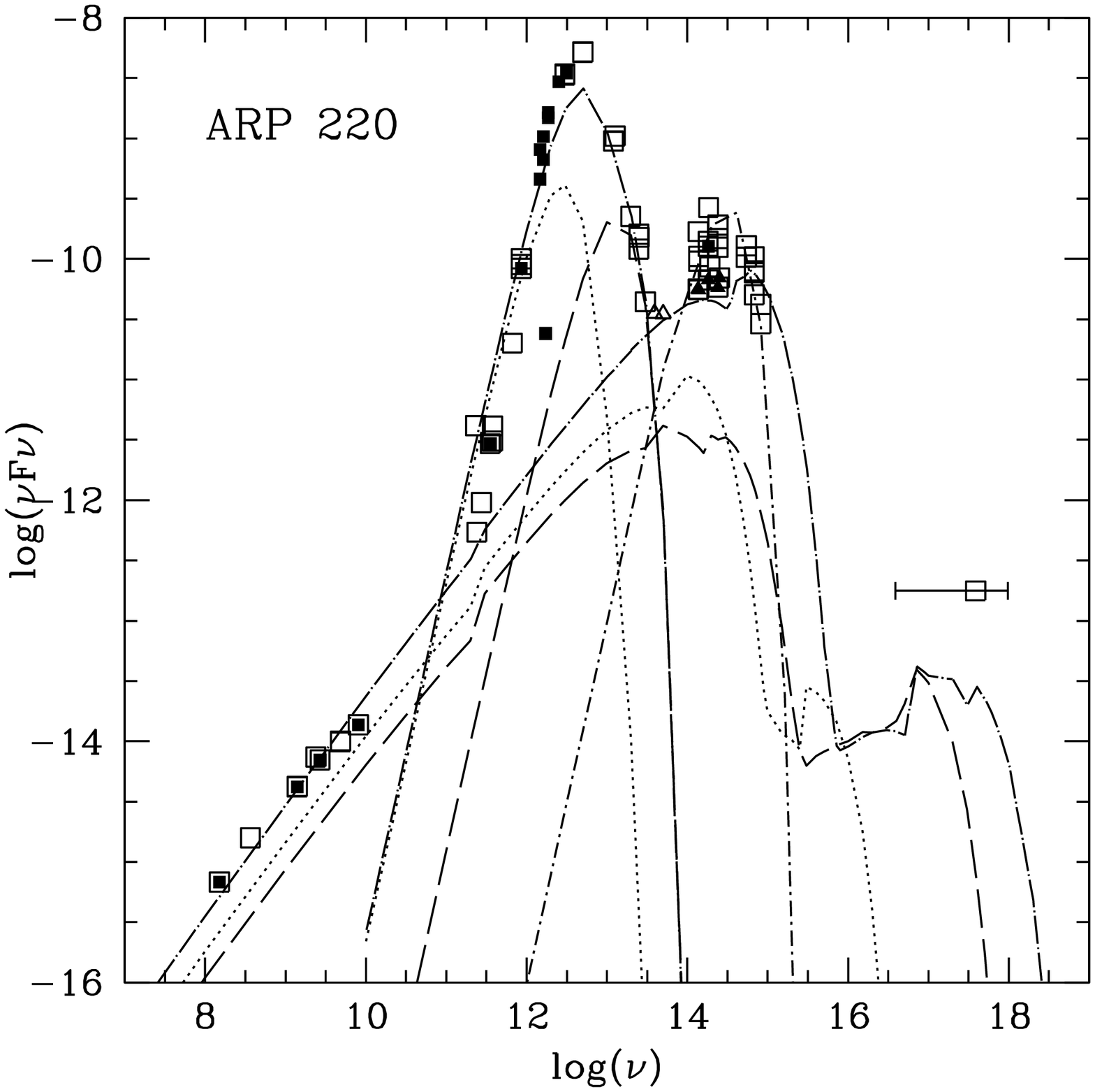} 
\includegraphics[width=56mm]{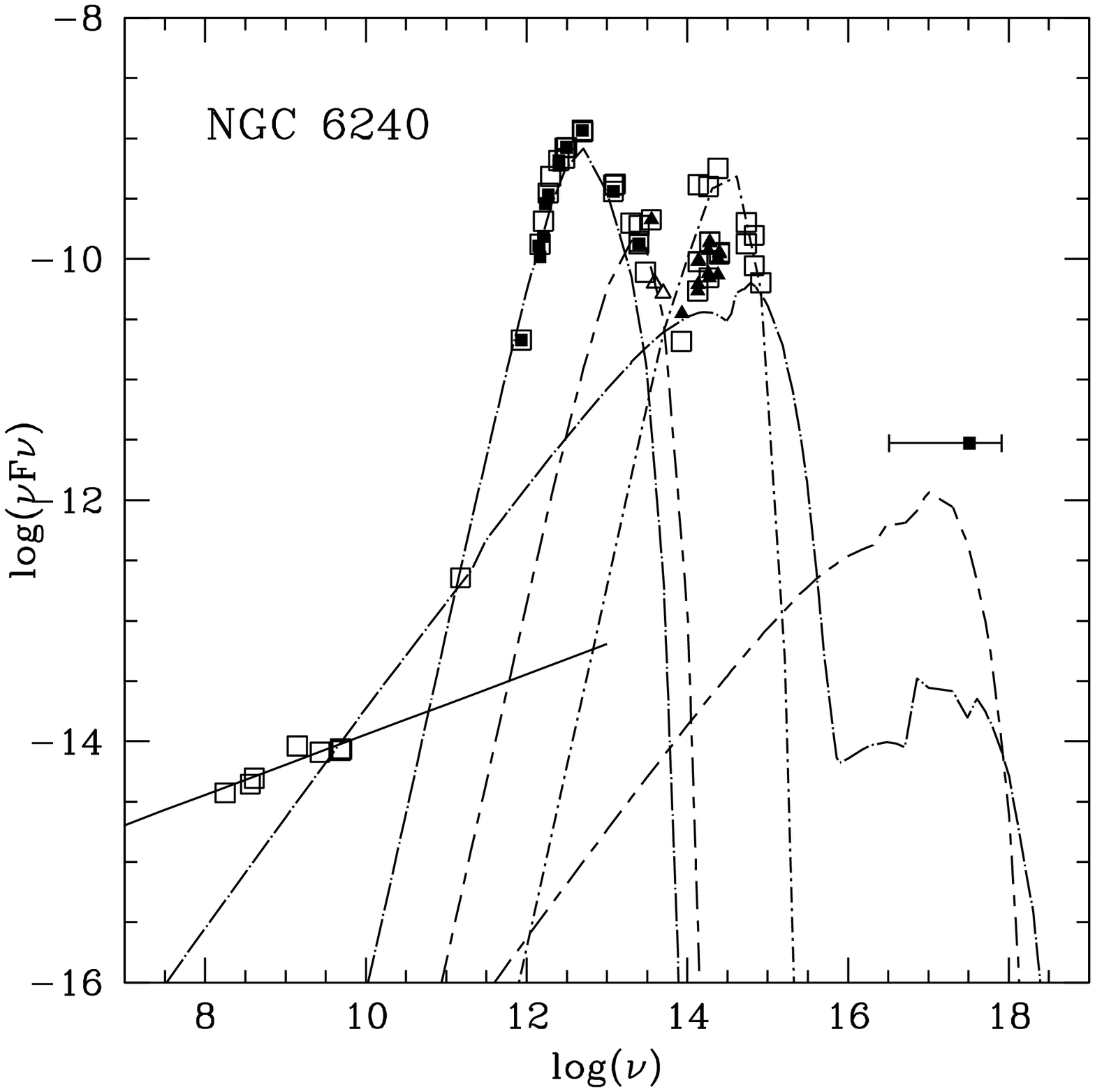}  
\caption
{Modelling the SED of the ULIRGs from the Rigopoulou et al. (1999) sample.
Symbols as in Fig. 5. Empty triangles: Rigopoulou et al. (1999) IR data at
5.9 and 7.7 \mum
}
\end{figure*}

\begin{figure*}
\centering
\includegraphics[width=56mm]{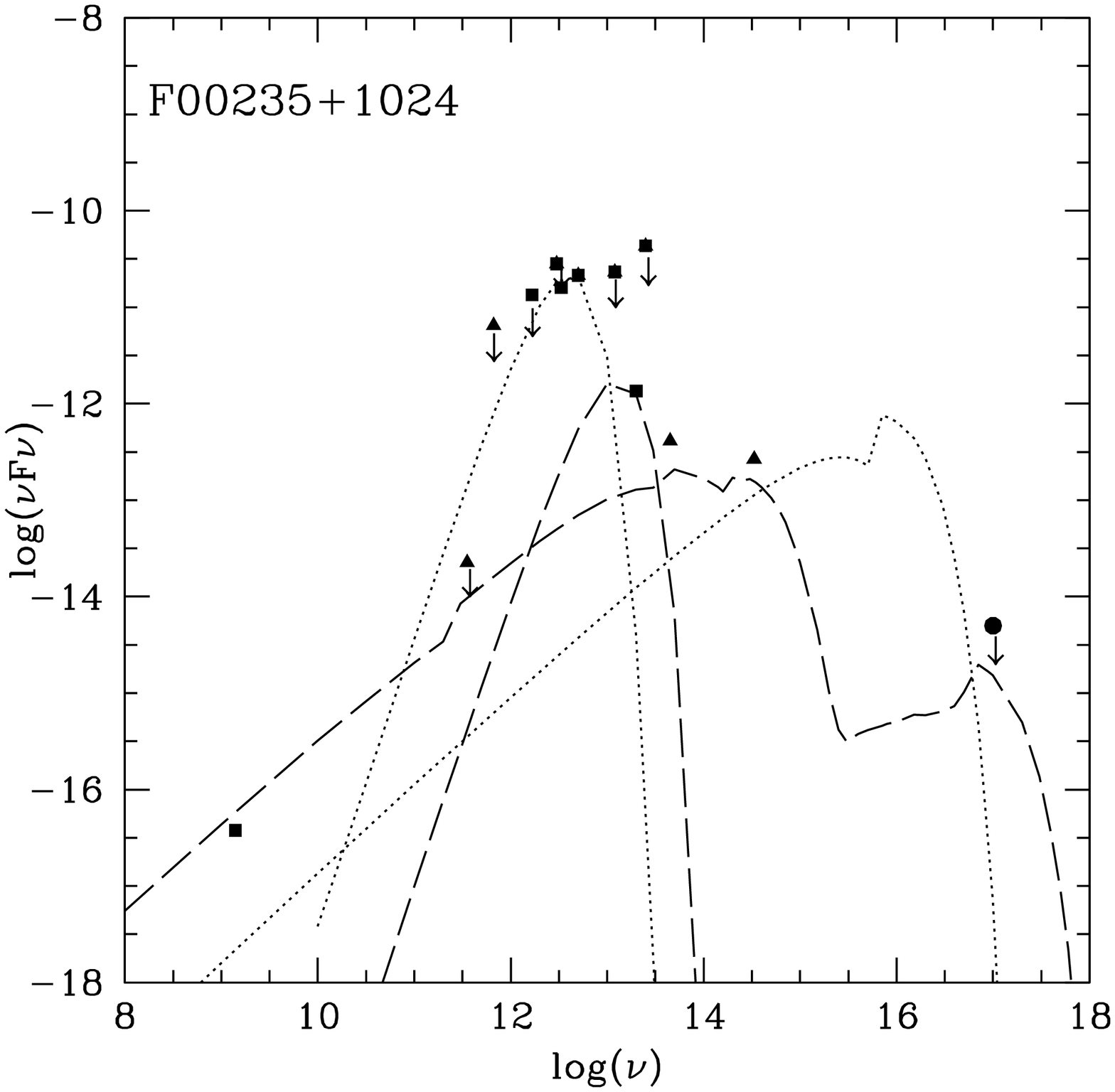}   
\includegraphics[width=56mm]{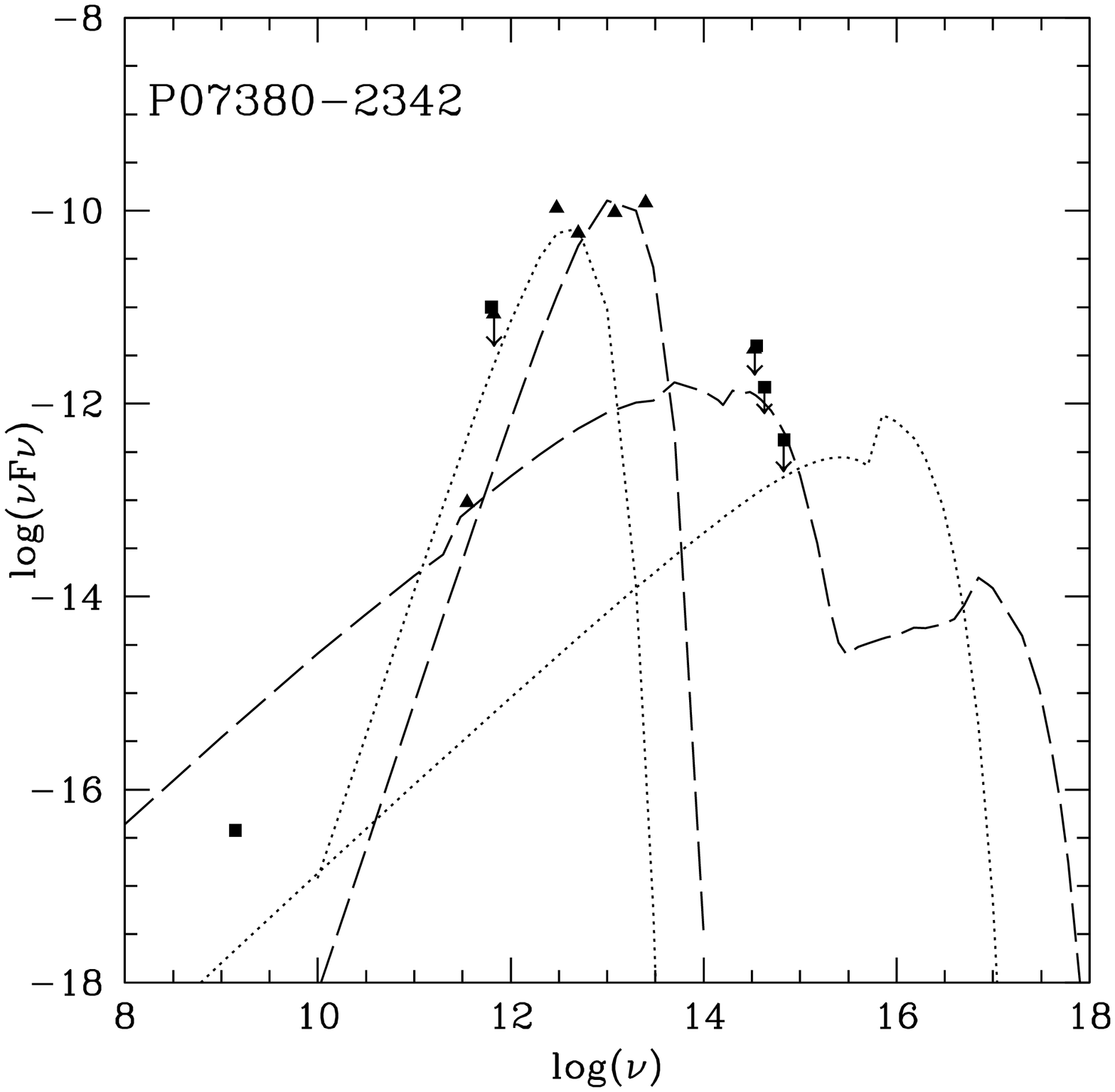}   
\includegraphics[width=56mm]{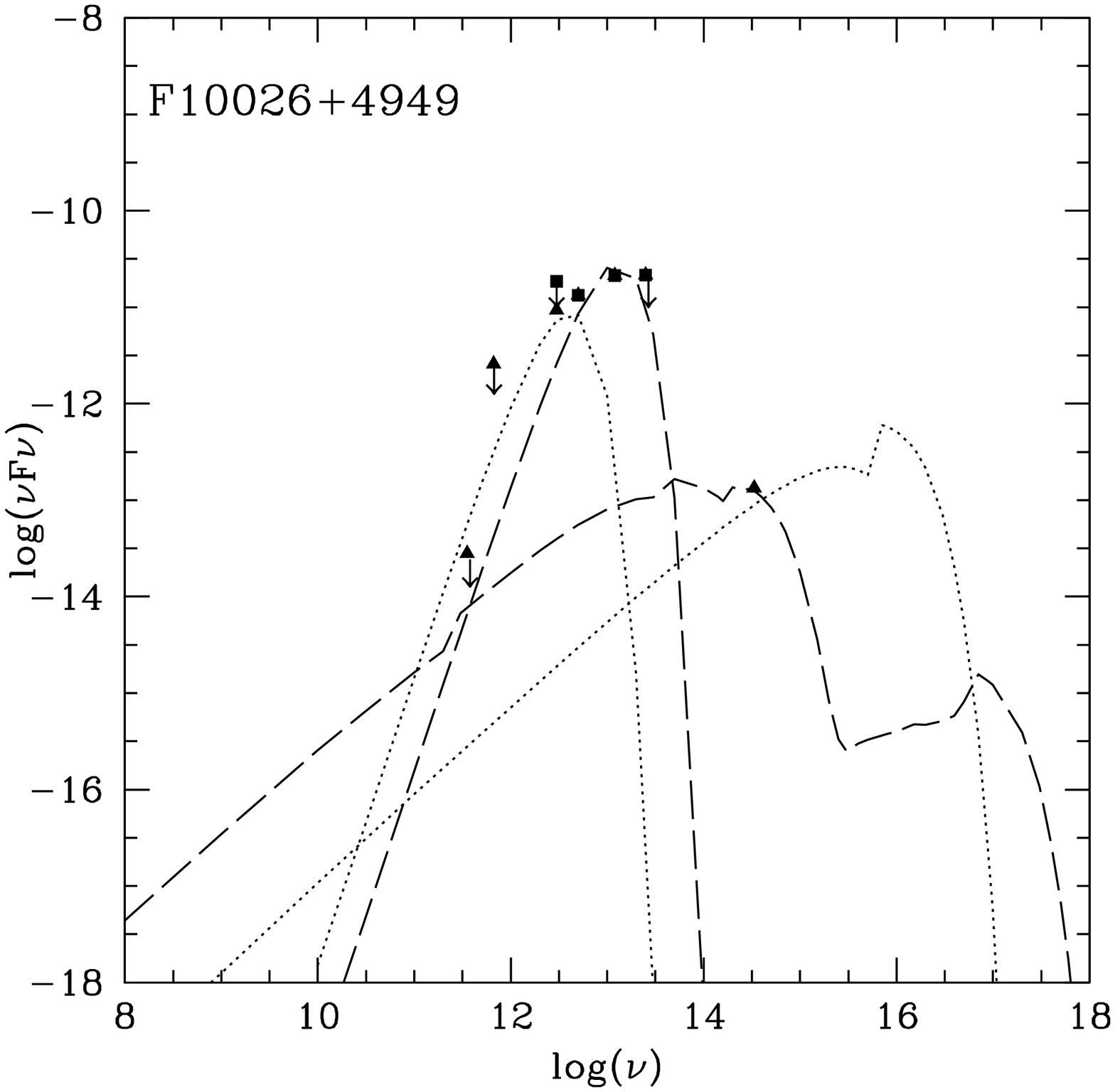}   
\includegraphics[width=56mm]{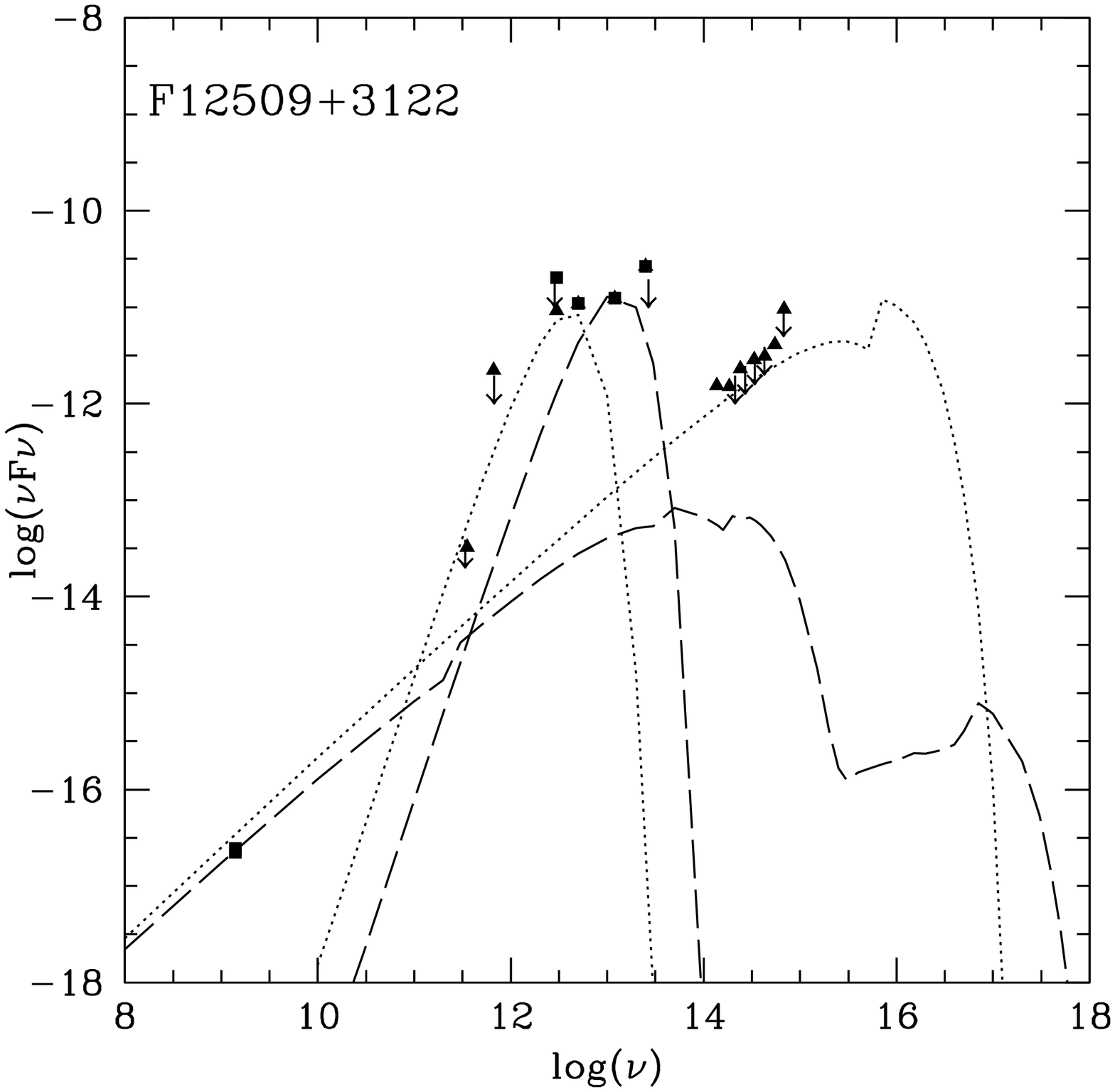}   
\includegraphics[width=56mm]{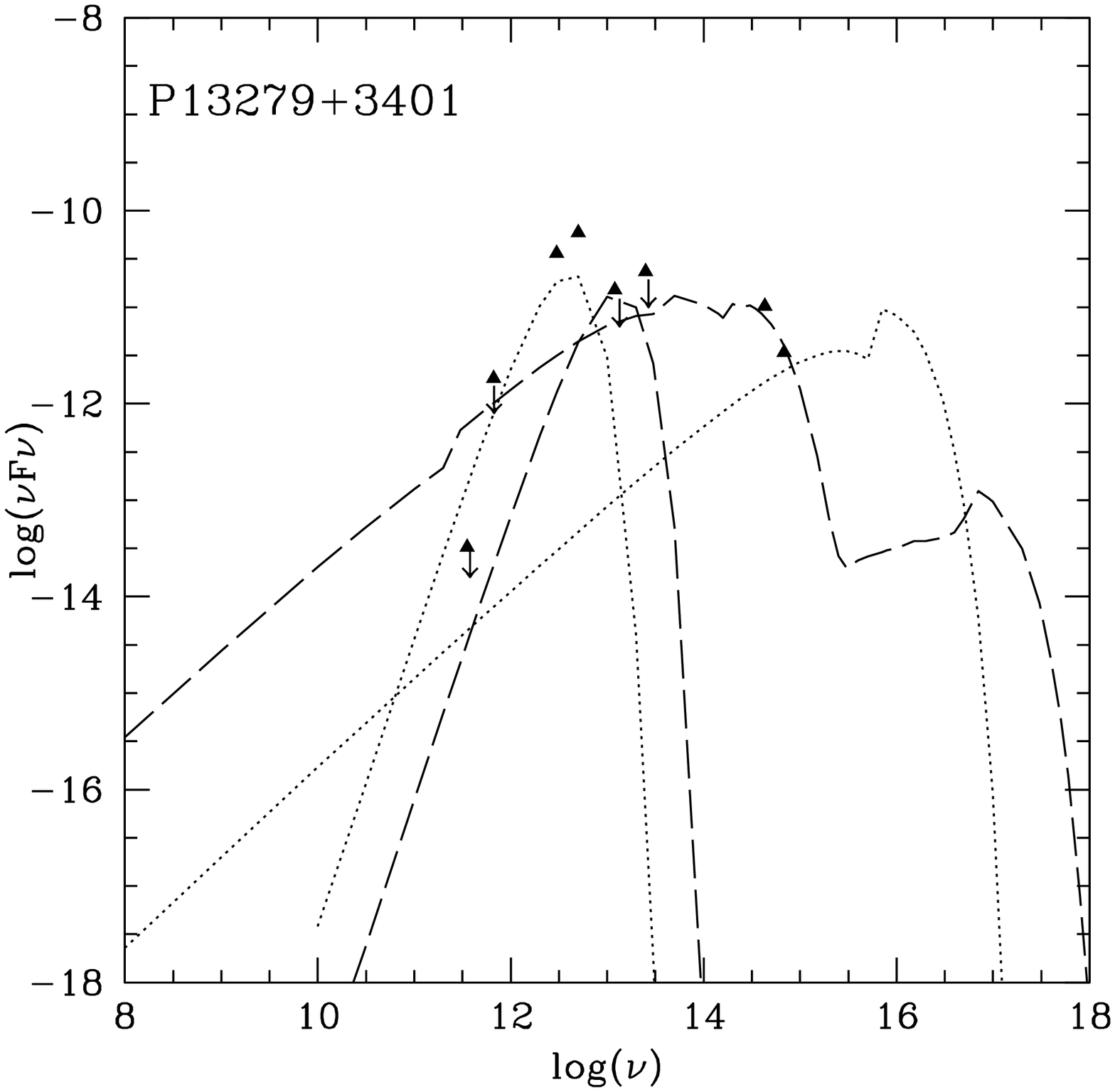}   
\includegraphics[width=56mm]{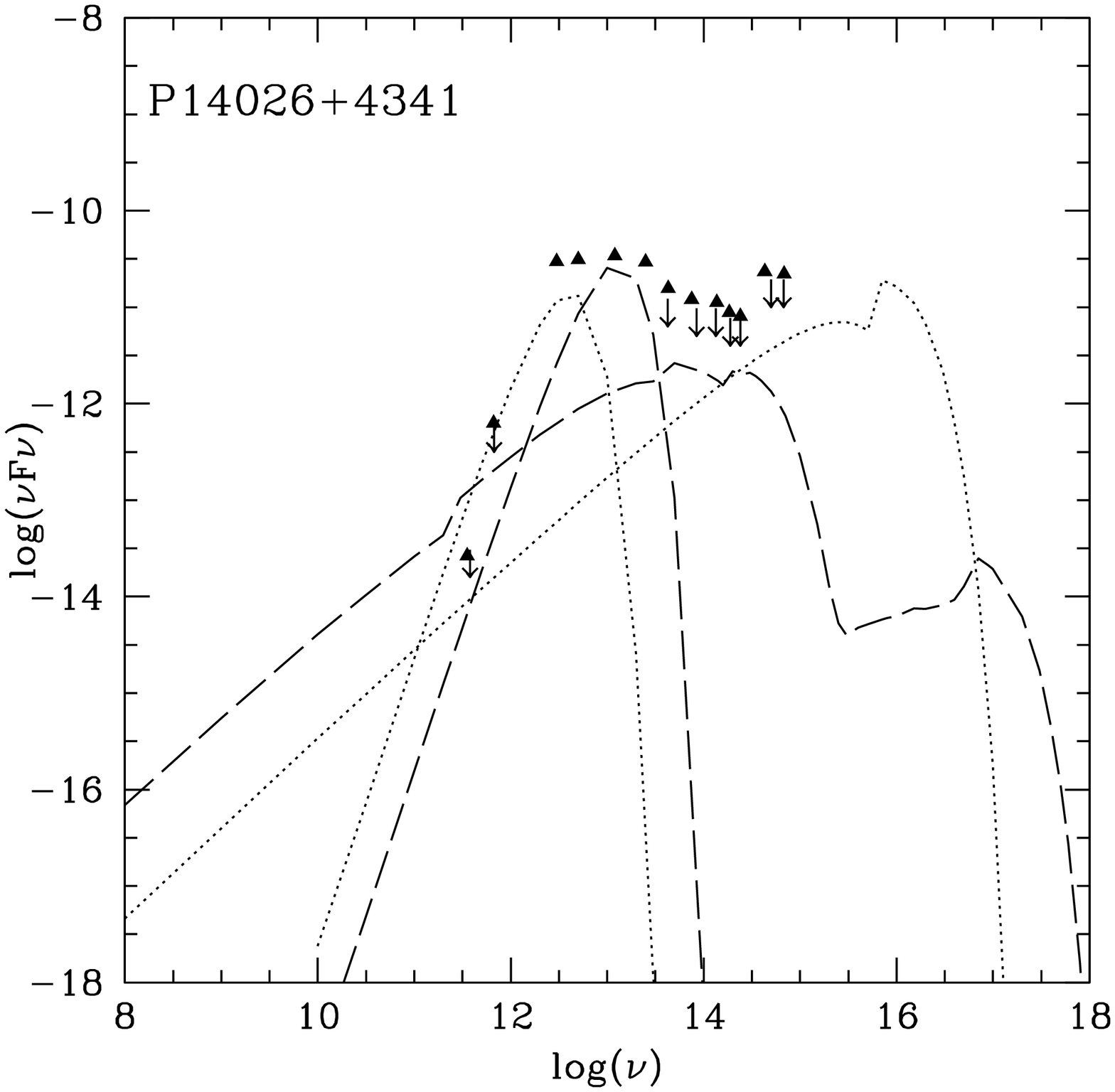}   
\centering
\includegraphics[width=56mm]{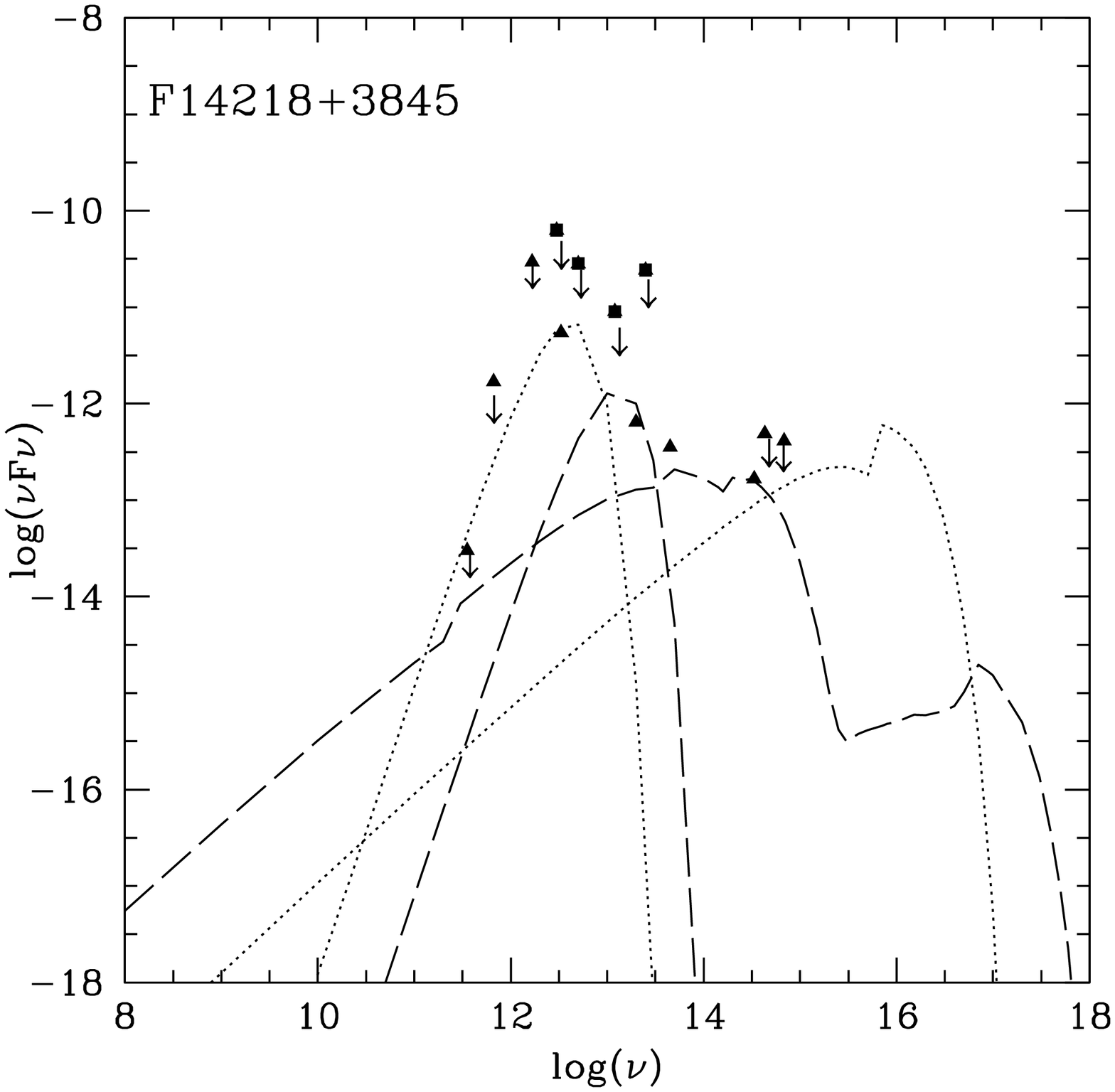}   
\includegraphics[width=56mm]{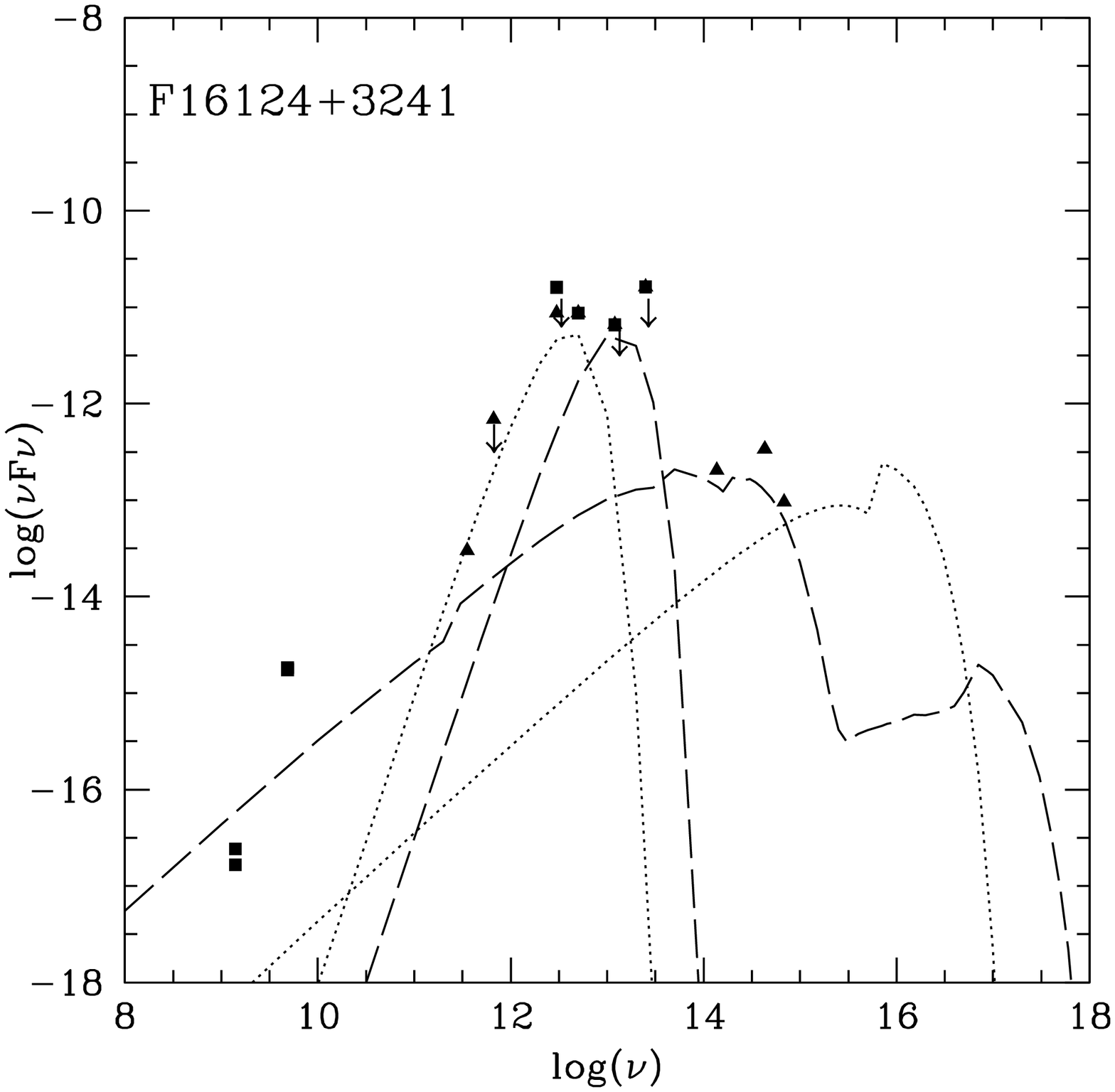}   
\includegraphics[width=56mm]{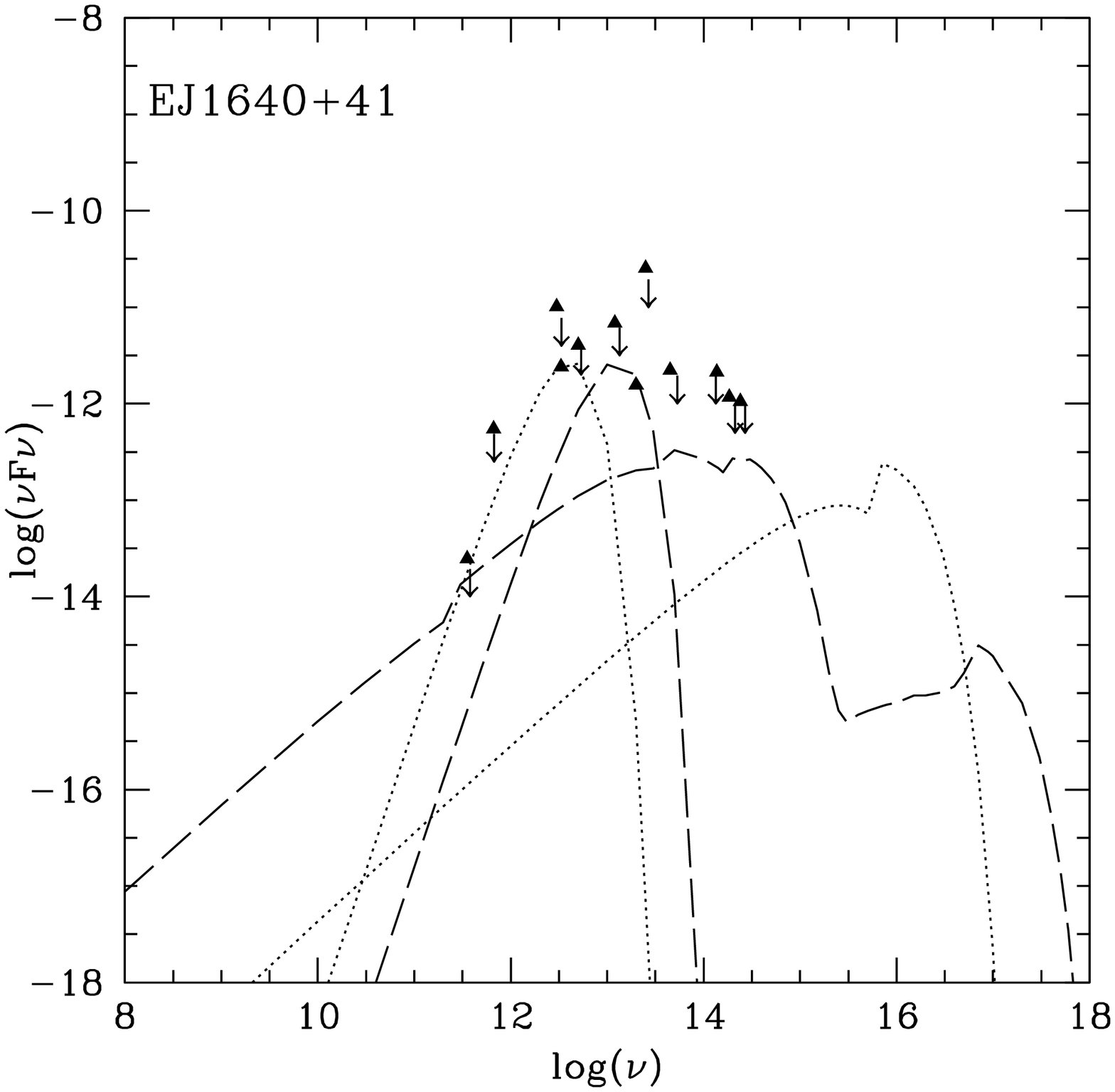}   
\includegraphics[width=56mm]{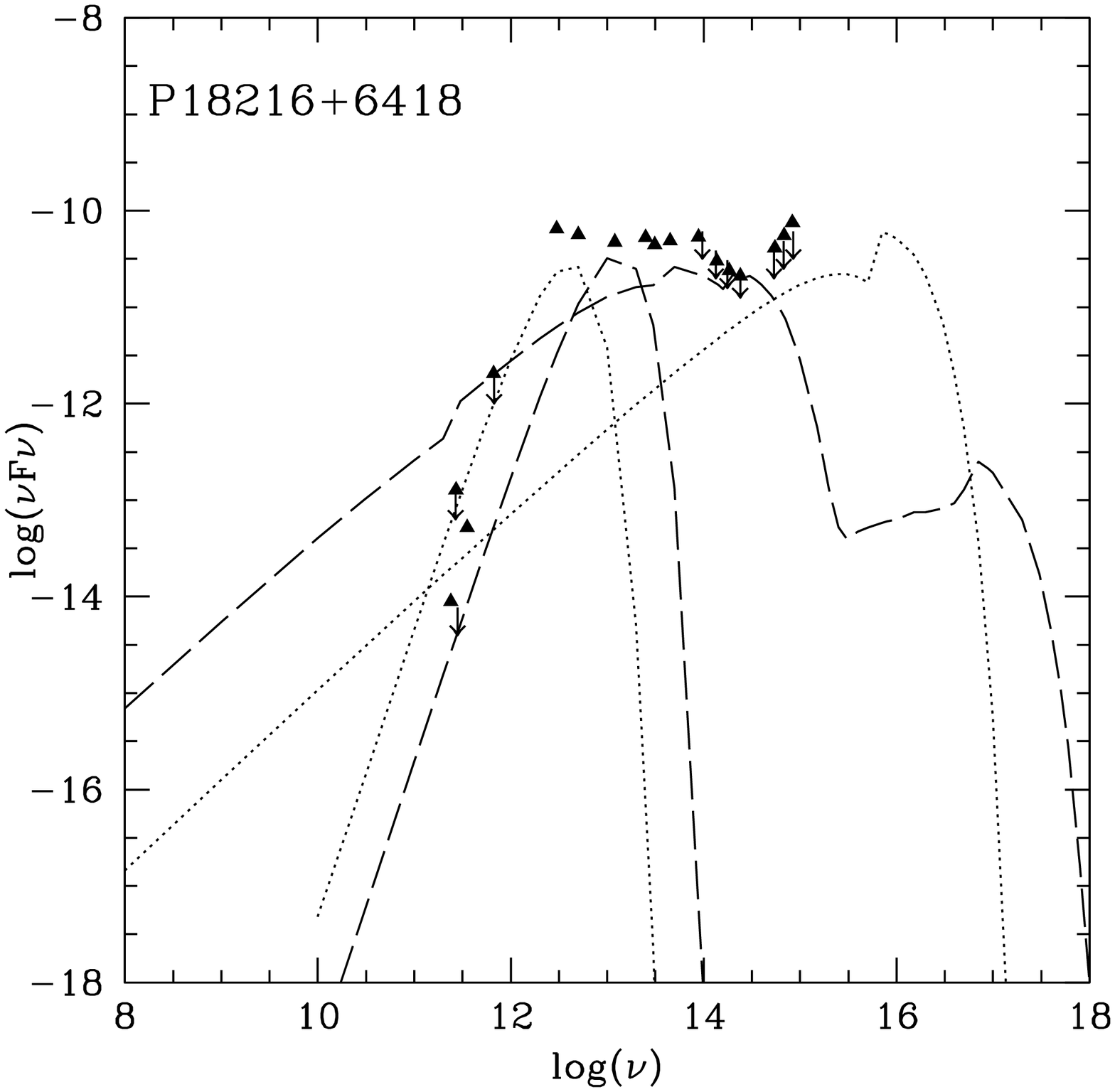}    
\caption
{Modelling the SED of the HLIRGs from the Farrah et al. (2002) sample
(all fluxes at rest-frame wavelengths shortward of 4 \mum ~ are treated as upper limits
due to unquantified contribution from the host galaxy).
Symbols as in Fig. 5. The filled circle represents the upper limit datum in the X-ray
range by Wilman et al. (1998).
}
\end{figure*}

\subsection{The grid}

From the current literature and from  diagnostic diagrams (Fig. 1) it  became evident that
many mechanisms coexist in each galaxy. Therefore,  the grid of  selected models (Table 2)
accounts for starbursts, AGN, HII regions, SNR winds, etc.
We reduced the full set of models to a few prototypes to have clear results. 

It is generally believed that starbursts  characterize luminous IR galaxies
because dust is formed in the atmosphere of massive stars. 
So  the models  suitable to the starburst clouds (m2-m5) show  different
 physical parameters which were  selected  cross-checking the SEDs.
An AGN, 
is often discovered in starburst galaxies, dominating the galactic central region
where  densities and velocities are relatively high,  while the
spectra emitted from shocked clouds in the external NLR of AGN is hardly distinguished
from those emitted by clouds in the neighbourhood of  starbursts.
Therefore,  we  included in the grid only one model (m7)  corresponding to the AGN,
with  large grains  ($a_{\rm gr}$ =1 \mum)
which can survive sputtering downstream at relative high shock velocities and
densities. 

SNR are strong emitters of radio and X-ray downstream  of  strong shocks 
created by  the  SN explosion. We  include in the grid model m6, 
a shock dominated model (U=0) with a relatively
high  velocity   suitable to explain soft-X-ray observations.

Finally, the underlying  galaxy spectra  must be taken into account, represented
by models showing a lower \Vs, a low geometrical thickness, $D$ due to the fragmentary
character  of the clouds, and  a diluted photoionizing flux from the source. 
We have added to the grid  model m0 ($d/g$ = 4 10$^{-4}$) which
was found necessary to reproduce the MIR continuum
of most of the galaxies in the  Alonso-Herrero et al. (2003) sample
(Contini, Viegas \& Prieto 2004). 
The $d/g$ values in the underlying galaxy  are typical of the ISM.

The models
(Contini \& Viegas (2001a,b) adopted to explain the line ratios,
were calculated assuming  dust-to-gas
ratios of   4 10$^{-5}$ by mass, which are  low,
even compared to the Galactic ISM  ($d/g$ $\sim 4\times 10^{-4}$).
By combining these  models to  fit the continuum SED, more realistic  $d/g$ ratios  will
be obtained.

The comparison of calculated   with observed   SEDs
 indicates that  the best agreement  is obtained by
 $d/g$ higher by some orders of magnitude in all the models, except in  those corresponding to
low \Vs (100-150 \kms) and low \n0 (100 \cm3).
The models with \Vs $\geq$ 200 \kms  $recalculated$  adopting a  more   suitable $d/g$
however,  could not explain the observed line ratios
(Figs. 1), indicating that the line spectra are emitted from gas less obscured
by dust and that dust is not distributed homogeneously throughout a galaxy.
Lines emitted  from  the  dusty regions are  weak compared
to those emitted  from  dust-free  ones
because a high $d/g$  speeding up the cooling rate,  reduces the ion fractional
abundances by  recombination.

\subsection{The single-cloud models}

The  calculated bremsstrahlung integrated throughout the radio, IR, optical, UV, and X-ray
ranges for the different models, are given in rows 7, 9, 10, 11, and 12 of Table 2, respectively.
In row 9   the calculated flux emitted in the IR by  reprocessed dust, $F_{\rm IR_d}$ is shown 
for comparison.    Notice that both 
bremsstrahlung and reradiation by dust increase with   
\Vs\ , \n0\, and  $U$. 
Particularly, shock-dominated models (calculated with $U=0$) such as  m6 
show very low bremsstrahlung radiation.

The maximum temperatures of the grains downstream are given in the last row  of Table 2 
(bottom).  The temperatures increase with \Vs. However,  density and  grain sputtering
also affect the heating and cooling of dust.
 This can be noticed by
comparing the maximum temperature of grains calculated with models m0 and m1.

The profiles of the temperature and density of the gas, and of the temperature 
and radius of dust grains
calculated for models m1, m5, and m7 are shown in Fig. 2 
(left, middle, and right diagrams, respectively)
for comparison. 

The left edge of the diagrams
represents the shock front.  Here collisional processes dominate the heating of dust
and gas up to a certain distance downstream which is determined by the shock
velocity and by the density. 
The grains are destroyed by sputtering in the region close to the 
shock front (middle and right diagrams).
The photoioning flux reaches the opposite edge,
heating and ionizing gas and dust.

The  continuum SEDs  calculated by models m0-m7  are compared in Fig. 3.
Two curves correspond to each model, one representing  dust reradiation and the other
bremsstrahlung. 
The bremsstrahlung peak  between 10$^{13}$ and 10$^{15}$ Hz depends on
photoionization,
while the maximum  at higher frequencies is determined  by the shock.
For each model, the peak intensity  of  dust reprocessed radiation depends on $d/g$
and the peak frequency on \Vs.
However, Fig. 3  shows
 that sputtering  may change the distribution of the grains
and shift the peak at lower frequencies (e.g. model m7).
The MIR flux  emitted by dust   corresponding to  \Vs = 300 \kms ~and 500 \kms
~(models m4 and m5, respectively)  show very similar SEDs, while the corresponding  bremsstrahlung
at frequencies $\geq$ 10$^{13}$ Hz are different.  
Therefore, we will constrain the models by both the IR dust reradiation flux and the
bremsstrahlung in the IR, optical, UV, and X-ray range.

The bremsstrahlung of models m6 and m7, both calculated with \Vs =1000 \kms,
show different distributions  because m6  is shock dominated   while m7 is radiation dominated.
Moreover, they are calculated by   different \n0\ and    different grain sizes.

The accuracy of the data in the optical-UV range is not high enough to select
between power-law and black-body fluxes by modelling the continuum SED only
(see also  Contini \& Viegas 2000, Fig. 1).
Shock velocities instead  can be estimated by
 the peak  at high frequencies of bremsstrahlung  and by the frequency 
of the  IR emission peak (Contini et al. 2004).
A comparison of different models is shown in Fig. 4.

\subsection{The continuum SED}

\subsubsection{Selection of the data}

The  data  from the NED (NASA Extragalactic Database) are adopted for
LIRGs,  complemented with data from the literature
   to  fulfil the dataset of ULIRGs and HLIRGs by Rigopoulou et al
and Farrah et al., respectively.
 The data  correspond to different apertures.
In the diagrams we will distinguish between  those  taken with the smallest aperture
(up to 14 arcseconds, black triangles), those with
large apertures  (white squares), and those integrated from maps
(black squares).
The observational errors are not shown in the diagrams
in order to   obtain less confused pictures.

All the data are important in modelling.
Those taken with a large aperture
cover different regions characterised by  radiation from different sources in the same galaxy,
therefore they correspond to multi-cloud models.
This method led  for example to  the identification of the starbursts in the region close to
the nucleus of the low luminosity AGN NGC 4579 (Contini 2004), or in the  outskirts of NGC 7130 AGN
(Contini et al. 2002).
Data obtained with different  apertures   at frequencies close to  silicate,
 PAH, graphite , ice, etc. bands   may include different
absorption/emission fluxes (Angeloni et al. 2007, in preparation).

We try to  reproduce ("by eye")  the largest  number of data, 
constraining the modelling by upper limits. 
The curves representing dust emission and bremsstrahlung from different 
clouds are not summed up in the diagrams
in order to  compare directly each model with the dataset trend.

The old stellar population background is  represented by a black body (bb) with
T$\sim$ 3000-5000 K. The data  contaminated by different contributions  of
light from the stars, are  nested  within the bb curve,
 e.g. NGC 2110 (Contini \& Viegas 2000), Arp 220 (Contini et al. 2004).
The  less contaminated data  show the  very bremsstrahlung from the clouds and 
constrain the models.

For all the galaxies,
X-rays are upper limits because the fluxes are integrated on a large frequency range.
The power-law trend of the data  in the  radio  continuum of many  galaxies
is explained  by  synchrotron radiation created by the Fermi mechanism at the shock front.

 The IR  dust reradiation  "bump" is generally complex,
accounting for  the contribution of  different grains in different conditions.

Finally, the   IR  bands on top of the continuum
 are  not   included in the calculations of dust,
 therefore, they can be recognised comparing calculated with observed SEDs.

\subsubsection{Modelling approach}

Models are calculated at the nebula, while  data  are observed
at Earth. Therefore to  suite model results  to the observed SED, the results must be
multiplied by a factor $\eta$ which accounts for the distance to Earth of the galaxy,
for the distance of clouds from the galaxy centre, and  for  the filling factor.

To determine the dust-to-gas ratios in each cloud of a galaxy, we will use
the method  presented  for  the starburst regions in luminous infrared galaxies
by Contini \& Contini (2003).
Dust and gas are coupled throughout the shock front and downstream  in each cloud
and the IR  flux  from dust is  consistently calculated with bremsstrahlung, so
 the $\eta$ factors corresponding to IR emission by dust and to bremsstrahlung 
 must be equal.
The $\eta$  is determined by  fitting the bremsstrahlung which extends from radio to X-rays.
To recover the same $\eta$ for the IR bump, we change the $d/g$ input parameter,
which affects directly dust reradiation.
These changes, however, must  account for the multi-cloud model.

This procedure  is straightforward, but not trivial because
changing  the  $d/g$  ratio by  a factor  $\geq $ 10 leads to a different
SED of the  bremsstrahlung.
Therefore the  $d/g$ ratios  result  from cross-checking the dust
reprocessed radiation  by  the bremsstrahlung,  for  each cloud in  every galaxy.

\section{Results}

\begin{table*}
\centering
\caption{Log($\eta$)$^a$  and the relative contribution of the different clouds, W$_i$ for LIRGs}
\begin{tabular}{lllllllll}\\ \hline\hline
 &\multicolumn{2}{|c|}{\em under-gal}&\multicolumn{4}{c|}{\em starbursts}&\multicolumn{1}{c|}{\em winds}& \multicolumn{1}{c|}{\em AGN}\\
\     &  m0     &   m1   & m2  & m3 & m4  & m5  & m6  &  m7 \\ \hline
\ {\bf NGC 253} &-&-4.35&-&-&-11.4&-&-11.9&-12.1\\
\ opt&0.0&   0.0831&0.0&   0.0 &   0.191&0.0    &0.0&0.726\\
\  UV&0.0&   0.170&0.0&   0.0 &   0.088&0.0&    0.0&0.742\\
\  X-ray  &0.0 & 0.611&0.0&   0.0 & 0.0228&  0.0&    0.331 & 0.0344\\
\  IR$_d$&0.0&   0.238&0.0&   0.0 &   0.515&0.0&    0.0&0.231\\
\ {\bf IC 342} &-&-&-11.&-&-11.1&-&-11.8&-12.5\\
\ opt&0.0&   0.0&      0.0173&0.0 &     0.115&0.0 &     0.0&0.868\\
\  UV&0.0&   0.0&      0.0012&0.0 &     0.0561&0.0 &     0.0&0.943\\
\  X-ray &0.0 & 0.0&   0.0016&0.0& 0.0104 & 0.0   & 0.957 & 0.0314\\
\  IR$_d$&0.0&   0.0&      0.499&0.0 &     0.264&0.0 &     0.0      &0.237\\
\ {\bf IIZW40} &-&-6.6&-&-&-12.4&-&-&-\\
\ opt&0.0&   0.196&0.0&      0.0 &     0.804&0.0 &     0.0 &     0.0 \\
\  UV&0.0&   0.521&0.0&      0.0 &     0.479&0.0 &     0.0 &     0.0  \\
\  X-ray &0.0 &0.938    &0.0  &0.0    &0.0622& 0.0  & 0.0  & 0.0 \\
\  IR$_d$&0.0&   0.207&0.0&      0.0 &     0.793&0.0 &     0.0 &     0.0   \\
\ {\bf M82}&-&-5.&-&-&-11.8&-&-11.5&-\\
\ opt&0.0&   0.709&0.0 &     0.0&      0.290&0.0 &     0.0&0.0 \\
\  UV&0.0&   0.915&0.0 &     0.0&      0.0841&0.0 &     0.001&0.0   \\
\  X-ray & 0.0 &0.141 &   0.0  & 0.0    & 0.001 &  0.0   & 0.858  & 0.0 \\
\  IR$_d$&0.0&   0.722&0.0 &     0.0&      0.278&0.0 &     0.0&0.0    \\
\ {\bf NGC3256}&-&-6.&-&-&-11.&-&-12.2&-12.5\\
\ opt&0.0&   0.0055&0.0 &     0.0&      0.142&0.0 &     0.0&0.853\\
\  UV&0.0&   0.012 &0.0 &     0.0&      0.0688&0.0 &     0.0&0.919\\
\ X-ray  &  0.0 & 0.0687 &0.0    & 0.0   & 0.0288  &0.0  & 0.834&  0.0687\\
\  IR$_d$&0.0&   0.0235&0.0 &     0.0&      0.571&0.0 &     0.0&0.406\\
\ {\bf NGC3690}&-&-6.&-&-&-11.2&-&-13.&-12.8\\
\ opt&0.0&   0.0105&0.0 &     0.0&      0.171&0.0 &     0.0&0.818\\
\  UV&0.0&   0.023&0.0 &     0.0&      0.0841&0.0 &     0.0&0.893\\
\  X-ray & 0.0  &0.271  & 0.0  & 0.0  & 0.0716 &  0.0  & 0.521  & 0.136  \\
\  IR$_d$&0.0&   0.0401&0.0 &     0.0&      0.613&0.0 &     0.0&0.347\\
\ {\bf NGC4945}&-&-&-10.2&-&-10.7&-&-12.2&-\\
\ opt&0.0&   0.0&      0.27&0.0&      0.72&0.0 &     0.0&0.0\\
\  UV&0.0&   0.0&      0.050&0.0&      0.94&0.0 &     0.0&0.0  \\
\  X-ray &  0.0  &0.0  & 0.0243 &  0.0  & 0.0628 &  0.0  & 0.913  & 0.0\\
\  IR$_d$&0.0&   0.0&      0.82&0.0&      0.17&0.0 &     0.0&0.0    \\
\ {\bf NGC5236}&-&-4.8&-&-&-10.9&0&-12.1&-12.3\\
\ opt&0.0&   0.0538& 0.0&      0.0&      0.11&0.0 &     0.0&0.83\\
\  UV&0.0&   0.10&0.0&      0.0&      0.050&0.0 &     0.0&0.85\\
\  X-ray &  0.0  & 0.477  & 0.0  & 0.0  & 0.0159 & 0.0  & 0.460  & 0.0477\\
\  IR$_d$&0.0&   0.21&0.0&      0.0&      0.41&0.0 &     0.0&0.37\\
\ {\bf NGC5253}&-&-5.1&-&-&-11.7&-&-13.2&-\\
\ opt&0.0&   0.607&0.0&      0.0&      0.39&0.0 &     0.0&0.0\\
\  UV&0.0&   0.87&0.0&      0.0&      0.12&0.0 &     0.0&0.0 \\
\  X-ray & 0.0 & 0.860 & 0.0  & 0.0  & 0.00904  & 0.0  & 0.131  & 0.0\\
\  IR$_d$&0.0&   0.62&0.0&      0.0&      0.37&0.0 &     0.0&0.0  \\
\ {\bf NGC7552}&-&-&-&-11.5&-10.8&-&-13.&-12.7\\
\ opt&0.0&   0.0&      0.0&      0.011&0.29&0.0 &     0.0&0.69\\
\  UV&0.0&   0.0&      0.0&      0.0&0.15&0.0 &     0.0&0.84\\
\  X-ray &  0.0 & 0.0  & 0.0  & 0.0059 & 0.205  & 0.0  & 0.594  & 0.195\\
\  IR$_d$&0.0&   0.0&      0.0&      0.47&0.41&0.0 &     0.0&0.11\\
\hline\\
\end{tabular}

$^a$ in the first row of each galaxy
\end{table*}

\begin{table*}
\centering
\caption{Log($\eta$)$^a$ and the relative contribution of the different clouds, W$_i$ for ULIRGs}
\begin{tabular}{lllllllll}\\ \hline\hline
  & \multicolumn{2}{|c|}{\em under-gal}&\multicolumn{4}{c|}{\em starbursts}&\multicolumn{1}{c|}{\em winds}&
\multicolumn{1}{c|}{\em AGN}\\ \hline
\     &  m0     &   m1   & m2  & m3 & m4  & m5  & m6  &  m7 \\
\ {\bf MRK1014} &-&-7.&-&-&-&-12.9&-12.&-\\
\ opt&0.0&   0.16&0.0 &   0.0&      0.0 &     0.83&0.0044&0.0\\   
\  UV&0.0&   0.33&0.0 &   0.0&      0.0 &     0.65&0.012&0.0  \\  
\  X-ray & 0.0  & 0.0051 & 0.0  & 0.0  & 0.0  & 0.0149  & 0.98  & 0.0 \\
\  IR$_d$&0.0&   0.019&0.0 &   0.0&      0.0 &     0.98&0.0&0.0    \\
\ {\bf UCG5101} &-9.1&-&-&-&-&-13.3&-&-13.7\\
\ opt&0.0033&0.0&  0.0&    0.0&      0.0  &    0.02&0.0&      0.97\\
\  UV&0.0047&0.0&  0.0&    0.0&      0.0  &    0.015&0.0&      0.98\\
\  X-ray & 0.0  & 0.0  & 0.0  & 0.0  & 0.0  & 0.649  & 0.0  & 0.351 \\
\  IR$_d$&0.0078&0.0&  0.0&    0.0&      0.0  &    0.64&0.0&      0.35\\
\ {\bf MRK231}  &-8.&-&-&-&-&-11.8&-&-13.4\\
\ opt&0.016&0.0&  0.0&      0.0&      0.0 &     0.24&0.0&      0.74\\
\  UV&0.024&0.0&  0.0&      0.0&      0.0 &     0.19&0.0&      0.78\\
\  X-ray & 0.0  & 0.0  & 0.0  & 0.0  & 0.0  & 0.967  & 0.0  & 0.033\\
\  IR$_d$&0.0046&0.0&  0.0&      0.0&      0.0 &     0.96&0.0&      0.033\\
\ {\bf MRK273} &-&-&-&-&-&-12.9&-&-13.5\\
\ opt&0.0 &0.0  & 0.0 &   0.0  &    0.0  &    0.031&0.0 &     0.97\\
\  UV&0.0  &0.0 & 0.0 &  0.0  &    0.0  &    0.024&0.0 &     0.97\\
\  X-ray & 0.0  & 0.0  & 0.0  & 0.0  & 0.0  & 0.745  & 0.0  & 0.255\\
\  IR$_d$&0.0  &0.0 & 0.0 &   0.0  &    0.0  &    0.74&0.0 &     0.25\\
\ {\bf Arp 220}&-&-9.1&-&-&-&-13.3&-&-12.9\\
\ opt&0.0&   0.0001&   0.0&   0.0  &    0.0 &     0.007 &     0.0&0.99\\
\  UV&0.0&  0.0003&   0.0&   0.0  &    0.0 &     0.003 &     0.0&0.997\\
\  X-ray & 0.0  & 0.0  & 0.0  & 0.0  & 0.0  & 0.23 & 0.0  & 0.77\\
\  IR$_d$&0.0&  0.001&   0.0&   0.0  &    0.0 &     0.22&   0.0&0.78\\
\ {\bf NCG6240} &-&-&-&-&-&-&-12.4&-13.\\
\ opt&0.0&   0.0&   0.0&   0.0  &    0.0 &     0.0 &     0.0&1.0\\
\  UV&0.0&  0.0&   0.0&   0.0  &    0.0 &     0.0 &     0.0&1.0\\
\  X-ray & 0.0  & 0.0  & 0.0  & 0.0  & 0.0  & 0.0  & 0.960  & 0.0397 \\
\  IR$_d$&0.0&  0.0&   0.0&   0.0  &    0.0 &     0.0 &     0.15&0.85\\
\hline\\
\end{tabular}

$^a$ in the first row of each galaxy
\end{table*}

\begin{table*}
\centering
\caption{Log($\eta$)$^a$ and the IR$_d$ relative contribution of the different clouds, W$_i$ for HLIRGs}
\begin{tabular}{lllllllll}\\ \hline\hline
  & \multicolumn{2}{|c|}{\em under-gal}&\multicolumn{4}{c|}{\em starbursts}&\multicolumn{1}{c|}{\em winds}&
\multicolumn{1}{c|}{\em AGN}\\ \hline
        &m0&  m1  & m2& m3&m4     &m5 & m6 & m7  \\ \hline
\ {\bf F00235+1024} &-  &-7.9& -   & -  &   -    &  -14.6 &   -      & -     \\
\ IR$_d$       &0.&  0.11 & 0.&   0.  &    0. &      0.89&0.&0. \\
\ {\bf 07380-2342}  & - & -7.9  & -  &  -     &  -     & -13.7    &-  &-  \\
\ IR$_d$      &0.&  0.01 & 0.&   0.  &    0. &      0.98&0.&0.\\
\ {\bf F10026+4949} & - & -8.   & -  & -      & -      &  -14.7   &-  &-  \\
\ IR$_d$      &0.&  0.11 & 0.&   0.  &    0. &      0.89&0.&0. \\
\ {\bf F12509+3122} &-  & -6.7  & -  &  -     & -      & -15.     &-  &-  \\
\ IR$_d$      &0.&  0.84 & 0.&   0.  &    0. &      0.16&0.&0. \\
\ {\bf 13279+3401}  &-  &  -6.8 & -  &  -     & -      &  -12.8   &-  &-  \\
\ IR$_d$      &0.&  0.05 & 0.&   0.  &    0. &      0.95&0.&0. \\
\ {\bf 14026+4341}  & - &  -6.5 &  - &  -     &   -    &   -13.5  &-  &-  \\
\ IR$_d$      &0.&  0.21 & 0.&   0.  &    0. &      0.79&0.&0. \\
\ {\bf F14218+3845} &  -& -8.   & -  &   -    &   -    &  -14.6   &-  &- \\
\ IR$_d$       &0.&  0.09 & 0.&   0.  &    0. &      0.91&0.&0. \\
\ {\bf F16124+3241} &-  & -8.4  & -  &  -     &  -     & -14.6    &-  &-   \\
\ IR$_d$   &0.&  0.04 & 0.&   0.  &   0. &      0.96&0.&0. \\
\ {\bf EJ1640+41}   &-  &  -8.4 & -  &  -     &  -     &  -14.4   &-  &-  \\
\ IR$_d$       &0.&  0.02 & 0.&   0.  &    0. &      0.98&0.&0. \\
\ {\bf 18216+6418}  & - & -6.   & -  &  -     & -      & -12.5    &-  &-  \\
\ IR$_d$      &0.&  0.07 & 0.&   0.  &    0. &      0.93&0.&0. \\
\hline\\
\end{tabular}

$^a$ in the first row of each galaxy

\end{table*}

\begin{table*}
\centering
\caption{The  distances  of the clouds (in pc) from the center  of each galaxy}
\begin{tabular}{lllllllllll}\\ \hline\hline
 &\multicolumn{2}{|c|}{\em under-gal}&\multicolumn{4}{c|}{\em starbursts}&\multicolumn{1}{c|}{\em winds}& \multicolumn{1}{c|}{\em AGN}&\multicolumn{1}{c|}{\em radio}\\

        &m0&  m1  & m2& m3&m4     &m5 & m6 &m7&  m$_{syn}$  \\
\hline
\ LIRGs :&&&&&&&&\\
\ NGC 253    &-     &220      &    -  & - &20.4    &      &3.63  &2.9 & 141.3\\
\ IC 342    &-      &-& 1.4   &-& 1.26& -&0.56  &0.28  &6.3  \\
\ II Zw 40  &-      &52      &-&-&6.6  &-&-&-  &-\\
\ M82       &-      &85       &-&-&10.7 &-&4.8&-   & 141. \\
\ NGC 3256  &-      &370         &-&-&85.1   &-&30.2  &21.4   &631.\\
\ NGC 3690  &-      &420         & -&-& 100. & - &13.2& 16.6    &933.  \\
\ NGC 4038/4039  &-      & -&-&-&-&-&12.3    & -     & 275.4\\
\ NGC 4945  &-      & -&59.6   &-&33.1   & -&5.95   &-         &331. \\
\ NGC 5236  &-      &270      & -&-&24.5  & - & 6.16  &4.9             &173.7\\
\ NGC 5253  &-      &150       & - & - &7.58   & - &1.35   &-          & -\\
\ NGC 7552  &-      & -&-& 37.6  & 84.1  &-& 6.68&9.44         &237.1\\
\ ULIRGs :&&&&&&&&\\
\ Mrk 1014       & -         & 2000.      &-&-&-&229. & 653.    &58.2         &518.8\\
\ UGC 5101       & 45       &-      &-&-&-&35.5   &-     &22.38             &-\\
\ Mrk 231        &170        &-       &-&-&-& 211.  &-      &33.5           &1.7e3\\
\ Mrk 273        & -         &-       &-&-&-& 52.5     &-  & 26.3         &1.05e3 \\
\ Arp 220        &-           &2029.  &-&-&-&16.11        &-       & 25.          &-\\
\ NGC 6240       & -         &-       &-&-&-&-& 60.25  & 30.2  &-\\
\ HLIRGs : &&&&&&&&\\
\ F00235+1024  &-  &2600       &-&-&-&113.5&-& -&-\\
\ 07380-2342   &-  &1300       &-&-&-&164. &-&-&-\\
\ F10026+4949  &-  &4500     &-&-&-& 199.&-&-&-\\
\ F12509+3122  &-  &14000     &-&-&-& 100.  &-&-&-\\
\ 13279+3401   &-  &6760       &-&-&-& 676.&-&-&-\\
\ 14026+4341   &-  &7200     &-&-&-&229.  &-&-&-\\
\ F14218+3845  &-  &4800     &-&-&-&245. &-&-&-\\
\ F16124+3241  &-  &2100     &-&-&-&173.8&-&-&-\\
\ EJ1640+41    &-  &2780       &-&-&-&278. &-&-&-\\
\ 18216+6418   &-  &11800      &-&-&-&668. &-&-&-\\
\hline\\
\end{tabular}
\end{table*}

The results of modelling of the continuum SEDs of LIRGs, ULIRGs, and of HLIRG   are presented in the 
following sections. 
For most objects one type of clouds is  not enough to explain the whole dataset from  FIR to  
near-IR.

\subsection{LIRGs}

The modelling  of  LIRG   
is shown in Fig. 5. The data (see Appendix) come from the NED.
These data are generally corrected for galactic reddening, however, background stars are included
in the galaxy field and corrections  have not been applied 
(see e.g. Code \& Welch, 1982).

 A deep feature at $\sim$ 10 \mum\ indicates strong absorption by silicates, particularly for
in NGC 253, IC 342, M 82, NGC 5236, and NGC 5253.
For NGC 5236 and NGC 7552 there is also evidence of  PAH emission in the mid-IR.

The wide peak in the IR, observed in all  the galaxies except 
 NGC 4038/4039, can be explained by the sum of at least two models corresponding to 
different shock velocities. 
The  emission  which peaks   longward of 95 \mum ~ corresponds to low-velocity shocks (\Vs\ = 100-200 \kms), 
which leads to maximum grain temperatures of $\sim$ 50 K, while the emission 
which peaks at  $\sim$ 30 \mum ~corresponds to 
relatively high velocity shocks ($\geq$ 300 \kms) with maximum grain temperatures of 
$\sim$ 120 K.  
We have tried to obtain an agreement with the data within the observational error ($\leq $ 30 \%).
For some galaxies the data are  insufficient to constrain the models  and  clouds  with \Vs = 500 \kms
 cannot be excluded. However,  they   correspond to lower relative contributions.
These results are consistent with  those found for  starbursts
and LIRGs by  Contini \& Contini (2003). 

Fig. 5 shows that the IR reradiation by dust of 
NGC 253, IC 342, NGC 3256, NGC 3690,  NGC 5236, and NGC 7552 can be  also modelled 
by a unique model
corresponding to a AGN with \Vs\ = 1000 \kms, \n0\ = 1000 \cm3, and a power-law flux with 
log\Fh\ = 11 (model m7) as was found in  Arp 220 (Contini et al. 2004, Fig. A1).
Indeed, the IR emission extends all over the IR wavelength range because the grains
are heated to high temperatures in the immediate post-shock region. The grains are large
enough ($a_{\rm gr}$ = 1 \mum) to survive sputtering,
and they  are heated to temperatures of $\sim$ 50 K by the radiation flux from the active center 
on the opposite side of the cloud (Fig. 2).
Consider for instance NGC 253, besides the good    agreement to the small aperture data  
by the  AGN model  in the optical range and  by the low velocity 
model  (m1) in the UV, we   must account also for the data 
observed with large apertures which are explained  by a starburst model (m4).

Synchrotron emission in the radio is observed in all the galaxies,
except IC 342, II ZW 40, and NGC 5253. 
Free-free  continuum absorption  is evident in 
   a few objects (NGC 3256, NGC 3690, NGC 7552).

The hybrid nature of LIRGs (starburst+AGN)  appears   in  about half of the objects.
(see Table 3).

\subsection{ULIRGs}

Rigopoulou et al. (1999) present, in their Table 1, ISO PHOT-S spectroscopy for a 
sample of ULIRGs. We have selected those which have a 
rich number of data from the  NED (see Appendix), and we have added the two data points at 
5.9 and 7.7 \mum\ reported by Rigopoulou et al. (1999).

In the absence of MIR spectroscopy
for ULIRGs and HLIRGs  
 we assume the same  grid of models used for LIRGs
are suitable for modelling systems of higher luminosity.
The  results  are shown in Fig. 6.

For all the galaxies of the ULIRG sample,  except Mrk 1014, the low $d/g$ clouds  (m1)
which significantly contribute in the LIRGs are not found,
while an AGN nucleus is revealed  by  modelling  the IR  prominent flux.
Free-free continuum self-absorption  is evident in NGC 1014  and Mrk 231.

\subsection{HLIRGs}

At the brightest end of the luminous IR galaxy population lie the HLIRGs, 
with L$_{\rm IR}>10^{13}$ h$^{-2}_{65}$ L$_{\odot}$ (Rowan-Robinson 2000; Farrah et al. 2002;
Sanders \& Mirabel 1996; Tran et al. 2001; Verma et al. 2002).
They claim that these objects may reveal an entirely new class of IR-emitter objects.

Farrah et al. (2002) present sub-millimetre photometry for eleven HLIRGs and use models 
by Efstathiou, Rowan-Robinson, \& Siebenmorgen (2000) for starbursts and by 
Efstathiou \& Rowan-Robinson (1995) for AGNs to examine the nature of the IR emission.
Dust composition for starbursts is given by the dust grain model of Siebenmorgen \& Krugel (1992)
and by the multigrain dust model of Rowan-Robinson (1992).
Farrah et al. (2002) found that AGN and starburst luminosities correlate, suggesting 
that a common physical factor, plausibly the dust mass, governs the luminosities of starbursts 
and AGNs in HLIRGs.

The models we use here are different due to the important role of collisional processes.
The results of our modelling are shown in Fig. 7. We found that the IR  flux can be explained 
by two types of clouds, one corresponding to low \Vs\ - \n0 values, low $U$, and low $d/g$, (m1)
while the second type of clouds corresponds to rather high velocities \Vs\ = 500 \kms,
high $d/g$ values, and a high ionization parameter $U$, (m5). 
The models correspond to $T_{\rm d}$(max) 
of $\sim$ 50 K and 120 K, respectively. 
The  X-ray  upper limit for the  HLIRG  F00235+1024 (Wiman et al 1998)
integrated in the 0.1-2.4 keV band, is well explained by  model m5
calculated by \Vs=500 \kms.
Free-free continuum absorption  is evident in P070380-2342 and  P18216+6418.

\section{Discussion}

\begin{table*}
\centering
\caption{Dust-to-gas ratios (in units of $4\times 10^{-4}$) in each galaxy}
\begin{tabular}{lllllllllll}\\ \hline\hline
  & \multicolumn{2}{|c|}{\em under-gal}&\multicolumn{4}{c|}{\em starbursts}&\multicolumn{1}{c|}{\em winds}& \multicolumn{1}{c|}{\em AGN}\\
           &    m0&m1  & m2    & m3   &m4   &m5  & m6  &m7 \\
\hline
LIRGs :&&&&&&&&\\
NGC 253  &-    & 0.70&   -           & -    & 100     & - & 100 &    200\\
IC 342      &-    &-       & 100            &-     & 100     & -   & 100      & 250\\
II Zw 40   &-    &0.75    &-               & -    & 158     &  -     &-       &-\\
M82       &-    &6.    & -              & -    & 340    &-    & 100      &-\\
NGC 3256   &-    & 4.7  &-               &-     &200   &-    & 100      &63.\\
NGC 3690   &-    &3    & -              &-     & 100     & -   & 100      & 79.\\
NGC 4038/9 & -    & -      &-                &-    &-      &-    &100       &-\\
NGC 4945    &-    & -      & 100              &-     &100      & -   & 100     &-\\
NGC 5236    &-    & 0.75 & -              &-     &100      & -   & 100     &126.\\
NGC 5253   &-   & 0.24 & -               & -   & 100     &   -     & 100    &-\\
NGC 7552  &-    & -      &-               & 8. &20. &-    &100      &79.  \\
ULIRGs : &&&&&&\\
Mrk 1014   & -  &0.95 &-                &-      &-     &5.&-       &-         \\
UGC 5101    &6.  &-    &-                &-       & -   &30.  & -      &100.      \\
Mrk 231   &0.16  &-    &-                &-       &   - &4.&-       &126.      \\
Mrk 273   & -     &-    &-               &-        &-    & 5.&-  &100       \\
Arp 220    &-   &  0.15  &-               &-        &-    &50.&-       &100       \\
NGC 6240  &-      &-    &-               &-        &-    &-& 20 &31.6      \\
HLIRGs :&&&&&&\\
F00235+1024  &-      &2.4 &-               &-       &-    &5.&-      &-\\
07380-2342 &-      &7.5 &-              &-        &-    &50. &-      &-\\
F10026+4949&-      &1.2 &-              &-        &-    &100.    &-     &-\\
F12509+3122 &-     &0.06 &-             &-        &-    &100. &-     &-\\
13279+3401 &-     &0.19 &-             &-        &-    &0.6 &-     &-\\
14026+4341  &-  &0.06 &-             &-        &-    &6.3&-     &-\\
F14218+3845&-     &0.95 &-             &-        &-    &4. &-     &-\\
F16124+3241 &-      &1.9  &-             &-        &-    &16. &-     &-\\
EJ1640+41 &-&0.95 &-             &-        &-    &5. &-     &-\\
18216+6418 &-   &0.04 &-             &-        &-    &0.8 &-     &-\\
\hline\\
\end{tabular}

\flushleft
$^a$ adopting H$_0$=75 \kms Mpc$^{-1}$

$^b$ 1-1000 \mum ~ luminosities obtained from the best-fitting combined starburst-AGN models (Farrah et al. 2002, table
 3)

\end{table*}

\subsection{The relative importance of the different energy  sources in each galaxy}

The $\eta$ factors (Sect. 3.5.2)  derived  by modelling  the SED,
allow  to calculate the relative contribution 
 of  AGNs,  starbursts,
 high velocity winds, and   the underlying galaxy,
 in the different frequency ranges $\Delta_{\nu}$
(${radio_{br}}$,
 ${IR_{br}}$,
 ${opt_{br}}$,
 ${UV_{br}}$,
 ${X_{br}}$,
 ${IR_d}$,
 Table 2, bottom) 
 for each galaxy, by
$W_i$ = ($F_{\Delta_{\nu}}$)$_i$ ($\eta$)$_i$/ $\Sigma$ ($F_{\Delta_{\nu}}$)$_i$ ($\eta$)$_i$ .
  They are given in Tables  3 and 4 for LIRGs and ULIRGs,  respectively.
The IR bremsstrahlung is lower by  a large factor than  reradiation from dust,
therefore  is neglected.

The {IR$_d$} contribution 
 from the starburst  clouds with different velocity is  generally high,  except for M82 which 
shows a large relative contribution from the background galaxy.
The strong  IR$_d$ from the starburst  dust relative to  AGN  depends on  
the different character of the flux from the external source: a power-law in the AGN
and a multiple black body in the starburts.
Tables 3  and 4  show, on the other hand,  that when an AGN   is present, its 
contribution prevails
in the optical and UV range, while in the X-ray the high velocity  winds 
dominate. 

Tables 3   shows that
relatively low \Vs - \n0\ clouds (\Vs\ $<$ 200 \kms)  dominate in
  M82 and NGC 5253. Starburst clouds with \Vs $>$ 300 \kms\ are less indicated
in LIRGs.
An AGN  contributes to the IR  by $\sim$ 23 \% in NGC 253  and IC 342, 
by $\sim$ 40\% in NGC 3256, NGC 3690, and NGC 5236. In ULIRGs (Table 4)
the AGN is always present (except for Mrk 1014) and dominates by $\geq $ 75 \% 
in Arp 220 and NGC 6240.

In Table 5 the relative contribution  to the IR  fluxes are given for HLIRGs. Most of
the data in Fig. 6   are upper limits, so we focus on the IR.
HLIRGs are modelled by two types of clouds. The relative  contribution to IR$_d$ of clouds with 
\Vs=500 \kms\ (m5) is  $\geq$  90\%  for all the galaxies of the sample,  except for 
 F12509+3122 which shows a large contribution from the underlying galaxy (m1).

\subsection{The distance of the emitting clouds from the galaxy center}

We have found  by modelling the SEDs that   the $\eta$ factors  corresponding to 
models  m0 and m1 are higher than those corresponding to models
m2-m7 by factors of $10^6-10^7$. This  can be explained by larger distances of the clouds from the
galaxy centre, and/or by a large number ($N$ $\sim$ 1/\ff\ where \ff\ is the filling factor)
of low \Vs - \n0\ clouds.
The explanations are both valid. Indeed, low \Vs-\n0\ clouds (models m0 and m1) with
$d/g$ ratios  even lower than those of the   Galactic  ISM (Table 2) are characteristic 
of the extended underlying galaxy and
low \Vs\ ($<$ 200 \kms)  - \n0\ ($\sim$ 100 \cm3) clouds were found in the outer narrow line
region of AGNs which may also contain starburst regions (Contini et al. 2002). 

 We  define  $\eta$=(d$_c$/d)$^2$ (1/$\it ff$),
where d is the distance of the galaxy to Earth and d$_c$ the distance of  clouds from 
the galaxy  center.
Fragmentation  by  shocks leads to  clouds
  with different geometrical thickness, $D$ (Table 2), particularly low
 in the underlying galaxy.
We   adopt  \ff $\sim$  $D_{min}$/$D_{max}$ = 10$^{-4}$
($D_{min}$ and $D_{max}$ are the minimum and maximum geometrical thickness 
of the clouds  in Table 2)
for the underlying  galaxy clouds (m0 and  m1),  and  \ff $\sim$1  for the starburst 
and  AGN clouds (m2-m7).

Models m6 and m7 with a high shock velocity (\Vs=1000 \kms) explain the observed soft X-rays,
which are upper limits.
Dust  emission  calculated by  model m7, which appears in most of the ULIRGs, explains
the   mid- to far-IR flux observed in some galaxies.
 The  $\eta$  of   models m6 and m7  are relatively low
indicating that the high \Vs\ clouds are located closer to the centre.

The distances of the clouds from the galaxy center are shown in Table 6.
We have found by fitting the continuum SED of  LIRGs that a  
large number of low \Vs, \n0\ clouds are
located in  the outskirt of the galaxies.
Even adopting the low \ff\ indicated above, the  distance of these clouds  from the center results
 higher by a factor  $\geq$ 10  than 
that of clouds corresponding to higher velocities and densities.  
The large regions of low \Vs - \n0\ clouds with low $d/g$ are less
evident  in ULIRGs, as  high and low velocity clouds coexist
because  merging leads to fragmentation and mixing.  
Indeed, the galaxies selected from the Rigopoulou et al. (1999) sample  are few but they all show
an irregular morphology due to merging (e.g. Leech et al 1994). 
Table 6 shows that the average distances of the  starburst clouds from the centre is  similar for
LIRGs and ULIRGs,  within 100 pc for almost all galaxies, except for Mrk 1014. 
The  luminous IR galaxies of the Verma et al sample, with L$_{IR}$ $> 10^{11}$ L$_{\odot}$,
  NGC 3256 and NGC 3690  which are identified as a colliding pair and  a 
peculiar merger,  respectively,  show starburst cloud distances from the centre $\sim$ 100 pc.
There are not enough data to model the NCG 4038/4039 interacting complex.

  HLIRG redshifts (Farrah et al. 2002, table 1 and references therein)  are  higher 
than those of LIRGs, 
leading to distances  larger by  factors of 100- 1000.
We have found 
(Table 6, bottom)  that  d$_c$ should be 
higher for HLIRGs than for LIRGs by about a factor  $>$ 10 in average, 
adopting  the same \ff\ which were  used for  the other  galaxy samples.
To recover the dimensions of the extended region of  the other galaxies,
lower  \ff\ should be adopted.
It seems that Mrk 1014  with log (L$_{IR}$/L$_{\odot}$) = 12.5
has characteristics similar to HLIRGs. 

Regions   of 200-900 pc are generally   characteristic of the NLR of objects such as
QSOs, Seyfert 1, and narrow line galaxies.  This indicates that also 
HII regions are located within the  NLR of these galaxies.
Synchrotron emission in the radio range corresponds to  larger distances from
the centre, which is reasonable enough considering that it is generally created in
the lobes of extended jets.

\subsection{The dust-to-gas ratios}

The $d/g$ ratios relative to those  of the Galactic ISM ($\sim 4\times 10^{-4}$ by mass)
 for each type of clouds in each galaxy are given in Table 7.
More particularly, $d/g$   as high as 0.1  were found in  
 M 82,  and $d/g$ $\geq$ 0.04 in the starburst   clouds of most of the LIRGs.
This  may indicate    that dust in the starburst regions  of LIRGs
is of SN origin. Indeed high velocity winds were found modelling the X-ray in Figs. 5.
SN corresponding to explosions of 10-30 \msol stars, 
  are potentially the most important source of interstellar dust (Dwek 2004).
The total mass of condensable elements (C, Mg, Si, S, Ca, Fe, Ni and associated O)
ejected by a  30 \msol mass star is 0.3-2 \msol leading to  $d/g$ $\sim$ 0.01-0.07.
Dust from red giants and AGB is less suitable to high velocity clouds.
For comparison, $d/g$ between $\sim$ 2.5 10$^{-3}$ and 6.6 10$^{-3}$  
were determined for
actively star-forming galaxies by Calzetti (2000) and for  PG  quasars by   
Haas et al (2003)  adopting the Hildebrand (1983)  classical method.

In  Table 8  dust masses, $M_{\rm d}$, calculated in the present work  are compared with
those calculated by the models of Farrah et al. (2002). Discrepancies within a factor $\leq$ 3
are found for all the galaxies of the sample considering the errors,
except for F10026+4949 and 18216+6418  where the discrepancies are  greater than a factor of  10.
The discrepancies depend on the fact that the dust masses were calculated by Farrah et al.
with the Hildebrand (1983) method.
In 18216+6418 we predict very low $d/g$,  because large  slabs of gas are deduced
from the  strong   self-absorption of free-free radiation at  long wavelengths (Fig. 7, last diagram).
Adopting  a temperature T=1000 K  and  a density n=10$^4$ \cm3 downstream  of 
 clouds in physical conditions corresponding to   \Vs=500 \kms and \n0=300 \cm3,
the gas is optically thick at $\nu$ $\leq$  300 GHz for a layer  $\geq$ 242.2 pc 
thick (Osterbrock 1988),
which is actually of the order of the calculated  radius of the NLR  (Table 6).

\section{Summary}

In this work we have modelled consistently the IR dust emission and
the bremsstrahlung.  
From the analysis of  dust-to-gas ratios in galaxies with different IR luminosities,
we conclude that
the high infrared  luminosities generally  depend on  high  amount  of dust relative to  
gas  close to the galaxy centre.  
This is  revealed by  the  continuum  emitted
downstream of  relatively  high  velocity clouds.
Relatively  high $d/g$ (0.04 by mass) were found in the LIRG clouds 
with \Vs\ = 300-500 \kms,
higher   by a factor of $\sim$ 100  than in the ISM ($\sim$ 4  10$^{-4}$).

Half of the LIRGs of the sample contain an AGN, which is found in almost all  ULIRGs
and in none of the HLIRGs.

Interestingly, we could not  demonstrate that higher  $L_{\rm IR}$ as those observed in HLIRGs
correspond to higher dust-to-gas ratios. 
On the contrary, the  highest $d/g$  appear in LIRGs.
Notice that the HLIRG types correspond to narrow-line galaxies, Seyfert 1, and QSO.
It was found in previous modelling (Contini et al. 2002) that $d/g$ ratios in the narrow 
line region of
these galaxies are lower by a factor of $\sim $ 10 than for Seyfert 2 and LINERs.
It seems that in  some HLIRGs the high IR luminosities include the contribution of bremsstrahlung 
from cool gas.
The relatively high bremsstrahlung  and  the  large distances  of the emitting clouds from the galaxy centre
suggest that  the  HLIRGs of the  Farrah et al sample   were selected  among large, massive objects.

\begin{table}
\caption{Comparison of $M_{\rm d}$ calculated in the present work with $M_{\rm d}$ 
reported by Farrah et al. (2002) in HLIRGs}
\begin{tabular}{lllllllll}\\ \hline
        & $M_{d}^1$&$M_d$(Farrah et al.) \\
        &(10$^8$\msol) &(10$^8$\msol) \\
\hline
F00235+1024 &1.86&1.95  \\
07380-2342 &17.8& 6.17\\
F10026+4949&35.&$<$ 5\\
F12509+3122  &6.4 &1.58\\
13279+3401 &0.22&-  \\
14026+4341&2.0 &1.74   \\
F14218+3845&1.4& 1.07\\
F16124+3241 &6.1& 3.47\\
EJ1640+41 &1.9& $<$1.58\\
18216+6418  &0.3    & 3.24\\
\hline\\
\end{tabular}

\flushleft
$^1$ adopting $M_{\rm gas} = 10^{11}$ \msol\  (Farrah et al. 2002)

\end{table}

\section*{Acknowledgments}

We are very grateful to an anonymous referee for helpful and valuable comments.
We thank R. Angeloni for interesting conversations.
This paper makes use of observations by the "Two micron all sky survey team"
from  NASA/IPAC Extragalactic Database.

\section*{Appendix : The references for the NED data}

\small{

Fabbiano et al. (1992),
Brinkmann et al. (1994),
Code \& Welch (1982),
De Vaucouleurs et al  (1991), 
Lauberts \& Valentijn (1989),
Spinoglio  et al. (1995),
Glass (1973),
Aaronson (1977),
Jarrett et al. (2003),
Aaronson et al. (1980),
Rieke et al. (1975),
Rieke \& Lebofsky (1978),
Rieke \& Low (1972),
Rice  et al. (1988),
Soifer et al. (1989),
Moshir et al. (1990),
Rieke et al.  (1973),
Hildebrand et al. (1977),
Chini et al. (1994),
Gelzhaler \& Witzel (1981),
Schimmins \& Wall (1973),
Wright et al. Otrupcek (1990),
Whiteoak (1970),
Kuhr et al. (1981),
Wall et al. (1976),
Condon et al. (1998),
Large et al. (1991),
Douglas et al. (1996),
Slee (1995),
Israel et al.  (1990),
Zwicky \& Herzog (1968),
Becklin  et al. (1980),
Becker et al. (1991),
Gregory et al.  (1991),
Gregory \& Condon (1991),
White  \& Becker (1992),
Thuan (1983),
Heisler et al. (1996),
Huchtmeier  \& Richter (1989),
Rifatto et al. (1995),
Donas et al. (1987),
Johnson (1966),
Kleinmann  \& Low (1970a),
Kleinmann \& Low (1970b),
Rieke  et al. (1980),
Golombek  et al. (1988),
Laing  \& Peacock (1980),
Genzel et al. (1976),
Kuhr (1980),
Pauliny-Toth  et al. (1978),
Kellermann  et al. (1969),
Kuhr et al. (1981),
Condon (1983),
White \& Becker (1992),
Pauliny-Toth  \& Wade (1964),
Cohen et al. (1977),
Kinney et al. (1993),
Gregory  et al. (1994),
Huchra (1977),
Joyce  \& Simon (1976),
Allen (1976),
Lebofsky \& Rieke (1979),
Soifer et al. (2001),
Carico  et al. (1972),
Griffith et al. (1994),
Mathewson \& Ford (1996),
Wright  et al. (1994),
Jones \&  McAdam (1992),
Mauch et al.  (1992),
Mauch  (2003),
Dyck  et al. (1978),
De Vaucouleurs \& Longo (1988),
Glass (1976),
Ward et al. (1982),
Roussel et al. (2001),
Frogel et al. (1982),
Wright et al. (1966),
White et al. (2000),
Moorwood \& Glass (1982),
Rieke (1976),
Kleinmann \& Wright (1974),
Knapp (1994), 
Griersmith  et al. (1982),
Surace \& Sanders (2000),
Maddox et al. (1990),
Scoville et al. (2000),
Neugebauer et al. (1987),
Haas  et al. (2003),
Sanders  et al. (1989),
Barvainis \&  Antonucci (1989),
Barvainis  et al. (1996),
Rigopoulou et al. (1996),
Maiolino et al. (1995),
Benford (1999),
McAlary et al. (1979),
Joyce (1975),
Rieke (1976),
Rieke (1978),
Stein \&  Weedman (1976),
Klaas et al. (2001),
Patnaik et al. (1992),
Sramek \&  Tovmassian (1986),
Zwicky \& Herzog (1966),
Zwicky \& Herzog (1963),
Eales et al (1989),
Dunne  et al. (2000),
Chini  et al. (1986),
Condon  et al. (1983),
Dressel \& Condon (1978),
Waldram  et al. (1996),
Zwicky   et al. (1961),
Rudy  et al. (1982),
Griffith  et al. (1995),
Gower  et al. (1967),
Becker  et al. (1995),
}

\end{document}